\documentclass[a4paper,11pt]{article}
\usepackage{jcappub} 
\usepackage{lipsum}
\usepackage{url}

\usepackage[T1]{fontenc}
\usepackage{subcaption}
\usepackage{xspace}
\usepackage{booktabs}
\usepackage{multirow}
\usepackage{rotating}

\usepackage{listings}
\newcommand{\ucat}{\texttt{galsbi.ucat}\xspace}
\newcommand{\ufig}{\texttt{UFig}\xspace}
\newcommand{\galsbi}{\texttt{galsbi}\xspace}
\newcommand{\sextractor}{\texttt{SExtractor}\xspace}

\usepackage{xcolor}
\usepackage{pifont}

\newcommand{\greentick}{\textcolor[HTML]{34BA09}{\ding{51}}}

\arxivnumber{2412.08701} 
\title{GalSBI: Phenomenological galaxy population model for cosmology using simulation-based inference}

\author[a,1]{Silvan~Fischbacher,} 
\note[1]{Corresponding author.}
\author[b]{Tomasz~Kacprzak,}
\author[c]{Luca~Tortorelli,}
\author[a]{Beatrice~Moser,}
\author[a]{Alexandre~Refregier,}
\author[c]{Patrick~Gebhardt,}
\author[c,d]{Daniel~Gruen}

\vspace{0.2cm}

\affiliation[a]{Institute for Particle Physics and Astrophysics, ETH Zurich, Wolfgang-Pauli-Strasse 27, CH-8093 Zurich, Switzerland}
\affiliation[b]{Swiss Data Science Center, Paul Scherrer Institute, Forschungsstrasse 111, 5232 Villigen, Switzerland}
\affiliation[c]{University Observatory, Faculty of Physics, Ludwig-Maximilian-Universit{\"a}t M{\"u}nchen,
Scheinerstrasse 1, 81679 Munich, Germany}
\affiliation[d]{Excellence Cluster ORIGINS, Boltzmannstr. 2, 85748 Garching, Germany}

\emailAdd{silvanf@phys.ethz.ch}

\abstract{
We present GalSBI, a phenomenological model of the galaxy population for cosmological applications using simulation-based inference.
The model is based on analytical parametrizations of galaxy luminosity functions, morphologies and spectral energy distributions.
Model constraints are derived through iterative Approximate Bayesian Computation, by comparing Hyper Suprime-Cam deep field images with simulations which include a forward model of instrumental, observational and source extraction effects.
We developed an emulator trained on image simulations using a normalizing flow.
We use it to accelerate the inference by predicting detection probabilities, including blending effects and photometric properties of each object, while accounting for background and PSF variations.
This enables robustness tests for all elements of the forward model and the inference.
The model demonstrates excellent performance when comparing photometric properties from simulations with observed imaging data for key parameters such as magnitudes, colors and sizes.
The redshift distribution of simulated galaxies agrees well with high-precision photometric redshifts in the COSMOS field within $1.5\sigma$ for all magnitude cuts.
Additionally, we demonstrate how GalSBI's redshifts can be utilized for splitting galaxy catalogs into tomographic bins, highlighting its potential for current and upcoming surveys.
GalSBI is fully open-source, with the accompanying Python package, \galsbi\footnote{\url{https://cosmo-docs.phys.ethz.ch/galsbi/}}, offering an easy interface to quickly generate realistic, survey-independent galaxy catalogs.
}

\begin{document}
\maketitle
\flushbottom

\section{Introduction}
\label{sec:introduction}
Cosmological large-scale structure surveys probe the late Universe by measuring the correlations of the positions and shapes of millions of galaxies.
Stage-III surveys such as the Dark Energy Survey \cite[DES,][]{dark_energy_survey_collaboration_dark_2016}, the Kilo-Degree Survey \cite[KiDS,][]{de_jong_kilo-degree_2013}, and the Hyper Suprime-Cam Subaru Strategic Program \cite[HSC,][]{aihara_hyper_2018} have provided competitive constraints on cosmological parameters, showcasing the immense potential of large-scale structure studies.
In the near future, Stage-IV surveys, such as the Vera Rubin Observatory's Legacy Survey of Space and Time \cite[LSST,][]{the_lsst_dark_energy_science_collaboration_lsst_2021}, Euclid \citep{laureijs_euclid_2011}, and the Nancy Grace Roman Space Telescope \cite[NGRST,][]{spergel_wide-field_2015}, will start a new era by increasing the statistical power of these observations by orders of magnitude.
However, this dramatic increase in data volume comes with a corresponding rise in the complexity of systematic challenges that must be addressed since their errors will dominate over the statistical uncertainties.

One of the key systematic uncertainties is the estimation of photometric redshift distributions \cite[see][for reviews]{salvato_many_2019,newman_photometric_2022}.
While Stage-IV surveys have much tighter requirements on the photometric redshift uncertainties (e.g.\ \cite{the_lsst_dark_energy_science_collaboration_lsst_2021,fischbacher_redshift_2023}), the challenge of accurately estimating redshifts for more distant galaxies grows due to limited spectroscopic calibration data.
Even current Stage-III surveys struggle with redshift estimation at higher redshifts.
For instance, DES had to discard the two highest redshift bins in the MagLim sample due to poor fits \citep{des_collaboration_dark_2022}, and HSC applied flat priors on the shifts of their highest two redshift bins after detecting signs of systematic bias in their redshift calibration \citep{li_hyper_2023,dalal_hyper_2023}.
Given the vast amount of cosmological information carried by high-redshift galaxies, addressing these challenges is critical to fully exploit the unprecedented data volume that upcoming surveys will provide.

Furthermore, the galaxy density per unit area will significantly increase for upcoming surveys, posing additional challenges such as source blending.
Source blending occurs when light from multiple objects overlaps, making it difficult to separate individual sources (see \cite{melchior_challenge_2021} and references therein).
This is particularly challenging for galaxy surveys, as galaxies, unlike stars, are extended objects with non-trivial light profiles rather than point sources.
While source blending is already a notable issue for Stage-III surveys, it is expected to become even more problematic for the deeper Stage-IV surveys (see e.g.\ \cite{samuroff_dark_2018,bosch_hyper_2018,sanchez_effects_2021}).
If blending is not properly accounted for, it affects the precision of shape measurements, galaxy photometry and photometric redshift estimation (see e.g.\ \cite{samuroff_dark_2018,huang_characterization_2018,dawson_ellipticity_2016,hoekstra_study_2016,euclid_collaboration_euclid_2019,maccrann_y3_2021,nourbakhsh_galaxy_2022,sanchez_effects_2021}), and if blended galaxies are rejected from the analysis, the statistical power of surveys is strongly limited \citep{chang_effective_2013}.
A way to statistically mitigate blending effects is the use of simulated images where the number of sources, their position, shape and size is known.
However, this does not only require realistic modelling of the instrumental effects, but also an accurate galaxy population model which describes the distribution of magnitudes, sizes, shapes and light profiles.

Several approaches have been developed to address these challenges through image simulations.
\cite{li_kids-legacy_2023} introduced the SKiLLS image simulation suite for the KiDS-Legacy analysis, building upon previous work on image simulations within the KiDS collaboration \cite{fenech_conti_calibration_2017, kuijken_fourth_2019}.
The galaxy catalogs are generated using the SURFS N-body simulation \citep{elahi_surfs_2018} and the semi-analytical \textsc{Shark} model of galaxy evolution \citep{lagos_shark_2018}.
The corresponding nine-band photometry is obtained with \textsc{ProSpect} \citep{robotham_prospect_2020}.
The image simulations incorporate all relevant observational systematics, including background variations and the point-spread function (PSF) and are used to simultaneously calibrate photometric redshifts and shear measurements.

Rather than relying on external models for the galaxy population, image simulations themselves can be used to constrain the population model.
\cite{carassou_inferring_2017} demonstrates on synthetic data how a parametric galaxy population model, based on luminosity functions, can be inferred using image simulations.
However, image simulations are often computationally expensive, and constraining a high-dimensional model can quickly become infeasible.
To address this, fast image simulators are essential.
\cite{herbel_redshift_2017} constrains the galaxy population model using \ufig \citep{berge_ultra_2013}, employing Approximate Bayesian Computation (ABC) as implemented by \cite{akeret_approximate_2015}.
The parametric galaxy population model from \cite{herbel_redshift_2017} was further refined in \cite{tortorelli_measurement_2020,tortorelli_pau_2021,moser_simulation-based_2024}, leading to the developments presented in this work.

The development of galaxy population models extends beyond improving image simulations to strengthening the connection between galaxy evolution and cosmology.
These models enable accurate uncertainty propagation into cosmological observables, facilitate photometric redshift estimation, and allow for a detailed study of selection effects.

A key aspect of accurate galaxy population modelling is the precise representation of spectral energy distributions (SEDs).
These are typically modelled with templates \citep{blanton_k-corrections_2007,brown_atlas_2014} or by using stellar population synthesis (SPS) codes (e.g.\ \texttt{FSPS}; \cite{conroy_propagation_2009,conroy_propagation_2010}).
SPS-based SEDs are more flexible than templates, as they model various emission sources within a galaxy, including stars, gas, dust or active galactic nuclei (AGNs).
Examples of such SPS-based models include \texttt{Prospector-$\alpha$} \citep{leja_deriving_2017,leja_hot_2018,leja_older_2019,johnson_stellar_2021} and \texttt{Prospector-$\beta$} \citep{wang_inferring_2023,wang_quantifying_2024}.

Given that SPS codes can be computationally intensive and may limit inference, there is ongoing work aimed at emulating SPS codes \cite{alsing_speculator_2020,melchior_autoencoding_2023} or at creating differentiable SPS models which allow the use of gradient-based inference methods.
Examples of such efforts include the differentiable SPS code \texttt{DSPS} \cite{hearin_dsps_2023}, the Diffstar model for star formation history \cite{alarcon_diffstar_2022}, and the Diffmah model for halo mass assembly \cite{hearin_diffmah_2021}.

Galaxy population parameters not only influence uncertainty propagation but can also introduce biases into cosmological measurements, such as photometric redshifts.
\cite{sudek_sensitivity_2022} investigates the sensitivity of various luminosity function parameters from \cite{herbel_redshift_2017} on the redshift distribution using \texttt{SkyPy} \cite{Amara_skypy_2021}.
Similarly, \cite{tortorelli_impact_2024} examines the impact of SPS-based SED modelling choices on forward modelling based tomographic redshift distributions.

The rise of simulation-based inference (SBI) (see \cite{cranmer_frontier_2020} for a comprehensive review) offers new opportunities for both utilizing and constraining galaxy population models.
SBI has become increasingly important in large-scale structure surveys in recent years, proving useful in a variety of applications.
These range from leveraging non-Gaussian features beyond the 2-point correlation function \citep{kacprzak_cosmology_2016,fluri_cosmological_2019,fluri_full_2022,zurcher_cosmological_2021,zurcher_dark_2022,zurcher_towards_2023,kacprzak_deeplss_2022,gatti_dark_2021,cheng_weak_2021,anagnostidis_cosmology_2022,lu_cosmological_2023,massara_sc_2024,valogiannis_precise_2024,jeffrey_dark_2024}, to handling non-Gaussianities in the likelihood function \citep{lin_simulation-based_2023,von_wietersheim-kramsta_kids-sbi_2024}.

While galaxy population models can be directly integrated into SBI approaches, e.g.\ for populating halos from N-body simulations with galaxies, they can themselves be constrained using SBI-based techniques through various simulation approaches.
\cite{alsing_forward_2022} introduces a forward modelling framework where an SPS-based galaxy population model is used for photometric redshift estimation.
Instead of image simulations as in our approach, they use a data-driven noise model, which, while omitting certain effects such as source blending or image-based selection biases, offers a computationally efficient alternative.
This effort is further refined in \cite{alsing_pop-cosmos_2024,leistedt_hierarchical_2023,thorp_data-space_2024,thorp_pop-cosmos_2024}.
Unlike the parametric galaxy population model used in our work, their method, referred to as \texttt{pop-cosmos}, employs a score-based diffusion model, optimized by minimizing the Wasserstein distance between data and observations.

Complementary to these approaches, the Bayesian SED modelling framework PROVABGS introduced in \cite{hahn_desi_2023} and later applied in \cite{hahn_provabgs_2023} infers the probabilistic stellar mass function of galaxies in the DESI BGS sample.
By forward modelling DESI photometry and spectroscopy, they constrain SPS parameters for single galaxies.
This model is subsequently used in the PopSED framework \cite{li_popsed_2024} to constrain the population distribution, utilizing a normalizing flow with Wasserstein distance loss to model the distribution.

Focusing on high-redshift galaxies, \cite{Sabti_gallumi_2022} presents \texttt{GALLUMI}, a likelihood code designed for analyzing the UV luminosity function.
They model the luminosity function while accounting for various sources of uncertainty and systematics, including cosmic variance, dust extinction, scatter in the halo–galaxy connection, and the Alcock–Paczyński effect.
By forward modelling the luminosity functions, they are able to simultaneously constrain both astrophysical and cosmological parameters.

Our work builds on the Monte Carlo Control Loops (MCCL) framework \citep{refregier_way_2014}.
MCCL is a forward modelling framework where a galaxy population model is constrained by comparing realistic image simulations with real survey images.
By performing source extraction on the images generated by the constrained model, we can provide an accurate description of the galaxy distribution in real images, including their redshift distribution.
The simulated images can be used for calibrating shape measurements or evaluating the impact of various systematics, such as source blending.

The image simulations are performed using \ufig \citep{berge_ultra_2013,fischbacher_ufig_2024}.
The galaxy population model was first described by \cite{herbel_redshift_2017}, where it was used for photometric redshift estimation.
It has since been applied to measure cosmic shear with DES-Y1 data \citep{bruderer_calibrated_2016,bruderer_cosmic_2018,kacprzak_monte_2020} and further expanded to measure the luminosity function with the Canada-France-Hawaii Telescope Legacy Survey \citep{tortorelli_measurement_2020}, and to measure galaxy distributions in the Physics of the Accelerating Universe Survey (PAUS) \citep{tortorelli_pau_2018,tortorelli_pau_2021}.
It has also been used to simulate spectra from the Sloan Digital Sky Survey (SDSS) CMASS sample \citep{fagioli_forward_2018,fagioli_spectro-imaging_2020}.
Most recently, \cite{moser_simulation-based_2024} extended the galaxy population model to provide an accurate description for upcoming deeper surveys using HSC deep fields.

In this work, we test the robustness of the model and explore several extensions, leading to a new galaxy population model with improved diagnostics, particularly for the size and redshift distributions.
We introduce the publicly available model, GalSBI, alongside all relevant software \citep{fischbacher_galsbi_2024,fischbacher_ufig_2024}.
Using the \galsbi Python package, users can easily generate catalogs with intrinsic properties of galaxies.
Additionally, realistic images based on this galaxy population model can be generated using \ufig, and survey-specific galaxy distributions can be produced using either source extraction software like \sextractor \citep{bertin_sextractor_1996}, or pretrained emulators, which directly predict measured galaxy distributions with survey-typical noise and PSF.
Besides the phenomenological GalSBI model, \citep{galsbi_sps} presents a first version of the GalSBI model based on stellar population synthesis.

This paper is structured as follows.
In Section~\ref{sec:methods}, we explain the methodology of this work including galaxy population model, inference and emulator architecture.
We describe the data to constrain the model in Section~\ref{sec:data}.
Section~\ref{sec:results} presents the results of our model exploration including the performance of our new fiducial model before we conclude in Section~\ref{sec:conclusion}.
\section{Methods}
\label{sec:methods}
\begin{figure*}[t]
    \centering
    \includegraphics[width=1\linewidth]{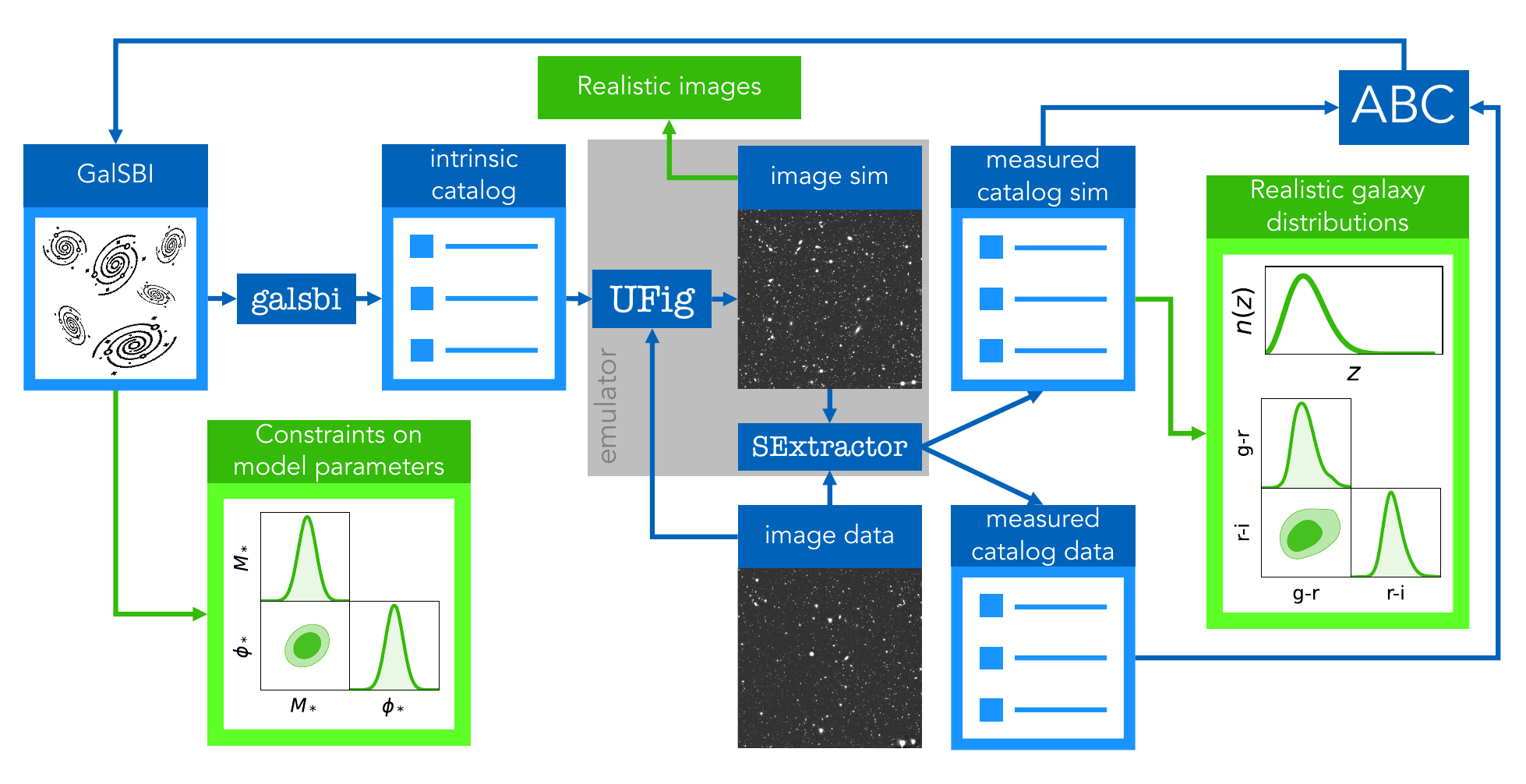}
    \caption{
        Overview of the pipeline to constrain the parameters of the GalSBI galaxy population model.
        Given certain parameters of the galaxy population model, we generate a catalog of galaxies with intrinsic properties. 
        This catalog is then used to generate a realistic astronomical image.
        By comparing the measured catalogs of the simulated image with the real image, we are able to constrain the model using Approximate Bayesian Computation (ABC).
        Finally, we obtain constraints on the parameters of the galaxy population model, can easily generate realistic images and can infer the photometric redshift distribution directly from the simulated catalog.
        An emulator is used for the transfer function from intrinsic catalog to output catalog in order to reduce the computational cost of the inference.
        }
    \label{fig:overview}
\end{figure*}

GalSBI is a parametric galaxy population model.
An overview of the parameter inference framework is shown in Figure \ref{fig:overview}.
For a given set of model parameters, galaxies are sampled with the GalSBI catalog generator \citep{herbel_redshift_2017,tortorelli_measurement_2020,tortorelli_pau_2021,kacprzak_monte_2020,moser_simulation-based_2024}.
This creates a catalog with intrinsic galaxy properties such as magnitudes, sizes, positions and light profiles.
This step is described in more detail in Section \ref{sec:galpop_model}.

The catalog is then used to create a realistic astronomical image using the ultrafast image generator \ufig \citep{berge_ultra_2013, fischbacher_ufig_2024}.
Apart from the galaxy catalog, image systematics such as the background level or the size and shape of the point-spread function (PSF) have to be known to generate a realistic image.
These quantities are estimated from the real observational data.
To ensure consistency, we run \sextractor on both the simulated and real images using identical settings.
This process generates catalogs of observed properties such as magnitudes or sizes, for both the simulation and the data.
This part of the pipeline is explained in more detail in Section \ref{sec:image_sims}.

The catalogs are then compared using distance metrics which allow us to constrain the model parameters using Approximate Bayesian Computation (ABC).
We employ an iterative implementation of ABC to significantly reduce the computational cost of inference.
By using ABC, we ensure that the posterior faithfully represents the true uncertainty, avoiding the risk of underestimation.
The details of the inference are given in Section \ref{sec:abc}.

The inference process is computationally intensive due to the high dimensionality of the galaxy population model requiring a large number of simulations.
Consequently, accelerating the simulations becomes crucial for exploring various model choices and parametrizations.
The primary bottleneck of the analysis is the image rendering by \ufig and the subsequent source extraction by \sextractor. 
We have developed an emulator for the transfer function from intrinsic to measured catalog to address this.
The emulator takes intrinsic properties from GalSBI and observational information such as background level and PSF as input and directly predicts which galaxies are detected, as well as the measured quantities needed for the ABC distances.
More details on the emulator architecture and training is given in Section \ref{sec:emu} and Appendix \ref{app:emu}.

Once the inference is complete, we can use the constrained model to create new galaxy catalogs independent of the survey that was used to constrain the model parameters.
By adapting the image simulations to the specifications of the new survey, we can produce realistic images.
These images can be used for calibration tasks such as the shape measurement for cosmic shear or to estimate the impact of blending.
By running \sextractor on the simulated images, we can directly measure the overall galaxy distribution in the real images; including the redshift distribution.
This is possible because the true redshift is known for each galaxy in the simulated catalog.

\subsection{Galaxy population model}
\label{sec:galpop_model}
GalSBI is a physically motivated, parametric, phenomenological galaxy population model first described in \cite{herbel_redshift_2017} and further extended and improved in \cite{moser_simulation-based_2024}.
Galaxies are drawn from two populations, star-forming (or blue) galaxies and quiescent (or red) galaxies, with (mostly) the same parametrizations but different values for the parameters.
By defining two populations, our approach simplifies inference compared to fitting a single, inherently bimodal distribution.
While a more flexible, unified model -- such as the one used in \cite{alsing_pop-cosmos_2024} -- could be advantageous for upcoming data, adopting such an approach would require a careful reassessment of our analysis choices to maintain consistency and reliability.
However, the inclusion of scatter parameters ensures that the two populations in our model can overlap, reducing the impact of rigid discretization. 

Sampling galaxies from a given galaxy population model follows these steps:
first, we sample galaxies from the luminosity function, providing a catalog of galaxies with redshift $z$ and absolute magnitude $M$ in the $B$ band.
We then compute the apparent magnitudes $m$ for each band by integrating over the spectral energy distribution (SED). The SED is assigned using a linear combination of the five spectral templates of \texttt{kcorrect} \cite{blanton_k-corrections_2007}.
Finally, we assign morphological properties such as light profile, size and ellipticity and assign the positions randomly within the image.
This way, we obtain a catalog of galaxies with all the properties needed for the image simulation.
We describe each of these modelling steps in more detail in the next sections.

\subsubsection{Luminosity function parametrization}
\label{sec:lumfunc}
\begin{figure*}[t]
    \centering
    \includegraphics[width=1\linewidth]{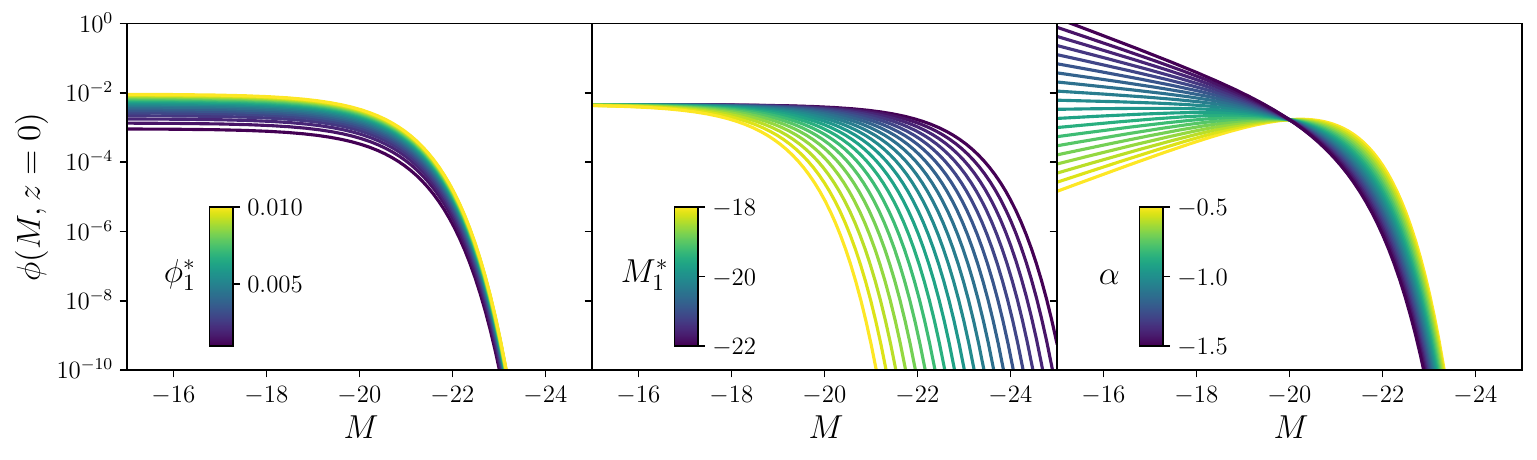}
    \caption{
        Impact of different parameters on the luminosity function given in Equation \ref{eq:lumfunc}.
        We illustrate the effect of different values of $\phi^*_1$, $M^*_1$ and $\alpha$ with values varied indicated by the colorbar at redshift $z=0$.
        At $z=0$, the other parameters have no impact on the luminosity function and the different parametrizations are consistent with each other.
        Intuitively, $\phi^*$ changes the normalization of the luminosity function, $M^*$ changes the knee of the distribution and $\alpha$ varies the slope of the bright end.
        An interactive version of this plot including the redshift evolution and live rendering of an astronomical image is available at \url{https://cosmo-docs.phys.ethz.ch/galsbi/galpop_pheno.html}
    }
    \label{fig:lumfunc}
\end{figure*}

We sample the absolute magnitude $M$ and redshift $z$ from a single Schechter luminosity function
\begin{equation}
    \label{eq:lumfunc}
    \phi(z, M) = \frac{2}{5} \ln(10) \phi^*(z) 10^{\frac{2}{5} (M^*(z) - M)(\alpha+1)} \exp\left(-10^{\frac{2}{5}\left(M^*\left(z\right)-M\right)}\right),
\end{equation}
where both $\phi^*(z)$ and $M^*(z)$ are a function of redshift and $\alpha$ a free parameter.
Figure \ref{fig:lumfunc} illustrates how the luminosity function changes for different choices of these parameters.
The redshift evolution can be parametrized in different ways.
In \cite{herbel_redshift_2017}, they use
\begin{equation}\label{eq:linexp}
    \phi^*(z) = \phi^*_1 \exp(\phi^*_2 z), \quad M^*(z) = M^*_1 + M^*_2 z,
\end{equation}
where $\phi_1^*$, $\phi_2^*$, $M_1^*$, $M_2^*$ are all free parameters of the model.
In \cite{moser_simulation-based_2024}, they use the following parametrization introduced in \cite{johnston_shedding_2011}
\begin{equation}\label{eq:logpower}
    \phi^*(z) = \phi^*_1 \left(1+z\right)^{\phi^*_2}, \quad M^*(z) = M^*_1 + M^*_2 \log(1+z).
\end{equation}
Note that for low redshift both parametrizations are equivalent with $\phi^*(z) = \phi^*_1 +\phi^*_2 z + O(z^2)$ and $M^*(z) = M^*_1 + M^*_2 z + O(z^2)$.
Another parametrization that we test in this work combines the two parametrizations above and adds an additional high redshift cut to suppress high-redshift tails in the overall redshift distribution.
It is given by
\begin{equation}\label{eq:trunc_logexp}
    \phi^*(z) = \phi^*_1 \exp(\phi^*_2 z), \quad 
    M^*(z) = 
    \begin{cases} 
        M^*_1 + M^*_2 \log(1+z) & \text{if } z < z_0 \\
        M^*_1 + M^*_2 \log(1+z_0) & \text{if } z \geq z_0
    \end{cases},
\end{equation}
and has an additional free parameter $z_0$.

\subsubsection{Spectral templates}
\label{sec:templates}
In order to compute the apparent magnitude of a galaxy in a specific band, we need to integrate its redshifted spectral energy density (SED) over the filter throughput.
We assign a SED to each galaxy by a linear combination of the five spectral templates from \texttt{kcorrect} the same way as in \cite{moser_simulation-based_2024}.
These templates are validated across ultraviolet, optical, and near-infrared bands, covering a broad wavelength and redshift range.
The coefficients are normalized to sum to 1 and correspond to the following templates:
\[
    c_0: \text{dusty red}, 
    c_1: \text{ELG with strong SRF}, 
    c_2: \text{ELG}, 
    c_3: \text{passive red}, 
    c_4: \text{post-starburst}.
\]
The normalized SED is rescaled so that integrating over the $B$-band yields the galaxy's sampled absolute magnitude.
The apparent magnitude in each filter band is then computed by assigning the redshift to the SED, applying reddening due to galactic extinction using the method outlined in \cite{herbel_redshift_2017} and integrating the spectrum over the filter throughput.

We use a flexible parametrization to describe the template distribution.
We enforce $\sum_i c_i=1$ by sampling the coefficients from a Dirichlet distribution.
The Dirichlet distribution is parametrized with two types of parameters:
the mode of each coefficient $\alpha_i$ and the standard deviation of the Dirichlet distribution $\alpha_\mathrm{std}$.
$\sum_i \alpha_i = 1$ is enforced when generating the prior.
This parametrization has the advantage that defining the distribution by its mode and standard deviation increases the interpretability of the parameters.
The resulting distribution of coefficients $c_i$ is centered around the specified modes $\alpha_i$, with $\alpha_\mathrm{std}$ governing the dispersion around the central values.

Since we expect the coefficients to evolve as a function of redshift, we use two sets of parameters; one fixed at $z=0$ and one fixed at $z=3$; and interpolate in between by
\begin{equation}
    \label{eq:template_redshift}
    \alpha_i (z) = \left(\alpha_{i,z=0}\right)^{1-\frac{z}{3}} \times \left(\alpha_{i,z=3}\right)^{\frac{z}{3}}.
\end{equation}
The choice of these two redshift anchors is flexible, but they are selected to encompass the range of galaxies used to constrain the model in this work.
The same is done for the scatter parameter $\alpha_\mathrm{std}$.
All these parameters are used the same way for the red and blue population, however we choose different priors (see Table~4 in \cite{moser_simulation-based_2024}).
These priors are derived by comparing simulated and observed galaxy colors and redshifts from the COSMOS2015 catalog \cite{laigle_cosmos2015_2016}.
We iteratively refine the Dirichlet parameter ranges by minimizing the Universal Divergence \cite{wang_divergence_2009} between simulations and real data, ensuring that the priors capture the redshift-color dependencies of higher-redshift galaxies.
For details we refer to Appendix A.1 in \cite{moser_simulation-based_2024}.

\subsubsection{Galaxy morphology}
\label{sec:morphology}
We assign the half light radius $r_{50}$ to each galaxy by sampling $\log r_{50}$ from a normal distribution with mean $\mu_{\log r_{50}}(M)$ and standard deviation $\sigma_{\log r_{50}}(M)$.
Both mean and standard deviations are functions of the absolute magnitude $M$.
Sampling from a log-normal distribution ensures that the half light radius is always positive and also matches with observations \citep{jong_local_2000,shen_size_2003}.

In \cite{herbel_redshift_2017,moser_simulation-based_2024}, they use a simple linear model for the absolute magnitude dependence of the mean and a constant standard deviation.
This parametrization is given by
\begin{eqnarray}
    \label{eq:size_simple}
    &&\mu_{\log r_{50}}(M) = A_{\log r_{50}} (M-M_0) + B_{\log r_{50}}, \nonumber\\ &&\sigma_{\log r_{50}}(M) = \sigma_{\log r_{50}},
\end{eqnarray}
where $M$ is the absolute magnitude and $A_{\log r_{50}}, B_{\log r_{50}}$ and $\sigma_{\log r_{50}}$ are free parameters of the model and $M_0=-20$ a fixed scaling parameter.
For this work, we additionally test an extended size model with different parametrizations for the red and blue population and a magnitude dependent standard deviation \cite{shen_size_2003}.
The mean of the distribution is then given by
\begin{equation}
    \label{eq:size_extended}
    \begin{aligned}
        \mu_{\log r_{50}, \mathrm{red}}(M) =& b_{\log r_{50}} -0.4 a_{\log r_{50}} M, \\
        \mu_{\log r_{50}, \mathrm{blue}}(M) =& (\beta_{\log r_{50}}-\alpha_{\log r_{50}}) \log \left(1+10^{-0.4(M-M_0)}\right) 
        -0.4 \alpha_{\log r_{50}} M + \gamma_{\log r_{50}}.
    \end{aligned}
\end{equation}
Note that this parametrization is the same as in Equation \ref{eq:size_simple} for the red population apart from different normalizations.
The standard deviation is parametrized by
\begin{equation}
    \label{eq:size_sigma}
    \sigma_{\log r_{50}}(M) = \sigma_{2, \log r_{50}} + \frac{\left(\sigma_{1, \log r_{50}}-\sigma_{2, \log r_{50}}\right)}{1+ 10^{-0.8(M-M_0)}},
\end{equation}
for both red and blue galaxies.
For this parametrization, $a_{\log r_{50}}, b_{\log r_{50}}, \alpha_{\log r_{50}}, \beta_{\log r_{50}}, \gamma_{\log r_{50}}$ are free parameters of the model, $\sigma_{1, \log r_{50}}$ and $\sigma_{2, \log r_{50}}$ are also free parameters but separate for red and blue galaxies and $M_0=-20.52$ is fixed to the measurement in \cite{shen_size_2003}.

In \cite{moser_simulation-based_2024}, they do not include an explicit redshift evolution of $r_{50}$ apart from the implicit redshift evolution through the dependence on the absolute magnitude.
Such a redshift evolution is found in observations and is well-described by a power law (e.g.\ \cite{van_der_walt_numpy_2011,ormerod_epochs_2023})
\begin{equation}\label{eq:size_evolution}
    r_{50} = r_{50,0} \left(1+z\right)^{\eta_{r_{50}}},
\end{equation}
where $r_{50,0}$ is sampled from the log-normal distribution, $z$ is the redshift of the galaxy and $\eta_{r_{50}}$ is a free parameter of the model, with separate values for red and blue galaxies.

The absolute ellipticities of the galaxies are sampled from a modified Beta distribution with two free parameters $e_\mathrm{mode}$ and $e_\mathrm{spread}$ which are related to the $\alpha$ and $\beta$ parameter of the standard Beta distribution via $e_\mathrm{mode}=\frac{\alpha-1}{e_\mathrm{spread} - 2}$ and $e_\mathrm{spread} = \alpha + \beta$.
Note that for $e_\mathrm{spread}=2$, the distribution is flat and $e_\mathrm{mode}$ has no impact.
For $e_\mathrm{spread}>2$, $e_\mathrm{mode}$ determines the mode of the distribution.
We assign a random phase to the absolute ellipticity to compute the two ellipticity components $e_1$ and $e_2$ which are required for the image simulation.
This corresponds to the same setup as in \cite{moser_simulation-based_2024}.

The light profile of the galaxies is modelled by a Sérsic profile.
The Sérsic index is sampled from a beta prime distribution with one free parameter $n_s$ which corresponds to the mode of the distribution.
It is related to the standard parameters $\alpha$ and $\beta$ of the beta prime distribution via $n_s=\frac{\alpha -1}{\beta + 1}$.
The parameter $\beta$ which determines the scatter of the distribution is fixed at $\beta=5$ for blue galaxies and $\beta=50$ for red galaxies to match the distribution from \cite{tarsitano_catalogue_2018}.
In \cite{moser_simulation-based_2024}, they use this parametrization without any redshift evolution.
We test the following parametrization to include redshift evolution of the Sérsic index
\begin{equation}\label{eq:ns_evolution}
    n_s = n_{s,0} \left(1+z\right)^{\eta_{n_s}}
\end{equation}
based on findings from galaxy evolution (e.g.\ \cite{krywult_vimos_2017}).

\subsection{Image simulations, source extraction and matching}
\label{sec:image_sims}
The galaxy population model described in Section \ref{sec:galpop_model} provides a catalog of galaxies with their intrinsic properties such as the absolute and apparent magnitude, redshift, angular size, light profile or ellipticity.
The following steps of simulating realistic HSC deep field images, performing source extraction on these images and matching the extracted objects to the objects in the intrinsic catalog are done the same way as in \cite{moser_simulation-based_2024}.
We therefore only give a short summary of the most relevant parts and refer to \cite{moser_simulation-based_2024} for more details.

We sample stars using the Besançon model of the Milky Way \cite{robin_synthetic_2003} to accurately represent stellar density variations in simulations.
The positions of the brightest stars are taken from the Gaia DR3 catalog \cite{prusti_gaia_2016,vallenari_gaia_2023} and matched to the sampled Besançon stars based on their magnitude.
This ensures that we can use the same star mask in both simulation and actual data.

We estimate the PSF, background and gain from the data such that we simulate the coadded images directly using \ufig \cite{berge_ultra_2013,fischbacher_ufig_2024,herbel_fast_2018}.
The point-spread function is estimated using the CNN-based approach presented in \cite{herbel_fast_2018} and further improved in \cite{kacprzak_monte_2020}.
In that work, the images were also used for shape measurement calibration, imposing even stricter requirements on the image simulations than in our case.
Additionally, \cite{moser_simulation-based_2024} validated the image simulations for HSC deep fields.
The background noise is estimated using varying Gaussian noise with mean from the image header and pixel-specific standard deviation estimated directly from the image.
We omit local background subtraction as it primarily affects regions around bright sources that are masked anyway.
The estimated background in our simulations shows very good agreement with the actual data \cite{moser_simulation-based_2024}.
We include simplified saturation treatment since it only affects bright stars that are already masked.

Source extraction is performed using \sextractor \cite{bertin_sextractor_1996} using the same settings for the data and the simulations, details on the hyper parameters can be found in \cite{moser_simulation-based_2024}.
The extracted objects are matched to the intrinsic catalog using the segmentation map by \sextractor.
Using consistent settings is essential because it ensures that all uncertainties in the source extraction process are consistently modelled.
As a result, the constraints on the galaxy population model should not be influenced by the choice of source extraction software.
Finally, we obtain a catalog of all detected objects with measured quantities such as apparent magnitudes or sizes for both simulation and data.

\subsection{Emulator}
\label{sec:emu}
Performing inference on a parameter space with dimensionality $d>40$ requires a lot of simulations at different points in parameter space.
The primary bottlenecks of the analysis presented in \cite{moser_simulation-based_2024} are the image simulation and the source extraction.
Since we only need the output catalog to compare with the data, we developed an emulator to learn the transfer function from the intrinsic catalog by GalSBI to the measured catalog of the simulated image generated by \sextractor (see the gray box in Figure \ref{fig:overview}).
The emulator is created using the \texttt{edelweiss} package which we developed for this work and is publicly available\footnote{\url{https://gitlab.com/cosmology-ethz/edelweiss/}}.
After training, the emulator generates the measured catalog 10 to 100 times faster than using image simulation and source extraction, depending on the image complexity.
This reduces the overall pipeline runtime by a factor of 5 to 6, with the emulator accounting for approximately 10\% of the remaining runtime.
The measured catalog for an HSC deep field image used in this work can then be generated in approximately 20 seconds.

\subsubsection{Overview of the different emulator components}
The general architecture of the emulator is based on two components.
In a first step, a detection classifier predicts for each object in the intrinsic catalog if it will be detected as a galaxy or not, i.e.\ if it will appear in the output catalog used to compute the distances. 
Then, a normalizing flow assigns all the necessary measured parameters to these objects in a second step.
A similar setup is also used in \cite{gandalf}.
In the following, we describe the two steps in more detail.

The input catalog contains the full list of galaxies and stars that would be rendered by \ufig including the observational conditions.
The detection classifier predicts for all objects if it is detected as a galaxy (label 1) or not (label 0).
For more details on the label assignment, we refer to Appendix \ref{app:emu_labels}.

All objects that are classified as detected are then used as input of a conditional normalizing flow.
The normalizing flow is conditioned on the same input parameters as the detection classifier.
Based on these conditions, the normalizing flow predicts all the output parameters that are necessary for the comparison with the real data.
Sampling with the normalizing flow naturally includes the statistical uncertainty coming from the rendering process and systematic and statistical uncertainty from the source extraction.
Details of the input and output parameters are given in the next section.

\subsubsection{Input and output parameters}
The output parameters of the normalizing flow depend on the parameters that are used for the comparison to the real data (see Section \ref{sec:abc} for more details on the inference).
These parameters include the measured magnitude \texttt{MAG\_APER3}, the measured size \texttt{FLUX\_RADIUS}, the absolute ellipticity (computed from \texttt{X2WIN\_IMAGE}, \texttt{Y2WIN\_IMAGE} and \texttt{XYWIN\_IMAGE}), the flux \texttt{FLUX\_APER3} (which is used to compute the flux fraction) and the signal-to-noise ratio (computed from \texttt{FLUX\_APER3} divided by \texttt{FLUXERR\_APER3}).

The choice of input parameters is more difficult since we do not know in advance which parameters actually impact the detection probability and the output parameters.
It is intuitively clear that parameters such as the intrinsic magnitude, the intrinsic size or ellipticity will impact the measured magnitudes, sizes and ellipticities.
These parameters are therefore used as input to the emulator.
We also include the object type with three classes; blue galaxy, red galaxy and star; mainly since galaxies are much more likely to be detected as a galaxy than a star.
Additionally, we use the Sérsic index, redshift and number of photons as input parameters.

Apart from the intrinsic galaxy properties above, the measured catalog also depends on some image systematics.
Although the HSC deep fields have a very stable PSF size and background level, we expect variations in PSF and background to impact the detection probability and some of the output parameters. We therefore include the PSF full width half maximum (FWHM) and the background amplitude and standard deviation.

Source blending significantly impacts detection probability, particularly given the deep data utilized in this work.
Our findings indicate that for a fixed galaxy population model, the classifier successfully learns the overall blending risk. 
However, during inference, we vary the galaxy population model, which consequently alters the blending risk.
If not accounted for, this varying blending risk introduces biases in the inference.
To address this issue, we incorporate parameters that correlate with blending risk, specifically the number of galaxies in the image at different magnitude cuts.
Our results demonstrate that the inclusion of these parameters effectively accounts for the varying impact of blending.
For more details on this, we refer to Appendix \ref{app:emu_blending_risk}.

The choice of parameters presented in this work are optimized for the simulation of HSC deep fields.
We expect that the same setup will also work for other surveys but with changing importance of the individual parameters.
For a shallower sample with less stable PSF and background, we expect that the importance of background and PSF parameters would be higher whereas the importance of the blending parameters would be reduced.

A summary of input and output parameters is given in Table \ref{tab:params}.
The input vector is 45-dimensional (8 band-dependent parameters with 5 bands plus 5 band-independent parameters), the output vector is 25-dimensional (5 parameters with 5 bands).
\begin{table*}[htbp]
    \centering
    \begin{tabular}{p{0.3\textwidth}p{0.3\textwidth}p{0.3\textwidth}}
         \toprule
         \textbf{Input parameters}  & \textbf{Input parameters} & \textbf{Output parameters}  \\
         (different for each band) & (the same for all bands) & (different for each band) \\
         \midrule
         - apparent magnitude $m$ & - size r50 & - \texttt{MAG\_APER}\\
         - PSF FWHM & - absolute ellipticity & - \texttt{FLUX\_RADIUS}\\
         - BKG noise ampl. and std & - object type & - \texttt{FLUX\_APER}\\
         - number of photons & - Sérsic index & - absolute ellipticity\\
         - $n_{\mathrm{gal},m<i}$ with $i=22,23,24$ & - redshift & - SNR\\
         \bottomrule
    \end{tabular}
    \caption{Overview of the input and output parameters for the emulator.}
    \label{tab:params}
\end{table*}

\subsubsection{Training}
As motivated above, the emulator needs to perform well across galaxy population models and across all images.
The training data is therefore generated from 10000 simulations with a different galaxy population model for each simulation and using all available images.
The 10000 different galaxy population models are sampled from the prior distribution of the first emulator iteration (see Section \ref{sec:abc} for details on the iterative inference process).

We increase the sensitivity of the classifier to both high and low blending regimes by adjusting the training data.
This is done by randomly multiplying the number of sampled galaxies by a factor between 0.05 and 3, ensuring that images with low blending risk are not underrepresented.
For the conditional normalizing flow, we upweight brighter galaxies by flattening the magnitude distribution in the training set.
Predicting galaxy detection labels directly leads to an underestimation at faint magnitudes, as galaxies with low detection probabilities are discarded, biasing the distribution. 
Instead, we predict detection probabilities and randomly accept or discard galaxies based on these probabilities, improving the overall distribution, i.e.\ instead of accepting all galaxies with probability $p>0.5$, we accept a galaxy based on the actual probability $p$.
The classifier's architecture has little effect on performance, we use a dense neural network in our fiducial setup.
For the normalizing flow, outlier handling is crucial for preventing numerical instability and loss divergence, and we find that a scaling transformation based on quantiles works well.
For more details on the emulator training we refer to Appendix \ref{app:emu_training}.

\subsection{Inference}
\label{sec:abc}
Although the run time starting from a set of galaxy population model parameters to the catalog of detected objects in a specific image is significantly reduced by using the emulator, simulating all images for enough samples in the parameter space would be computationally infeasible.
We are therefore using an iterative approach to reduce the computational cost while still probing the full survey area.

According to Bayes' theorem, the posterior of the model parameters $\theta$ given the data $d$ is given by
\begin{equation}
    p(\theta|d) = \frac{p(\theta) p(d|\theta)}{p(d)},
\end{equation}
with $p(\theta)$ the prior, $p(d|\theta)$ the likelihood and $p(d)$ the Bayesian evidence.
We start our inference with wide priors on all the free parameters.
In a first iteration, we only simulate a small patch of the sky $d_1\subset d$ and obtain a posterior given this data set $p(\theta|d_1)$.
This posterior is then used in a next iteration as the prior $p(\theta)$.
In iteration $n$, the posterior is evaluated on a patch $d_{n}\subset d$ with $d_n \cap d_i = \emptyset\,  \forall i \in {1, 2, ..., n-1}$ and can be written as
\begin{equation}
    p(\theta|d_{n}) =  \frac{p(\theta |d_{n-1}) p(d_{n}|\theta)}{p(d_{n})}.
\end{equation}
However, this iterative approach introduces potential risks of overfitting to specific sky regions due to cosmic variance and biasing due to limited data in each iteration.
To mitigate these risks, we employ two key strategies:
first, we select the tiles randomly such that the simulated tiles of iteration $i$ are scattered across the footprint, which reduces the effect of cosmic variance.
Second, we estimate the posterior using Approximate Bayesian Computation (ABC) which is a conservative estimate of the posterior and therefore reduces the risk of getting stuck in a local minima because of limited statistics in one iteration.

The conservative nature of ABC is essential, as Bayesian simulation-based inference methods can produce overconfident posterior estimates (e.g., \cite{hermans_trust_2022}).
This is due to the fact that inevitable errors in ML-based modelling of the high-dimensional posterior from small number of samples will often lead to pre-mature exclusion of feasible parameter space regions.
The incremental convergence of our iterative batch ABC approach avoids this problem.
The only modelling step in the algorithm is the density estimation of the posterior, in order to produce prior samples for the next iteration. 
The accuracy for this step is straightforward to evaluate at each iteration.
As demonstrated in previous work \cite{herbel_redshift_2017,kacprzak_monte_2020,tortorelli_measurement_2020,moser_simulation-based_2024} and Section \ref{sec:emu_val} of this work, we obtain robust and reliable posteriors.
However, more efficient methods may be useful in the future, though these will need extensive testing to avoid overconfident posteriors.

Estimating the posterior with ABC requires a distance $\delta$ between the simulated and real catalogs and a threshold $\varepsilon$.
The posterior can then be approximated by
\begin{equation}
    p(\theta|d) \approx p(\theta | \delta(d_\mathrm{sim}, d_\mathrm{data})<\varepsilon).
\end{equation}

In our framework, we sample 10000~different samples from the prior and simulate 10~tiles ($0.38 \,\mathrm{deg}^2$).
For each sample, we compute the distance (details on the computation of distances are provided below) and define the posterior as the distribution of the 2000~samples with the lowest distance.
Therefore, the threshold $\varepsilon$ changes for each iteration and is defined relative to the samples.
Next, we use a Gaussian mixture model with 20 components to resample 10000~new samples from this posterior, which are then used as the prior for the next iteration.
In each iteration, we increase the number of simulated tiles until we reach 30 tiles in iteration 21 ($1.15 \,\mathrm{deg}^2$).
From that point onward, we continue with 30 tiles per iteration until either all tiles are used or the distance threshold no longer improves.
The choice of distance measures is outlined in the next section.

\subsubsection{Distances}
\label{sec:distances}
The distances that are used in this work compare the simulated catalogs and the real catalogs.
A straightforward way to compare these two catalogs is the number of objects in the catalog.
We do this by computing the \emph{fractional difference in number of galaxies}
\begin{equation}
    d_{\mathrm{frac}} = \frac{n_{\mathrm{sim}} - n_{\mathrm{data}}}{n_{\mathrm{data}}},
\end{equation}
where $n_{\mathrm{sim}}$ is the number of galaxies in the simulation and $n_{\mathrm{data}}$ the corresponding number in the data.
Because we use \sextractor in forced photometry mode and only consider objects with photometric measurements in all bands in the final catalog, the number of galaxies is independent of the band.

If redshift information is available for the data catalog, one can compare the two catalogs by computing the \emph{difference in mean redshift} of the two catalogs
\begin{equation}
    \label{eq:distance_mean_redshift}
    d_{\mu_z} = |\mu_{z_\mathrm{data}} - \mu_{z_\mathrm{sim}}|,
\end{equation}
where $\mu_{z_\mathrm{data}}$ is the mean redshift of the galaxies in the data and $\mu_{z_\mathrm{sim}}$ the corresponding in the simulation.

Furthermore, we use multidimensional distance measures to compare the distributions of the two catalogs.
We analyze the same parameters as in \cite{moser_simulation-based_2024}: the size (measured by \texttt{FLUX\_RADIUS} in the $i$-band), the absolute ellipticity in the $i$-band, the magnitudes (measured by \texttt{MAG\_APER3} in all bands), the flux fractions in all bands and the redshift.
The flux fraction is calculated using
\begin{equation}
    f_b = \frac{\texttt{FLUX\_APER3}_b}{\sum_j \texttt{FLUX\_APER3}_j}
\end{equation}
where $b$ is the band and the sum goes over all bands.

The redshift distribution in the simulation is obtained by matching the objects of the measured catalogs with the objects of the intrinsic catalog following the methodology of Section 3.3 in \cite{moser_simulation-based_2024}.
For the data, we assign a photometric redshift to each galaxy using the approach described in Section 3.4 in \cite{moser_simulation-based_2024}.
This approach utilizes the photo-z codes LePhare \cite{arnouts_measuring_1999, ilbert_accurate_2006} and EAZY \cite{brammer_eazy_2008}, resulting in two redshift estimates that are both compared to the redshift in the simulation.
Since \cite{moser_simulation-based_2024} found that the mean redshifts for a bright galaxy selection can differ by almost $1\sigma$ between the two methods, we include both estimates in the distance computation to avoid overfitting to one of them.
The dimension of the catalog entering the distance measures is therefore 14.
Comparing these 14-dimensional distributions is done with three different distance measures that act on the normalized distributions.

\begin{itemize}
    \item 
\emph{Maximum Mean Discrepancy (MMD)} is a kernel two-sample test for multi-dimensional distribution \cite{gretton_kernel_2012} and was already successfully applied to this problem \cite{herbel_redshift_2017,tortorelli_measurement_2020,tortorelli_pau_2021,moser_simulation-based_2024}.
The distance is computed by
\begin{equation}
    \label{eq:mmd}
    \mathrm{MMD} = \frac{1}{N(N+1)} \sum_{i,j; i\neq j} \big[ k(x_i, x_j) + k(y_i, y_j)  - k(x_i, y_j) - k(y_i, x_j)\big],
\end{equation}

where $x_i$ and $y_i$ are elements of the two samples, $N$ is the size of the samples and $k$ is a Gaussian kernel of predefined width (see \cite{herbel_fast_2018,moser_simulation-based_2024} for details).

\item 
The \emph{universal divergence} estimates the distance between two samples using $k$-nearest-neighbor distances ($k$NN) \cite{wang_divergence_2009}.
Assume we have two samples ${x_1,\cdots,x_n}$ and ${y_1,\cdots,y_m}$.
The Euclidean distance between $x_i$ and its $k$NN in $x_{j,j\neq i}$ is defined as $\rho_k(i)$ and the distance to its $k$NN in ${y_j}$ is $\nu_k(i)$.
The distance estimator is then given by
\begin{equation}
    \label{eq:universal_div}
    D = \frac{d}{n} \sum_i^n \log \frac{\nu_k(i)}{\rho_k(i)} + \log \frac{m}{n-1}.
\end{equation}
For large $n$ and $n=m$, the second terms vanishes and we are left with the first term.
If both samples are coming from the same distribution, the Euclidean distances to the $k$NN should be the same and $D\to 0$ as expected.
If the two samples are not coming from the same distribution, we will have $\nu_k(i) > \rho_k(i)$ and the distance $D > 0$.
We choose $k=10$ as a default in this work.

\item 
The \emph{Wasserstein distance} (sometime also called Kantorovich-Rubinstein metric, optimal transport distance or Earth-Mover's metric) between two distributions corresponds to the minimal cost of transforming one distribution into the other one \cite{vaserstein_markov_1969,kantorovich_space_1958}.
The term Earth-Mover's metric originates from the intuitive idea of two dirt piles representing the two distributions.
The Wasserstein distance measures the minimal amount of work needed to transform the first pile into the second one.
This work can be visualized as the cost associated with moving the dirt (or "earth") from one location to another.
Assume we have two discrete distributions ${x_1,\cdots,x_n}$ and ${y_1,\cdots,y_n}$ with the same number of elements.
We can easily quantify the cost of moving one element of the first distribution to the position of an element in the second distribution, e.g.\ by computing the $\ell^p$-norm between the two positions.
We can define a cost matrix $M$ with $M_{i,j}=\| w^x_i x_i - w^y_j y_j \|_p$ the cost of moving $x_i$ to $y_j$ and $w^x, w^y$ the weights of each element.
In the intuitive picture of moving a pile of dirt from one position to another, the weights would correspond to the mass of each specific clump.
Transforming the first distribution to the second distribution can be expressed as a transport matrix $\gamma$ which maps each $x_i$ to a specific $y_j$.
The total cost of each transport solution $\gamma$ is therefore given as 
\begin{equation}
    C = \sum_i^n \sum_j^n \gamma_{i,j}M_{i,j}.
\end{equation}
The optimal transport solution $\gamma^*$ is the transport matrix that minimizes the total cost $C$.
The Wasserstein distance corresponds to the cost of this optimal transport solution $\gamma^*$.
We define the Wasserstein-$p$ distance as the Wasserstein distance associated to the $\ell^p$-norm
\begin{equation}
    \label{eq:wasserstein}
    W_p = \min \sum_{i,j} \gamma_{ij} \| w^x_i x_i - w^y_j y_j \|_p.
\end{equation}
In this work, we use $p=1$ and $p=2$.

\end{itemize}
\subsubsection{Combining distances into final distance}
\label{sec:combined_distances}

The final distance is computed as a weighted combination of the individual distance measures described above.
Specifically, in each iteration, we have 10000~different samples, for which we simulate $n$ tiles.
For each distance measure, we obtain one distance value per sample and per tile.
First, we calculate the median distance across the different tiles to make the calculation robust to outliers, yielding a single value per distance measure for each sample.
Since the dynamic ranges of the different distance measures vary significantly -- for instance, MMD may change by $O(10^{-2})$ across iterations, while the fractional difference in galaxy counts change by $O(1)$ -- normalization is essential prior to combination.
Each distance is therefore scaled such that its median value across all samples is set to 1, and its minimum value to 0.
While the specific scaling method is not critical, proper normalization ensures that each measure contributes according to its weight to the final result.
The normalized distances are then combined through a weighted sum, resulting in a single, aggregate distance value per sample.

The full analysis is performed on catalogs with a magnitude cut of $\texttt{MAG\_APER3}<25$ in the $i$-band.
\cite{moser_simulation-based_2024} introduced a reweighting scheme to upweight brighter galaxies and improve the perfomance of the method on shallower data sets.
For any of the above distances $d$, this can be done by computing the distance $d_m$ on a sample with a magnitude cut of $\texttt{MAG\_APER3}<m$ in the $i$-band.
For the Wasserstein distances, the reweighting can be done in a more elegant way by using the weights in equation \ref{eq:wasserstein} to rebalance the magnitude distribution.
For more details on how the weights are computed, we refer to Appendix \ref{app:wasserstein_weights}.
\section{Data}
\label{sec:data}

We use publicly available data from the Hyper-Suprime Cam (HSC) Subaru Strategic Program, in particular the deep and ultradeep fields of the third data release (PDR3) \cite{aihara_third_2022}.
Using deep HSC data has several advantages.
In order to constrain the galaxy population model, we do not require a particularly large survey area but more depth increases the constraining power for the higher redshift population significantly.
HSC can also be considered a precursor to upcoming Stage-IV surveys such as LSST due to the similarity in depth and image quality (e.g.\ \cite{nicola_tomographic_2020}).
Furthermore, the HSC deep fields overlap with the COSMOS2020 panchromatic photometric catalog \cite{weaver_cosmos2020_2022}.

The COSMOS2020 catalog is the latest public catalog of the COSMOS field.
The catalog contains nearly 1 million galaxy that are measured in multiple filters and are analyzed using two different source extraction tools.
Furthermore, the catalog provides high-quality photometric redshifts by two techniques: LePhare \cite{arnouts_measuring_1999,ilbert_accurate_2006} and EAZY \cite{brammer_eazy_2008}.

The HSC deep fields consist of roughly 1500 patches.
We perform the same blacklisting methodology as in \cite{moser_simulation-based_2024} and use the remaining 746~patches in the five broad band filters $g,r,i,z,y$ to constrain the model.
We rerun \sextractor on the HSC deep fields to ensure consistency in the output catalogs.
We assign a redshift to each galaxy in the created catalogs by the reweighting technique described in Section 3.4 of \cite{moser_simulation-based_2024}.
After constraining the model, we validate the photometric properties and the redshift distribution on the validation sample.
This consists of 56~deep fields that almost fully cover the COSMOS field.

The comparison is done on the objects that are both in COSMOS2020 and in the HSC deep fields and where we therefore have the high-quality photometric redshifts available for each galaxy.
We match objects in COSMOS2020 with our reprocessed HSC catalogs based on position and magnitude using the Classic software.
Since our used \sextractor configuration is very similar to the one used to obtain the Classic catalog, the expected magnitude shifts are minimal (see e.g., \cite{Desprez_2023}) and will therefore not bias our photometric redshift assignment by matching errors.
For more details to the data, the blacklisting, the selection, the catalog generation and the estimation of the sample variance in COSMOS2020, we refer to \cite{moser_simulation-based_2024}.
\section{Results}
\label{sec:results}
The fiducial model is obtained in three steps.
First, we validate the emulator-enhanced inference in Section \ref{sec:emu_val}.
Then, we use the emulator to test different adaptions and extensions to the model. 
The newly constrained models are compared to the fiducial model of \cite{moser_simulation-based_2024}.
This is described in detail in Section \ref{sec:explore}.
Finally, we define the fiducial model and present it in Section \ref{sec:fid_model}.

\begin{figure*}[t]
    \centering
    \includegraphics[width=1\linewidth]{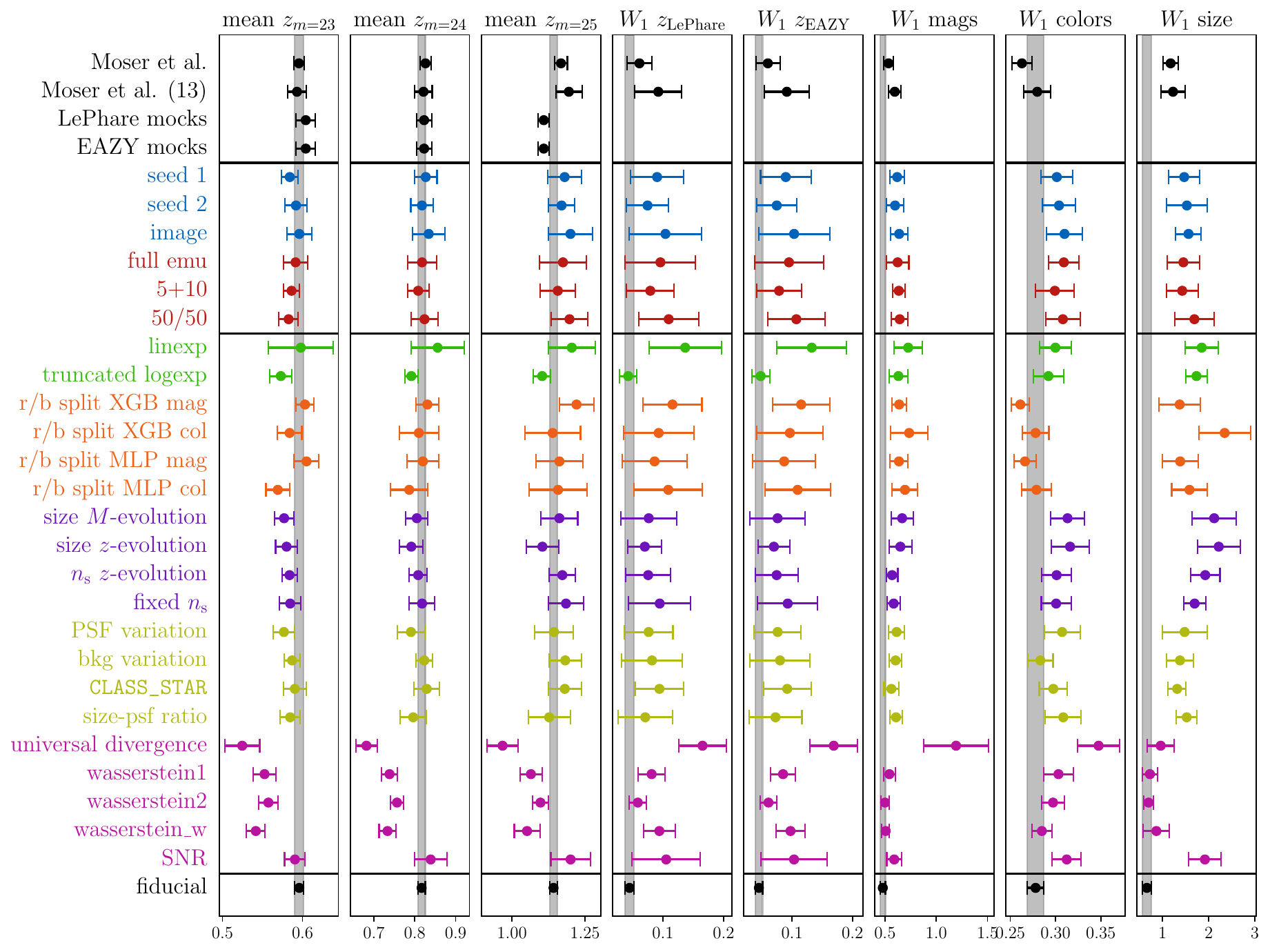}
    \caption{
    Summary of the different tested models.
    The fiducial model from \cite{moser_simulation-based_2024} is shown at the top both after final convergence and after reaching our convergence criterion for model comparison after iteration 13.
    Furthermore, the mean redshifts of the mocks of LePhare and EAZY derived in \cite{moser_simulation-based_2024} are shown.
    All the models below are from this work.
    They are grouped by subsections where they're discussed in detail (see Table \ref{tab:models} for descriptions).
    We evaluate the mean redshift at different magnitude cuts and Wasserstein-1 distance of the overall redshift distribution (lower is better).
    Additionally, multidimensional Wasserstein-1 distances for magnitudes, colors, and sizes in all bands are calculated (lower is better). Comparisons for model selection are made against iteration~13 of \cite{moser_simulation-based_2024}, but the posterior values are also presented for comparison.
    Finally, we show the fiducial model of this work at the bottom and with a gray band in all rows.
    }
    \label{fig:diagnostics}
\end{figure*}

For the various tests conducted, running the inference to full convergence is not necessary.
We define a convergence metric based on the contraction of the posterior distribution relative to the prior.
Specifically, for the set of parameters that are shared across all models, we compute the ratio of the posterior to prior standard deviation.
Convergence is achieved when the median of these ratios falls below 0.5, which we find to be sufficient to draw meaningful conclusions about the model evaluation.
This is typically achieved after 11 to 18~iterations.
However, for the fiducial model, we opt to run the inference to full convergence.
An overview of all models with a short description is given in Table \ref{tab:models} and their diagnostics are shown in Figure \ref{fig:diagnostics}.

\begin{table*}[t]
	\centering
    \begin{tabular}{p{0.005\textwidth} p{0.22\textwidth}p{0.6\textwidth}p{0.02\textwidth}p{0.01\textwidth}}
    	\toprule
         & \textbf{Model identifier} & \textbf{Description} & $n_\mathrm{par}$  &\greentick\\
        \midrule
        \multirow{1}{*}[0pt]{\begin{sideways}\end{sideways}} 
        & Moser et al. & Model explained and constrained in \cite{moser_simulation-based_2024} & 46 & \\
        
        \midrule
        \multirow{6}{*}[0pt]{
            \begin{sideways}
                \footnotesize\ref{sec:emu_val}
            \end{sideways}
        } 
        & seed 1 & Rerun of \cite{moser_simulation-based_2024} with different seed & 46 & \\
        & seed 2 & Rerun of \cite{moser_simulation-based_2024} with different seed & 46 & \\
        & image & Rerun of \cite{moser_simulation-based_2024} with different image selections & 46 & \\
        & full emu & all iterations with emulator & 46 & \\
        & 5+10 & 5~iterations with full images, 10 iterations with emulator & 46 & \greentick\\
        & 50/50 & every second iteration with emulator & 46 & \\
        
        \midrule
        \multirow{2}{*}[0pt]{
            \begin{sideways}
                \footnotesize\ref{sec:lumfunc_parametrizations}
            \end{sideways}
        } 
        & linexp & Parametrization: Equation \ref{eq:linexp} & 46 & \\
        & truncated logexp & Parametrization: Equation \ref{eq:trunc_logexp} & 48 & \greentick \\
        
        \midrule
        \multirow{4}{*}[-15pt]{
            \begin{sideways}
                \footnotesize\ref{sec:redbluesplit}
            \end{sideways}
        } 
        & r/b split XGB mag & red/blue separation by a XGBoost classifier using magnitudes & 46 & \\
        & r/b split XGB col & red/blue separation by a XGBoost classifier using colors & 46 & \\
        & r/b split MLP mag & red/blue separation by a MLP classifier using magnitudes & 46 & \\
        & r/b split MLP col & red/blue separation by a MLP classifier using colors & 46 & \\
        
        \midrule
        \multirow{4}{*}[0pt]{
            \begin{sideways}
                \footnotesize\ref{sec:evolution_morph}
            \end{sideways}
        } 
        & size $M$-evolution & include size evolution with luminosity from Equation \ref{eq:size_extended} & 49 & \greentick \\
        & size $z$-evolution & include size evolution with redshift from Equation \ref{eq:size_evolution} & 48 & \greentick\\
        & $n_s$ evolution & include Sérsic index evolution from Equation \ref{eq:ns_evolution} & 48 & \\
        & fixed $n_s$ & Sérsic index fixed to default values & 44 & \\
        
        \midrule
        \multirow{4}{*}[0pt]{
            \begin{sideways}
                \footnotesize\ref{sec:image_systematics}
            \end{sideways}
        } 
        & PSF variation & additional noise to the PSF parameters & 46 & \\
        & bkg variation & additional noise to the background parameters & 46 & \\
        & \texttt{CLASS\_STAR} & change sample selection to \texttt{CLASS\_STAR}<0.98 & 46 & \\
        & size-psf ratio & change sample selection to size-psf ratio>0.75 & 46 & \\
        
        \midrule
        \multirow{5}{*}[0pt]{
            \begin{sideways}
                \footnotesize\ref{sec:distances}
            \end{sideways}
        } 
        & universal divergence & universal divergence distance instead of MMDs & 46 & \\
        & wasserstein1 & Wasserstein 1 distance instead of MMDs & 46 & \\
        & wasserstein2 & Wasserstein 2 distance instead of MMDs & 46 & \\
        & wasserstein\_w & Wasserstein 1 distance with weights from App. \ref{app:wasserstein_weights} & 46 & \greentick \\
        & SNR & include SNR for each band in the distance & 46 & \\

        \midrule
        & fiducial & incorporating all adaptions marked by a \greentick, and mean redshift distances & 53 & \\
        
        \bottomrule
	\end{tabular}
 \caption{
    Overview of the different models tested in this work.
    Apart from the first model which is the result from \cite{moser_simulation-based_2024}, all models are developed and constrained in this work.
    All analysis choices that are not mentioned in the description of the latter models are the same as the first one from \cite{moser_simulation-based_2024}.
    All models that are not used for emulator validation use the emulator as described in Section \ref{sec:emu}.
    We group the models thematically and refer to the corresponding section in the main text.
    Additionally, we report the number of free parameters $n_\mathrm{par}$ that we constrain using ABC.
    The final column indicates which changes were implemented in the fiducial run.
    In the context of emulator validation, this signifies that the specified emulator strategy is employed across all model explorations.
    For the models below, it denotes that the indicated change is incorporated into the fiducial model.
    }
    \label{tab:models}
\end{table*}

\subsection{Emulator validation}
\label{sec:emu_val}
Both the classifier and the normalizing flow show excellent diagnostics when comparing catalogs of simulated and emulated images.
The classifier produces well-calibrated probability estimates, with the magnitude distributions of detected galaxies in image simulations and emulated data showing near-perfect agreement.
Similarly, the normalizing flow accurately reproduces the magnitude distributions of real image simulations across multiple quantiles and bands. More details are provided in Appendix \ref{app:emu_diag}.

However, most importantly using the emulator should not bias the inference process.
To ensure that the use of the emulator does not introduce any bias into the inference process, we demonstrate that it produces the same results as using image simulations.
Therefore, we repeat the fiducial analysis by \cite{moser_simulation-based_2024} using the emulator.
To begin, we compare different runs with full image simulations to test the inference robustness.
Repeating the fiducial analysis by \cite{moser_simulation-based_2024} with different seeds reveals no significant deviations in diagnostics.
To assess robustness against cosmic variance, we conduct runs with different image selections, again finding no significant deviations.

As described in Section \ref{sec:emu}, the emulator is trained once on a large training set generated from the prior distribution of the first emulator iteration.
We explore three strategies to incorporate the emulator in our inference.
\begin{itemize}
    \item Training the emulator on the prior distribution and using it for all iterations.
    \item Running five iterations with image simulations, then training the emulator on the posterior from iteration 5, and continuing the inference using the emulator.
    \item Running one iteration with image simulations, training the emulator on the first iteration's posterior, then alternating between emulator and images simulations for the remaining iterations (without retraining the emulator).
\end{itemize}
The diagnostics for all three approaches closely match those obtained from pure image simulation runs, see Figure \ref{fig:diagnostics}.
This agreement demonstrates that the methods effectively reproduce the results of the more computationally expensive image simulations.

We adopt the strategy of running five iterations with image simulations before training the emulator.
This precaution accounts for the possibility that a new parametrization could produce highly unusual galaxy populations, for which our emulator setup has not been validated.
However, such extreme cases would be rejected within the first few iterations with image simulations.
While this approach is intentionally conservative, we note that no such extreme models have been encountered in practice for the new parametrizations we tested.
Nonetheless, we maintain this setup to ensure robustness across a wide range of possible scenarios.
We therefore proceed by running five iterations with image simulations, generating the training dataset from the posterior of the fifth iteration, training the emulator, and using it for the subsequent iterations.
This approach produces unbiased parameter constraints and ensures robust diagnostics and -- unlike the alternating strategy -- allows for more emulator iterations, significantly reducing the overall computational cost of the analysis.

\subsection{Model explorations}
\label{sec:explore}

In this section, we test various adaptations and extensions of the model to assess the sensitivity of the inference to different aspects, from the galaxy population model to image systematics and details of the inference process.
The goal is to refine the fiducial model of \cite{moser_simulation-based_2024} by addressing discrepancies between simulations and data, capturing additional physical effects and test the robustness of the model.
Model complexifications are introduced primarily to improve the diagnostics, ensuring that key discrepancies between simulations and data are minimized.
Additionally, some effects present in the data may be constrained even if they do not directly impact the diagnostics, as they could become relevant in other regimes.
To ensure that modifications are both meaningful and reliable, we require that any newly introduced parameters can be constrained by the data without degrading the diagnostics.
At the same time, to avoid unnecessary complexity and overfitting, we only adopt extensions that are physically motivated and supported by other studies.
The changes ultimately adopted in the fiducial model are indicated by a green check in Table \ref{tab:models}.

The fiducial model of \cite{moser_simulation-based_2024} uses the following setup.
The luminosity function is parametrized by Equation \ref{eq:logpower}, the sizes are sampled using the magnitude dependence of Equation \ref{eq:size_simple} and no redshift evolution.
The Sérsic index is a free parameter in the model but without redshift evolution.
They use the fractional distance in number of galaxies and MMDs with the following parameters: apparent magnitude and flux fractions in all bands, size and absolute ellipticities in the $i$-band and the photometric redshifts by LePhare and EAZY.
The distances are computed at two magnitude cuts at $\texttt{MAG\_APER3}<23$ and $\texttt{MAG\_APER3}<25$.
The total distance is computed by 
\begin{equation}
    \label{eq:distance_moser}
    \delta = 0.1 d_\mathrm{frac,23} + 0.1 d_\mathrm{frac,25} + 0.2 \mathrm{MMD}_{23} + 0.6 \mathrm{MMD}_{25},
\end{equation}
with the scaling presented in Section \ref{sec:combined_distances}.
This setup serves as the baseline configuration for the following experiments.
In each subsequent model, only the explicitly mentioned parameters are altered; all unmentioned aspects remain consistent with this default setup.
This model constrains a total of~46 free parameters, evenly divided between red and blue galaxy populations (23 each).
For each population, these parameters are distributed across luminosity functions (5), template coefficients (12), size model (3), ellipticities (2), and light profile (1).

\subsubsection{Luminosity function parametrizations}
\label{sec:lumfunc_parametrizations}
We test the different luminosity function parametrizations given in Equation \ref{eq:linexp}, \ref{eq:logpower} and \ref{eq:trunc_logexp}.
The baseline analysis uses the parametrization from Equation \ref{eq:logpower}.
Switching to the parametrization from Equation \ref{eq:linexp} (linexp) results in a slight performance decline across all diagnostics, with the most noticeable impact on the redshift distribution.
On the other hand, using the parametrization from Equation \ref{eq:trunc_logexp} (truncated logexp) does not affect the photometric diagnostics (comparing magnitudes, colors, or sizes) when compared to the fiducial model.
However, there are noticeable changes in the redshift diagnostics with improved Wasserstein distances and a more consistent mean redshift for the faint sample with a magnitude cut of 25.
This improvement comes at the cost of slightly lower mean redshifts for the brighter samples.
Given the mixed effects on the redshift diagnostics, we consider the parametrization as a potential modification to the fiducial model, which will be tested alongside other adopted changes in Section \ref{sec:fid_model}.
Despite introducing two additional parameters, $z_{0,\mathrm{blue}}$ and $z_{0,\mathrm{red}}$, the model shows good convergence. Both parameters are well-constrained, though their impact on the galaxy distribution is relatively small due to their higher redshift values ($z_0 > 2$).

\subsubsection{Separate distance measures for red and blue galaxies}
\label{sec:redbluesplit}
In the fiducial setup, although the blue and the red populations are parametrized as fully independent; they are constrained simultaneously with the same distance measure.
Since blue galaxies dominate the overall population, the red galaxies might not be weighted enough in the distance measure for unbiased parameter constraints on this population.
To address this, we split both simulation and data in a red and a blue catalog and use separate distance measures with equal weights for the two populations.
The total distance is given by 
\begin{equation}
\label{eq:distance_redblue}
    \begin{split}
        \delta = & \ 0.05 d_\mathrm{frac,blue,23} + 0.05 d_\mathrm{frac,red,23} + 0.05 d_\mathrm{frac,blue,25} + 0.05 d_\mathrm{frac,red,25} \\
            & +0.1 \mathrm{MMD}_\mathrm{blue,23} + 0.1 \mathrm{MMD}_\mathrm{red,23} + 0.3 \mathrm{MMD}_\mathrm{blue,25} + 0.3 \mathrm{MMD}_\mathrm{red,25}.
    \end{split}
\end{equation}

The population split is done using a classifier which is trained on simulations from the posterior distribution of \cite{moser_simulation-based_2024}.
We test different classifier architectures; namely boosted decision trees and dense neural networks; and different input features; magnitudes and colors; but find no significant differences between the setups.

Overall, we find that splitting the populations leads to improved agreement in the color diagnostics.
However, this improvement comes at the expense of larger error bars in other diagnostics, such as the size distribution.
Since a change in the size distribution changes the detection probability of galaxies, a less constrained size model also leads to larger variability in the redshift diagnostics.
Running one of these models to full convergence balances out these discrepancies, and we find similar agreement in both color and redshift as \cite{moser_simulation-based_2024}.

In conclusion, performing inference on both populations combined does not introduce a bias in the constraints on the individual populations.
Instead, it affects the rate at which different parameters are constrained. 
Importantly, the final results remain robust and are not sensitive to changes in the population selection within the distance measures.

\subsubsection{Morphology}
\label{sec:evolution_morph}
Since \cite{moser_simulation-based_2024} found a slight discrepancy in the size distribution, exploring extended size models is one of the key goals of this work.
Specifically, we investigate the extended model for the dependence on the absolute magnitude given by Equation \ref{eq:size_extended} and \ref{eq:size_sigma}, adding two parameters for the blue population and one parameter for the red population to the model.
This results in a total of 49~parameters to constrain.
The additional parameters are well constrained by the data indicating that the data supports a magnitude-dependent scatter.
However, the diagnostics do not improve, mainly due to slower convergence due to the higher dimensionality.

We observe a similar trend when incorporating redshift evolution, as described in Equation \ref{eq:size_evolution}, introducing two additional parameters to our inference model ($\eta_{r_{50}}$ for the red and the blue population).
The new parameters are well constrained, indicating that a redshift evolution of the size is present in the data.
However, we again do not find an improvement in the diagnostics.

We adopt both modifications in our fiducial model because the additional parameters are both supported by the literature and can be well constrained by the data used in this work.
In both cases above, the parametrization used in \cite{moser_simulation-based_2024} is nested within the extended parametrizations.
It is important to note that the parameters are not constrained in a way that would reduce the more complex parametrization to the simpler one. This indicates that these effects are genuinely present in the data, providing justification for adopting the more complex extensions.

Apart from the size, we investigate model adaptions for the light profile.
First, we include redshift evolution in the Sérsic index $n_\mathrm{s}$ by adding two new parameters $\alpha_{n_s}$ for red and blue galaxies as given in Equation \ref{eq:ns_evolution}.
This has very little impact on the diagnostics apart from a slightly slower convergence due to the additional two parameters.
Since the additional parameters are barely constrained, we are not adopting this for our fiducial model.
We additionally test the impact of fixing the Sérsic index to the default parameters of the \galsbi package, reducing the parameter space by 2.
We find that the model is slightly faster converging but we find no improvement, e.g.\ by having a more stable convergence for the size parameters.
Since the free Sérsic index get constrained by the data, we are not adopting this simplification of the model.

\subsubsection{Image systematics}
\label{sec:image_systematics}
Variations in the PSF or the background values can impact the measured catalog values and have therefore the potential to bias the parameter results.
We investigate the robustness of the results towards small changes in these image systematics.
The PSF full width half maximum (FWHM) is varied for each pixel with a scatter roughly corresponding to the PSF FWHM scatter across all images.
We find that the parameter constraints and the diagnostics are not affected by this scatter.
Similarly, we vary both mean background value and the background scatter according to their scatter across images.
Again, neither parameter constraints nor the diagnostics are affected.

Furthermore, we change the galaxy selection functions.
The selection is based on \texttt{CLASS\_STAR} by \sextractor and the object-size-to-PSF-ratio which is computed as the ratio between FWHM of the object and the FWHM of the size.
In the fiducial case, we select objects with $\texttt{CLASS\_STAR}<0.95$ and size-psf ratio > 0.5.
We test the impact of $\texttt{CLASS\_STAR}<0.98$ and size-psf ratio > 0.75.
These changes significantly affect the total number of objects in the catalogs, but it impacts both simulation and data.
When applying the constrained galaxy population model to the validation sample, comparing the diagnostics using the fiducial selections, we find no difference in the diagnostics compared to the case where the galaxy population model was also constrained with the fiducial selection function.
Therefore, the inference is robust to changes in the galaxy selection function.

\subsubsection{Distance measures}
\label{sec:distance_measures}
The distance measure in an ABC is the summary statistic that compresses the two multidimensional catalogs to one number.
As every summary statistics, the distance measure can be more or less informative and can have different sensitivities.
We test the impact of changing the method of computing the multidimensional distance measure.

First, we replace the MMD distances with universal divergences.
The new distance is given by
\begin{equation}
    \label{eq:ud_test}
    \delta = 0.1 d_\mathrm{frac,23} + 0.1 d_\mathrm{frac,25} + 0.2 D_{23} + 0.6 D_{25},
\end{equation}
with $D_{23}$ and $D_{25}$ given by Equation \ref{eq:universal_div} with magnitude cuts at $\texttt{MAG\_APER3}<23$ and $\texttt{MAG\_APER3}<25$.
The weights are the same as in \cite{moser_simulation-based_2024} but with universal divergences instead of MMDs.
Using this distance in the ABC worsens all our diagnostics significantly except for the size distribution which is slightly improved.
We test if choosing a different value for $k$ helps improving the result and find no significant improvement.
The reason for the worse performance is the fact that due to finite sampling, the universal divergence can become negative, especially in high dimensions and if the second distribution has a lower spread than the target distribution.
We discuss this effect further in Appendix \ref{app:distance_toy_model} using a toy model.

Second, we use Wasserstein distances instead of MMDs.
The new distance is given by
\begin{equation}
    \label{eq:ud_test}
    \delta = 0.1 d_\mathrm{frac,23} + 0.1 d_\mathrm{frac,25} + 0.2 W_{p,23} + 0.6 W_{p,25},
\end{equation}
with $p=1$ or $p=2$ and uniform weights.
For both $p=1$ or $p=2$, the diagnostics improve significantly, especially for the size and the full redshift distribution.
This comes at the cost of slightly worse agreement in the redshift distribution for the bright samples.
We find that the Wasserstein-1 distance is more sensitive to the shapes of the distribution which is highly beneficial for photometric properties, especially the sizes; but lacks a bit of sensitivity regarding the mean of the distribution which is the most relevant quantity for the redshift measurement \cite{amon_dark_2022,dalal_hyper_2023,busch_kids-1000_2022,li_kids-1000_2023}, see Appendix \ref{app:distance_toy_model} for more details.

Third, we use a weighted Wasserstein-1 distance following the weighting scheme presented in Appendix \ref{app:wasserstein_weights}.
With this approach, the reweighting system using different magnitude cuts is no longer necessary, as the reweighting is directly incorporated into the weights.
The new combined distance consists therefore of one less distance
\begin{equation}
    \label{eq:ud_test}
    \delta = 0.1 d_\mathrm{frac,23} + 0.1 d_\mathrm{frac,25} + 0.8 W_{p,\mathrm{w}},
\end{equation}
where the weights are chosen such that the weight of the Wasserstein distance is the same as the sum of the two Wasserstein distances before.
We observe similar improvements compared to the Wasserstein distances with uniform weights and the same drawbacks.
Since the new weighting scheme simplifies the inference by reducing the number of distances and is more motivated than combining distances at two different magnitudes cuts, we adopt this distance for the fiducial model.

Finally, we investigate the impact of additional galaxy properties in the distance measure and add the logarithm of the signal-to-noise ratio (SNR) in each band to the parameters.
The distance is still computed with the MMD distances defined in Equation \ref{eq:distance_moser}.
Since the diagnostics do not improve, we do not include the SNR to the distances of the fiducial model.

\subsection{Fiducial model}
\label{sec:fid_model}
In Section \ref{sec:explore}, we tested different model extensions and their performance.
Based on these findings, we adopt the extended size model with redshift evolution and the weighted Wasserstein distance.

Furthermore, since the Wasserstein distance is lacking sensitivity in the mean of the redshift distribution, we test the impact of adding the difference in mean redshift (see Equation \ref{eq:distance_mean_redshift}) as an additional distance.
This approach performs significantly better in the redshift distribution, especially for the bright samples, which is why we adopt this distance for our fiducial model.
This new model is then tested to full convergence with the luminosity function parametrizations given by Equation \ref{eq:trunc_logexp} (truncated logexp) and Equation \ref{eq:logpower} (logpower) and we find consistently better diagnostics for the new parametrization (truncated logexp) and therefore adopt it.
We further test the impact of splitting all or some of the distances by population with this new setup and find no significant improvement.
Compared to the model in \cite{moser_simulation-based_2024}, we are therefore adopting the following changes:
\begin{itemize}
    \item using an emulator from iterations 6 on,
    \item new parametrization of the luminosity function given by Equation \ref{eq:trunc_logexp},
    \item extended size model incorporating the dependence on absolute magnitude given by Equation \ref{eq:size_extended} and \ref{eq:size_sigma} as well as the redshift evolution parametrized by Equation \ref{eq:size_evolution},
    \item replace the two MMD distances by one weighted Wasserstein-1 distance following the scheme outlined in Appendix \ref{app:wasserstein_weights}.
\end{itemize}

This new model consists of a total of 53 parameters with the combined distance given by
\begin{equation}
    \label{eq:final_distance}
    \delta = 0.1 d_\mathrm{frac,23} + 0.1 d_\mathrm{frac,25} + 0.1 d_{\mu_z,23} + 0.1 d_{\mu_z,25} + 0.6 W_{1,\mathrm{w}}.
\end{equation}
Since we extend the parameter space compared to \cite{moser_simulation-based_2024}, we also extend the prior for our final run.
The prior for each parameter is given in Table \ref{tab:prior} and is motivated by observations.

\begin{table}[tbp]
    \centering
    \begin{tabular}{llllll}
        \toprule
        & & Parameter & Definition & Prior\\

        \midrule
        \multirow{12}{*}{\begin{sideways}Luminosity functions\end{sideways}} &
        \multirow{12}{*}{\begin{sideways}(12)\end{sideways}}
        
        & $\phi^*_\mathrm{1,b}$ 
        & Equation~\ref{eq:trunc_logexp} 
        & $\mathcal{U}[1.1\times10^{-5}, 1.2\times10^{-2}]$
        \\

        && $\phi^*_\mathrm{2,b}$ 
        & Equation~\ref{eq:trunc_logexp} 
        & $\mathcal{U}[-2, 1.5]$
        \\

        && $\phi^*_\mathrm{1,r}$ 
        & Equation~\ref{eq:trunc_logexp} 
        & $\mathcal{U}[2\times10^{-8}, 2.5\times10^{-2}]$
        \\

        && $\phi^*_\mathrm{2,r}$ 
        & Equation~\ref{eq:trunc_logexp} 
        & $\mathcal{U}[-11, 7]$
        \\
        
        && $M^*_\mathrm{1,b}$ 
        & Equation~\ref{eq:trunc_logexp} 
        & $\mathcal{U}[-23, -16]$
        \\

        && $M^*_\mathrm{2,b}$ 
        & Equation~\ref{eq:trunc_logexp} 
        & $\mathcal{U}[-6, 1.5]$
        \\
        
        && $M^*_\mathrm{1,r}$ 
        & Equation~\ref{eq:trunc_logexp} 
        & $\mathcal{U}[-23, -17]$
        \\

        && $M^*_\mathrm{2,r}$ 
        & Equation~\ref{eq:trunc_logexp} 
        & $\mathcal{U}[-4, 3]$
        \\

        && $z_\mathrm{const,r/b}$ 
        & Equation~\ref{eq:trunc_logexp} 
        & $\mathcal{U}[0.1, 5]$
        \\

        && $\alpha_\mathrm{blue}$ 
        & Equation~\ref{eq:lumfunc}
        & $\mathcal{U}[-1.5, -0.5]$
        \\

        && $\alpha_\mathrm{red}$ 
        & Equation~\ref{eq:lumfunc} 
        & $\mathcal{U}[-0.8, -0.1]$
        \\ 
        
        \midrule
        \multirow{13}{*}{\begin{sideways} Galaxy morphology\end{sideways}} &
        \multirow{13}{*}{\begin{sideways} (17) \end{sideways}} 
        & $\alpha_{\log{r_{50}},b}$ 
        & Equation \ref{eq:size_extended}
        & $\mathcal{U}[0.001, 1]$ 
        \\

        && $\beta_{\log{r_{50}},b}$ 
        & Equation \ref{eq:size_extended}
        & $\mathcal{U}[0.001, 1]$ 
        \\

        && $\gamma_{\log{r_{50}},b}$ 
        & Equation \ref{eq:size_extended}
        & $\mathcal{U}[-5, 5]$ 
        \\

        && $a_{\log{r_{50}},r}$ 
        & Equation \ref{eq:size_extended}
        & $\mathcal{U}[0.001, 1]$ 
        \\
        
        && $b_{\log{r_{50}},r}$ 
        & Equation \ref{eq:size_extended}
        & $\mathcal{U}[-10, 0]$ 
        \\

        && $\sigma_{1/2,\log{r_{50}},b/r}$ 
        & Equation \ref{eq:size_sigma}
        & $\mathcal{U}[0.001, 1]$ 
        \\

        && $\eta_{r_{50},b/r}$ 
        & Equation \ref{eq:size_evolution}
        & $\mathcal{U}[-2, 0]$ 
        \\
        
        && $n_\mathrm{s,b}$ 
        & Section \ref{sec:morphology}
        & $\mathcal{U}[0.2, 2]$ 
        \\

        && $n_\mathrm{s,r}$ 
        & Section \ref{sec:morphology}
        & $\mathcal{U}[1, 4]$ 
        \\
        
        && $e_\mathrm{mode,b/r}^{\rm{blue/red}}$ 
        & Section \ref{sec:morphology}
        & $\mathcal{U}[0.01, 0.99]$ 
        \\

        && $e_\mathrm{spread,b/r}^{\rm{blue/red}}$ 
        & Section \ref{sec:morphology}
        & $\mathcal{U}[2,4]$ 
        \\
        
        \midrule
        \multirow{2}{*}{\begin{sideways}SED\end{sideways}} &
        \multirow{2}{*}{\begin{sideways}(24)\end{sideways}} 
        & $\alpha^{\rm{blue/red}}_{i, 0/3}$ & Section \ref{sec:templates}
        & \cite{moser_simulation-based_2024}, Section A.1
        \\

        && $\alpha^{\rm{blue/red}}_{ \mathrm{std}, 0/3}$ 
        & Section \ref{sec:templates}
        & $\mathcal{U}[10^{-4}, 0.16]$
        \\ 
        \bottomrule
   \end{tabular}
    \caption{
    We indicate for each set of parameters where they are defined in the text and give the prior.
    $\mathcal{U}$ denotes uniform distributions.
    If the priors are the same for the red and blue population, or for similar parameters, this is indicated in the parameter name by ''r/b'', ''1/2'', etc.
    }
    \label{tab:prior}
\end{table}

The fiducial model is run until all images of the data set are used, which is the case after 31~iterations.
The constraints on the parameters of the galaxy population are given in Figure \ref{fig:galpop_lumfunc}, \ref{fig:galpop_templates} and \ref{fig:galpop_morph}.
All parameters of the model are constrained and only a few of the template parameters, $z_\mathrm{const,b}$ and $\sigma_{1\log r_{50},b}$, are close to their prior limits.
Since all the approached limits are physical boundaries, we are not concerned about this.

In \cite{tortorelli_measurement_2020}, the luminosity function measured using the GalSBI methodology is compared to results from other studies at various redshifts \cite{beare_z_2015,giallongo_b-band_2005,ilbert_vimos-vlt_2006,zucca_zcosmos_2009,loveday_galaxy_2012,cool_galaxy_2012,fritz_vimos_2014}.
The findings of this work align well with these external measurements and those presented in \cite{tortorelli_measurement_2020}, but with notably reduced uncertainty, particularly at high redshift.

\begin{figure*}[t]
    \centering
    \includegraphics[width=1\linewidth]{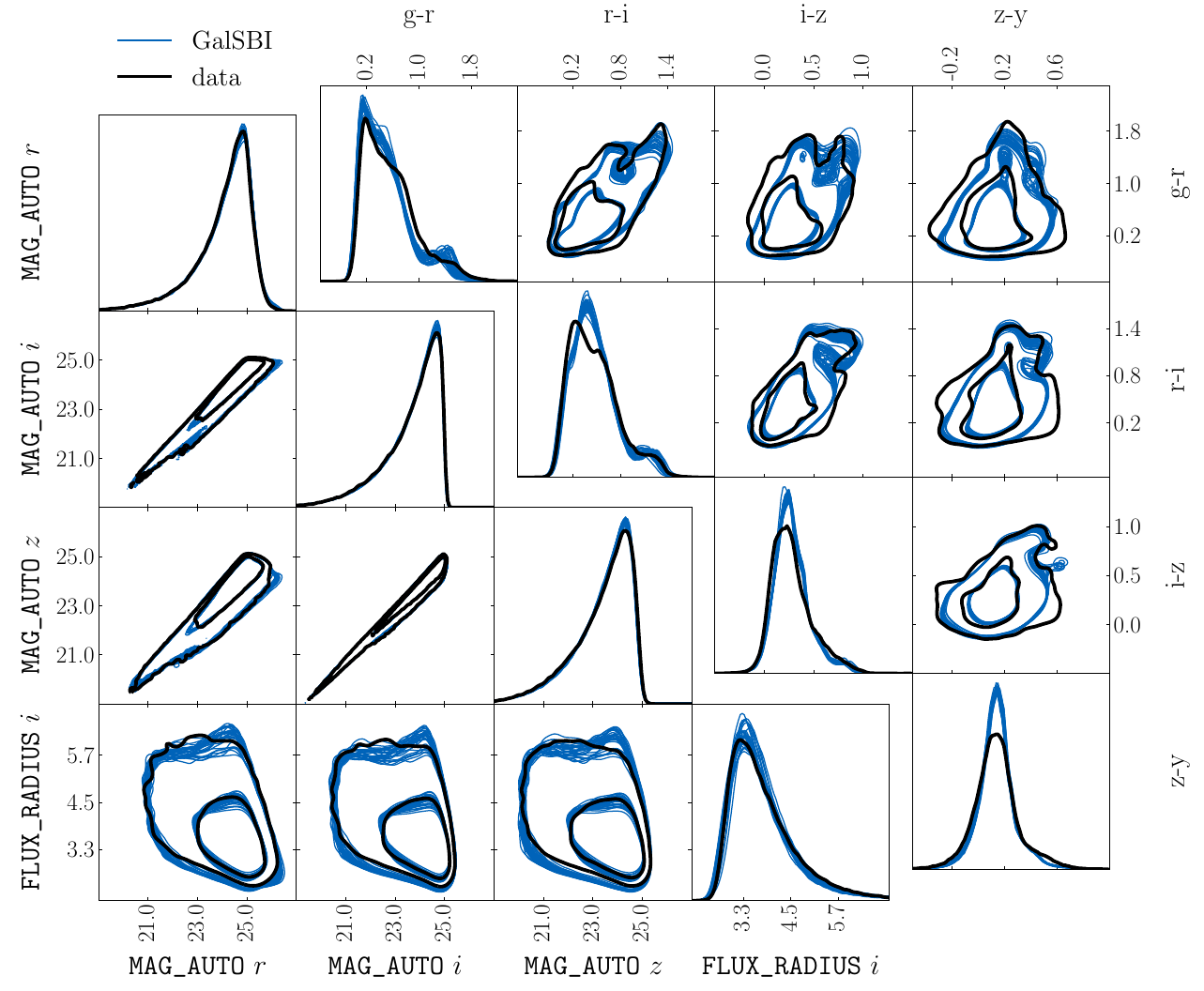}
    \caption{Comparison of photometric properties between simulation (blue) and data (black).}
    \label{fig:photo}
\end{figure*}

The comparison of some photometric properties between simulations and real data is given in Figure \ref{fig:photo}.
The agreement in magnitude, size and color distribution is excellent, especially the size distribution has clearly improved in comparison to \cite{moser_simulation-based_2024}.

\begin{figure}[t]
    \centering
    \includegraphics[width=1\linewidth]{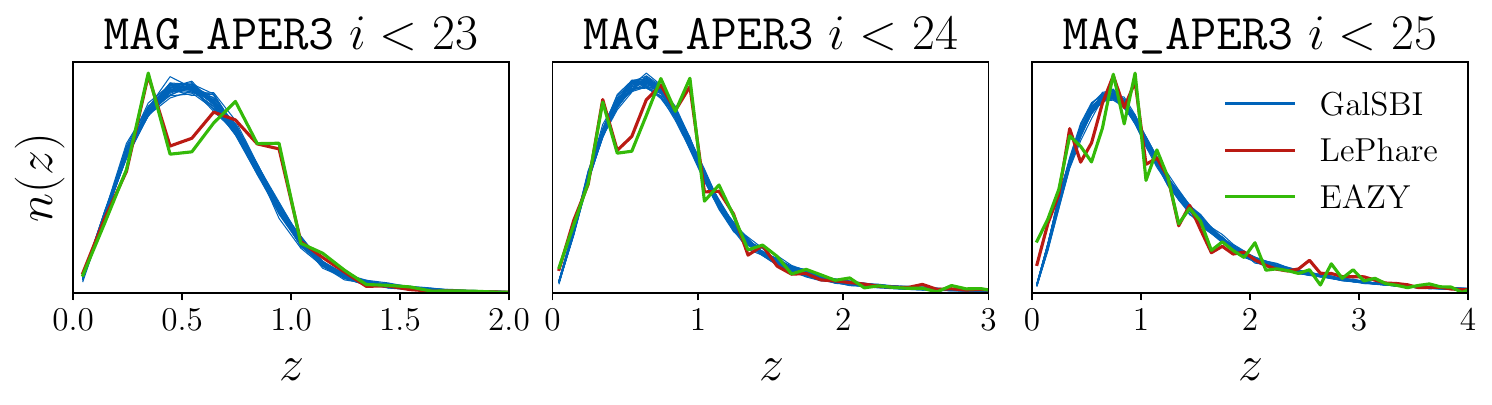}
    \caption{
    Comparison of redshift distribution for different magnitude cuts between simulation (blue) and data.
    The photometric redshift for the data is estimated using LePhare (red) and EAZY (green).
    }
    \label{fig:redshift}
\end{figure}

The photometric redshifts for different magnitude cuts are shown in Figure \ref{fig:redshift}.
We compare our redshift estimates with the photometric redshifts from LePhare and EAZY.
The means and uncertainties are shown in Table \ref{tab:z_estimates}.
Since our simulations do not contain clustering, they are not affected by cosmic variance which is why the distributions are much smoother than the distributions measured in COSMOS2020 by the photo-z codes.
Furthermore, this means that our reported uncertainty is underestimating the real uncertainty since it is only including the uncertainty of the galaxy population model.
This effect will be especially strong for the bright samples, where \cite{moser_simulation-based_2024} already estimated a systematic offset and uncertainty due to cosmic variance in the COSMOS2020 field.
Taking these effects into account, the total redshift distribution agrees at the $1.5\sigma$ level at all magnitude cuts, with better agreement with LePhare.
\begin{table}[t]
    \centering
    \begin{tabular}{lllll}
    \toprule
         & $m < 23$ &  $m < 24$ & $m < 25$\\
         \midrule
         $\bar{z}_{\mathrm{GalSBI}}$& $0.596 \pm 0.006$ & $0.817 \pm 0.009$ & $1.143 \pm 0.013$ \\
         $\bar{z}_{\mathrm{LP}}$ & 0.618 & 0.824& 1.110\\
         $\bar{z}_{\mathrm{LP,\ mocks}}$ & $0.604 \pm 0.012$ & $0.823 \pm 0.019$ & $1.109 \pm 0.019$ \\
         $\bar{z}_{\mathrm{E}}$ &0.632 & 0.842 & 1.110\\
         $\bar{z}_{\mathrm{E,\ mock}}$ & $0.617 \pm 0.012$ & $0.841 \pm 0.019$ & $1.108 \pm 0.018$ \\
         \midrule
         $\sigma$ vs.\ LP & 0.6 & 0.3 & 1.5 \\
         $\sigma$ vs.\ E & 1.5 & 1.14 & 1.6 \\
         $\sigma$ LP vs.\ E & 0.8 & 0.3 & 0.04\\
        \bottomrule
    \end{tabular}
    \caption{
    We report the mean redshift for GalSBI, LePhare (LP) and EAZY (E) at different cuts of the measured magnitude $m$ in the $i$ band for the validation sample.
    We further report the mean redshift and their uncertainty for LePhare and EAZY using the reweighted mocks from \cite{moser_simulation-based_2024}.
    Using these uncertainties, we estimate the agreement in $\sigma$ between the different methods assuming uncertainties add in quadrature.
    }
    \label{tab:z_estimates}
\end{table}

A good agreement for the redshift distribution for different magnitude cuts is a necessary condition for a galaxy population model to be applicable to both Stage-III and Stage-IV surveys.
However, in cosmic shear surveys, the redshift sample is typically split into tomographic bins.
Comparing tomographic bin assignments of the different methods is therefore a good test of how well the color-redshift relation is reproduced in the simulations.

\begin{figure}[t]
    \centering
    \includegraphics[width=1\linewidth]{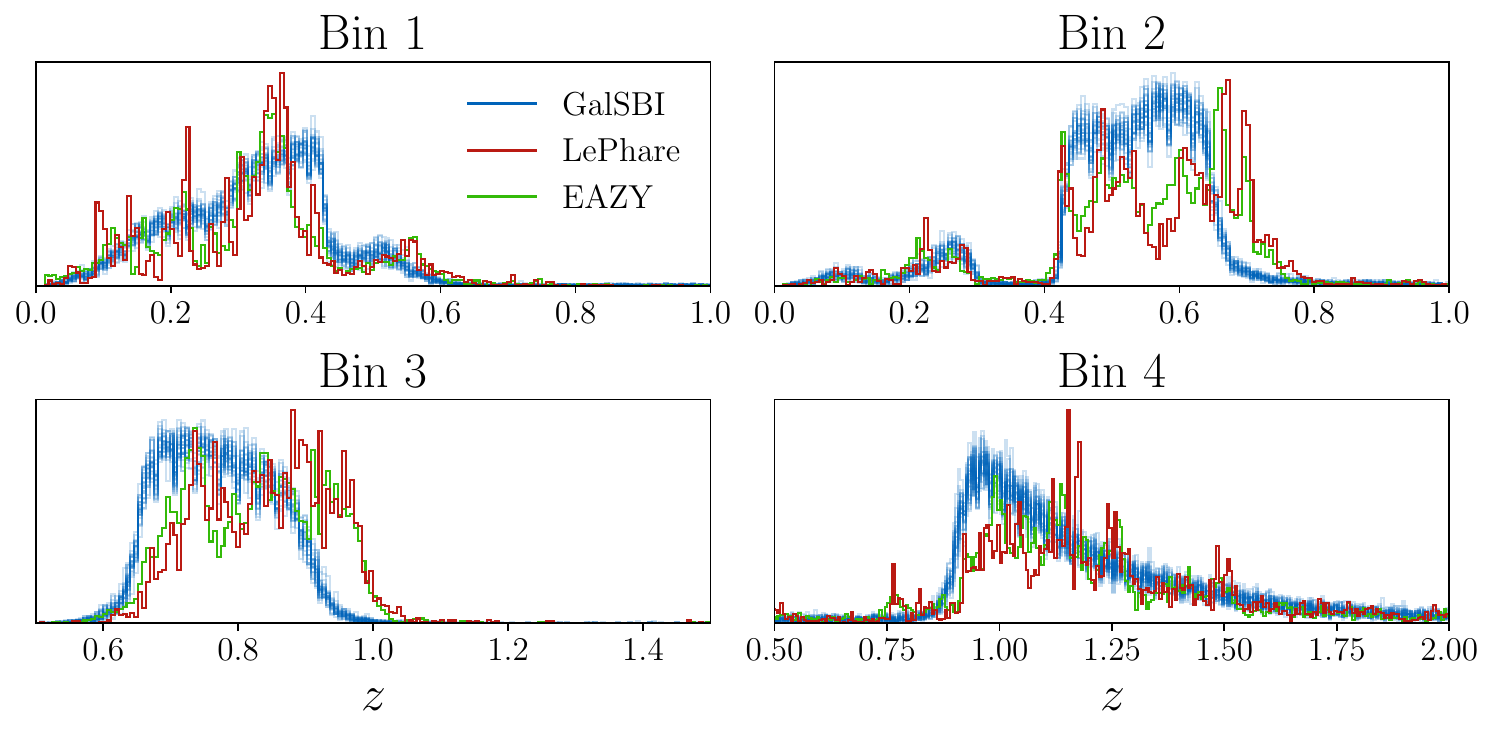}
    \caption{
    Tomographic bin assignment for the Stage-III survey setup.
    }
    \label{fig:tomographic235}
\end{figure}

\begin{figure}[t]
    \centering
    \includegraphics[width=1\linewidth]{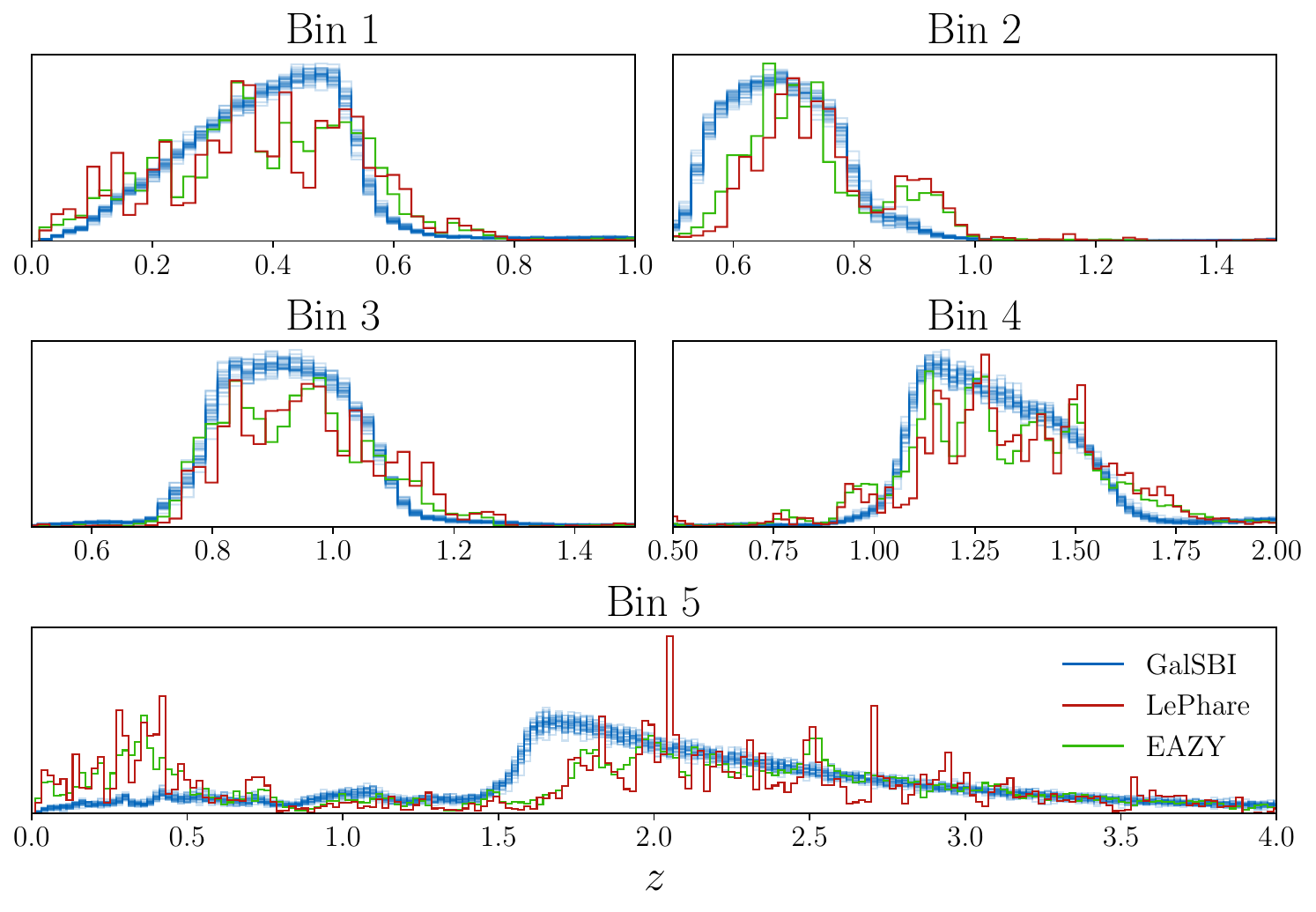}
    \caption{
    Tomographic bin assignment for the Stage-IV survey setup.
    }
    \label{fig:tomographic25}
\end{figure}

We perform simple magnitude cuts to both simulations and data.
We use $\texttt{MAG\_AUTO}<23.5$ for the Stage-III sample following the DES-Y3 analysis choice \cite{amon_dark_2022,secco_dark_2022} and $\texttt{MAG\_AUTO}<25$ for the Stage-IV sample.
We split the simulation into $n$ equally populated bins and train a boosted decision tree on the bin assignment using colors as input parameters.
This classifier is then applied to both simulation and data to obtain the tomographic redshift distribution.
We use $n=4$ for Stage-III following the DES-Y3 analysis \cite{amon_dark_2022,secco_dark_2022} and $n=5$ for Stage-IV following the LSST science requirement document \cite{the_lsst_dark_energy_science_collaboration_lsst_2021}.
The tomographic bins are shown in Figure \ref{fig:tomographic235} and \ref{fig:tomographic25}.
The mean redshifts for each bin and both setups are given in Table \ref{tab:tomographic_redshifts}.

\begin{table*}[t]
    \centering
    \begin{tabular}{lllllll}
    \toprule
         &  & Bin 1 & Bin 2 & Bin 3 & Bin 4 & Bin 5\\
         \midrule
         \multirow{3}{*}[0pt]{
            \begin{sideways}
            S-III
            \end{sideways}
        }
         & $\bar{z}_{\mathrm{GalSBI}}$ & $0.364 \pm 0.006$ & $0.547 \pm 0.005$ & $0.780 \pm 0.003$ & $1.225 \pm 0.017$ & -\\
         & $\bar{z}_{\mathrm{LePhare}}$ & 0.362 & 0.554 & 0.827 & 1.211 &-\\
         & $\bar{z}_{\mathrm{EAZY}}$ & 0.375 & 0.570 & 0.844 & 1.236 &-\\
         \midrule
         \multirow{6}{*}[0pt]{
            \begin{sideways}
            S-IV
            \end{sideways}
        }
         & $\bar{z}_{\mathrm{GalSBI}}$ & $0.515 \pm 0.018$ & $0.722 \pm 0.012$ & $0.993 \pm 0.009$ & $1.34 \pm 0.006$ & $2.20 \pm 0.05$\\
         & $\bar{z}_{\mathrm{LePhare}}$ & 0.554 & 0.754 & 0.998 & 1.33 & 1.90\\
         & $\bar{z}_{\mathrm{EAZY}}$ & 0.558 & 0.766 & 1.011 & 1.34 & 1.87\\
         & $\tilde{z}_{\mathrm{GalSBI}}$ & $0.408 \pm 0.004$ & $0.677 \pm 0.006$ & $0.934 \pm 0.005$ & $1.291 \pm 0.004$ & $2.06 \pm 0.03$\\
         & $\tilde{z}_{\mathrm{LePhare}}$ & 0.406 & 0.707 & 0.945 & 1.293 & 2.06\\
         & $\tilde{z}_{\mathrm{EAZY}}$ & 0.410 & 0.728 & 0.959 & 1.327 & 2.05\\
         \bottomrule
    \end{tabular}
    \caption{
    We report the redshift summaries for GalSBI, LePhare, and EAZY for each tomographic bin and for both the Stage-III (S-III) and Stage-IV (S-IV) setup.
    $\bar z$ denotes mean redshift where $\tilde z$ denotes median redshift.
    }
    \label{tab:tomographic_redshifts}
\end{table*}

Although we use a very simple tomographic bin assignment method, we find very good agreement between our simulations and the two photo-z codes for the Stage-III setup.
For the first and last bin, our mean redshift lies between the two mean redshifts of the photo-z codes.
For the second bin, we are slightly lower than both.
However, the difference between the mean LePhare redshift and the mean redshift by EAZY is larger than the difference between our mean and LePhare.
Only the third bin shows a slight disagreement.

For the Stage-IV setup, the requirements are tighter and this very simple tomographic bin assignment method might reach its limitations.
We find good agreement for the intermediate bins 3 and 4 where the difference between our mean redshift is of the same order than the difference between the other photo-z codes.
For the first two and the last bin, the mean deviates more.
For these bins, we have outliers either at very high ($z\gtrsim2.5$ for the first two bins) or very low redshift ($z\sim 0.5$) which are driving the discrepancy in the mean.
The median for both bins agrees with the photo-z code at the $1\sigma$-level.
We test how these outliers in the redshift distributions affect a typical observable such as the angular power spectrum $C_\ell$.
We find that the $C_\ell$ computed from LePhare or EAZY redshifts falls within the uncertainty band of the $C_\ell$ computed from the different GalSBI realizations for the Stage-III setup and the first four bins of the Stage-IV setup.
There is a slightly higher disagreement for the cross-correlations that include the last redshift bin.
The uncertainty of our redshifts are again underestimating the real uncertainty due to the absence of cosmic variance.
To apply GalSBI to estimate the photometric redshift distribution for upcoming surveys, this additional uncertainty should be estimated, either by including clustering in the simulations or by estimating the impact of clustering on the mean.
Furthermore, a more refined bin assignment methodology could help to reduce the outliers in the bin assignment, see e.g. \cite{tortorelli_impact_2024} where the authors used a self-organizing map (SOM) to perform tomographic bin assignment of simulated galaxies.
\section{Conclusions}
\label{sec:conclusion}
We have introduced GalSBI, a versatile phenomenological galaxy population model with broad applications in cosmology, constrained by observational data.
While previous work was limited by the computationally intensive inference, this work presents an emulator that directly predicts output catalogs of simulations based on a catalog of intrinsic properties.
The introduction of our emulator represents a significant methodological advancement in our modelling efforts.
By reducing the computational cost, we were able to explore and rigorously evaluate over forty different model configurations and analysis choices.
This represents a major increase compared to previous approaches, fundamentally expanding our ability to test and refine the galaxy population model.

It is crucial to note that achieving unbiased constraints from the emulator-enhanced inference requires more than just estimating the photometric noise arising from image systematics such as PSF and background level.
It is equally important to accurately include the impact of blending, which is not solely depending on the survey depth but also varies with the galaxy population model itself.
We have thoroughly validated our emulator's performance and found that our approach does not introduce a bias in the inference of the galaxy population model.

This highlights the power of an image-simulation-based approach, where effects such as blending, source extraction processes, and variations in background or PSF levels can be easily forward-modelled.
The strong dependence on blending demonstrates that image simulations may be necessary to fully exploit the depth of a survey.
This effect will become even more significant as future surveys continue to reach deeper depths.
Another advantage of image simulations is that by reproducing the corresponding measurement pipelines, a direct and consistent catalog comparison becomes straightforward.

We have explored and implemented several model extensions, resulting in a new fiducial model that shows improved performance compared to previous work.
Notable improvements are evident for the size distribution and for the redshift distribution with a high magnitude cut.
We find good agreement to established photometric redshift estimation methods, specifically the LePhare and EAZY photo-z codes.
This agreement is evident both in the overall distribution at different magnitude cuts and in the tomographic redshift distributions for Stage-III and Stage-IV-like survey scenarios.
These findings highlight the potential of this methodology for upcoming surveys, especially since it does not require spectroscopic calibration data overlapping with the targeted survey.

The underlying galaxy population model, parametrized by physically motivated components such as luminosity functions and observed size distribution scalings, is directly applicable to other datasets.
Given that the HSC deep fields already reach a depth comparable to that expected from Stage-IV surveys, the model should be well-suited to meet the requirements of both current and future surveys.

GalSBI can be applied to various tasks, including photometric redshift estimation, image-related calibrations, and forward modelling of selection effects in survey setups.
To facilitate its use by the broader community, we are releasing all the code necessary to generate realistic galaxy catalogs.
The Python package \galsbi provides an easy-to-use interface to generate intrinsic catalogs using the ultra fast catalog generator \ucat from a given galaxy population model with just a few lines of code.
An example how to generate a galaxy catalog for a given survey area with the fiducial model of this work is given in Listing \ref{lst:galsbi_example}.
Additionally, \galsbi can generate realistic images using \ufig \footnote{\url{https://cosmo-docs.phys.ethz.ch/ufig/}} (see \cite{fischbacher_ufig_2024} for a description of the \ufig package) or produce measured catalogs using a pretrained emulator.
For further description of the usage, we refer to \cite{fischbacher_galsbi_2024} and the code documentation\footnote{\url{https://cosmo-docs.phys.ethz.ch/galsbi/}}.
The published code is open source and adaptable to any research workflow.

\begin{figure}[t]
\centering
\begin{lstlisting}[
    language=python,
    caption={Example of how to generate a galaxy catalog using the fiducial model from this work with the \galsbi Python package. The code defines a survey area using a HEALPix mask, with \( N_{\text{pix}} = 12 \times \text{Nside}^2 \) and \( \text{Nside} = 64 \), resulting in 49,152 pixels. The first pixel (pixel 0) is marked to specify the survey region.
},
    label={lst:galsbi_example}
    ]
from galsbi import GalSBI

# Define survey area as healpix mask
mask = np.zeros(12 * 64**2)
mask[0] = 1

model = GalSBI("Fischbacher+24")
model(healpix_map=mask)
cats = model.load_catalogs()
\end{lstlisting}
\end{figure}

For researchers interested in adopting our galaxy population model, we recommend using the fiducial model developed in this work, which demonstrates improved performance in size and redshift distributions compared to the model presented in \cite{moser_simulation-based_2024} which is also available in the \galsbi package.
The model generates galaxy catalogs with realistic photometric and redshift distribution for any survey area.
This can for example be used to study the impact of selection functions or the photometric properties of the galaxies can be added to existing clustering simulations.
The intrinsic galaxy catalogs are independent of the survey.
To properly model the photometric noise of a survey, additional image simulations (e.g., with \ufig or pretrained emulators) are required.

Although we have improved the model in terms of accuracy and precision, upcoming Stage-IV surveys will drastically reduce the statistical uncertainties, making the analysis requirements extremely challenging to meet. 
Further improvements to the model may be necessary to address these demands.
One key adjustment is the addition of realistic clustering to our simulation, which will help us to better understand its impact on the redshift distribution and shed light on the sources of the deviations in mean redshift observed so far.
This enhancement is made possible by recent advancements in forward modelling of galaxy clustering \cite{berner_rapid_2022,berner_fast_2024,fischbacher_sham-ot_2025}.

The HSC deep fields were a well-suited data set to constrain the galaxy population model up to high redshift, however, they lack infrared coverage.
Applying GalSBI to a survey with infrared data, such as KiDS-VIKING \citep{de_jong_kilo-degree_2013,edge_vista_2013}, would offer a stringent test of the SED modelling presented here.
Performing an inference, either partly or fully, on such a data set could improve the constraints on the template parameters, both in terms of accuracy and precision.
Adapting the \ufig simulations to the optical bands of the KiDS survey requires only the estimation of the necessary image systematics in the KiDS images.
However, simulating VIKING images with \ufig will require some code modification due to the different detectors and associated systematic effects in infrared imaging.

Another promising approach is stellar population synthesis (SPS)-based modelling \cite{galsbi_sps} where the sampling of galaxies is not done based on luminosity functions but on stellar mass functions, and the template parameters are replaced by more physically motivated SPS parameters.
SPS-based galaxy population modelling has become increasingly prevalent in the field (e.g., \cite{alsing_forward_2022,alsing_pop-cosmos_2024,hahn_desi_2023,hahn_provabgs_2023,li_popsed_2024}), driven in part by advances in emulating SPS codes \cite{alsing_speculator_2020,melchior_autoencoding_2023}.
An additional advantage is the elimination of the need for template-based approaches, which might require improvements for the high-redshift regime that will be probed by future surveys.

Performing inference using this setup would not only provide a robust cross-check for both models but also strengthen confidence in the accuracy of the simulated spectra.
This would enable us to replace the image simulations in the inference with spectra simulations and use spectroscopic data sets such as DESI \cite{desi_collaboration_desi_2016,desi_collaboration_overview_2022} or 4MOST \cite{de_jong_4most_2019} to constrain the model.
Additionally, the ability to simulate both realistic spectra and images from the same galaxy catalog would enable us to forward model the complex target selections of spectroscopic surveys that rely on photometric data.

While the model presented in this work accurately describes the optical photometry from low to high redshift, the adaptions mentioned above have the potential to validate the method across a broader wavelength range, provide realistic uncertainties from cosmic variance, and enable the model to simulate spectra effectively.

Moreover, it would be valuable to conduct a comprehensive comparison of diverse galaxy population modelling approaches across multiple datasets.
Such analysis could include established spectral energy distribution (SED) fitting frameworks like \texttt{Prospector}-$\alpha$ and \texttt{Prospector}-$\beta$, alongside with SBI-derived models such as PROVAGS, PopSED, \texttt{pop-cosmos}, or GalSBI.
These methodologies differ significantly in their population modelling strategies -- ranging from parametric relations to diffusion models -- and in their treatment of photometric uncertainty, encompassing both forward modelling the full image simulation and source extraction process and data-driven noise models.
Consistency in results across these different approaches would provide robust validation for galaxy population modelling efforts and enhance confidence in their application to upcoming surveys.

\section*{Data availability}
The Python packages \texttt{galsbi} and \ufig are publicly released and described in two accompanying papers \cite{fischbacher_galsbi_2024} and \cite{fischbacher_ufig_2024}.
These packages can be used to create galaxy catalogs from the fiducial model of this work.
The code is open source, contributions and/or feedback is very welcome.

The code to reproduce this analysis is also available.
Pipelines, config files and plotting scripts can be found at \texttt{galsbi-cpvn}\footnote{\url{https://cosmo-gitlab.phys.ethz.ch/silvanf/galsbi-cpvn/}}.
The inference uses the package \texttt{legacy-abc}\footnote{\url{https://cosmo-gitlab.phys.ethz.ch/cosmo_public/legacy_abc_public/-/tree/fischbacher24}} which is also released publicly with this work.
The emulator uses the \texttt{edelweiss} package\footnote{\url{https://gitlab.com/cosmology-ethz/edelweiss/}} which is both developed and released with this work.

Please note that repeating the full inference analysis will require access to the processed HSC data.
If you need access to this data, feel free to contact us.

\acknowledgments

We thank Arne Thomsen, Dominik Zürcher, Jed Homer and Tilman Tröster for helpful discussions and critical questions and Uwe Schmitt and Diego Moreno for informatics support.
We acknowledge the support of Euler Cluster by High Performance Computing Group from ETHZ Scientific IT Services that we used for most of our computations.
This project was supported in part by grant 200021\_192243 from the Swiss National Science Foundation and by the Deutsche Forschungsgemeinschaft (DFG, German Research Foundation) under Germany's Excellence Strategy – EXC-2094 – 390783311.

We acknowledge the use of the following software packages: \texttt{numpy} \cite{van_der_walt_numpy_2011}, \texttt{POT} \cite{peyre_computational_2020}, \texttt{scikit-learn} \cite{pedregosa_scikit-learn_2018}, \texttt{pzflow} \cite{crenshaw_probabilistic_2024, crenshaw_jfcrenshawpzflow_2024}, \texttt{scipy} \cite{virtanen_scipy_2020}, \texttt{tensorflow} \cite{tensorflow_developers_tensorflow_2021} and \texttt{xgboost} \cite{chen_xgboost_2016}.
Jobarrays were submitted with \texttt{esub-epipe} \cite{zurcher_cosmological_2021,zurcher_dark_2022,zurcher_towards_2023}, plots were created using \texttt{matplotlib} \cite{hunter_matplotlib_2007} and \texttt{trianglechain} \cite{fischbacher_redshift_2023,kacprzak_deeplss_2022}.

This paper is based on data collected at the Subaru Telescope and retrieved from the HSC data archive system, which is operated by Subaru Telescope and Astronomy Data Centre (ADC) at NAOJ. 

COSMOS2020 is based on observations collected at the European Southern Observatory under ESO programme ID 179.A-2005 and on data products produced by CALET and the Cambridge Astronomy Survey Unit on behalf of the UltraVISTA consortium. 

This work has made use of data from the European Space Agency (ESA) mission {\it Gaia} (\url{https://www.cosmos.esa.int/gaia}), processed by the {\it Gaia} Data Processing and Analysis Consortium (DPAC, \url{https://www.cosmos.esa.int/web/gaia/dpac/consortium}).
Funding for the DPAC has been provided by national institutions, in particular the institutions participating in the {\it Gaia} Multilateral Agreement. 



\bibliographystyle{JHEP}
\bibliography{references}

\providecommand{\href}[2]{#2}\begingroup\raggedright\begin{thebibliography}{100}

\bibitem{dark_energy_survey_collaboration_dark_2016}
{Dark Energy Survey Collaboration}, T.~Abbott, F.B.~Abdalla, J.~Aleksić, S.~Allam, A.~Amara et~al., \emph{The {Dark} {Energy} {Survey}: more than dark energy - an overview}, \href{https://doi.org/10.1093/mnras/stw641}{\emph{Monthly Notices of the Royal Astronomical Society} {\bfseries 460} (2016) 1270}.

\bibitem{de_jong_kilo-degree_2013}
J.T.A.~de~Jong, G.A.~Verdoes~Kleijn, K.H.~Kuijken, E.A.~Valentijn and {KiDS and Astro-WISE Consortiums}, \emph{The {Kilo}-{Degree} {Survey}}, \href{https://doi.org/10.1007/s10686-012-9306-1}{\emph{Experimental Astronomy} {\bfseries 35} (2013) 25}.

\bibitem{aihara_hyper_2018}
H.~Aihara, N.~Arimoto, R.~Armstrong, S.~Arnouts, N.A.~Bahcall, S.~Bickerton et~al., \emph{The {Hyper} {Suprime}-{Cam} {SSP} {Survey}: {Overview} and survey design}, \href{https://doi.org/10.1093/pasj/psx066}{\emph{Publications of the Astronomical Society of Japan} {\bfseries 70} (2018) S4}.

\bibitem{the_lsst_dark_energy_science_collaboration_lsst_2021}
{The LSST Dark Energy Science Collaboration}, R.~Mandelbaum, T.~Eifler, R.~Hložek, T.~Collett, E.~Gawiser et~al., \emph{The {LSST} {Dark} {Energy} {Science} {Collaboration} ({DESC}) {Science} {Requirements} {Document}},  Tech. Rep. \href{http://arxiv.org/abs/1809.01669}{arXiv:1809.01669}, arXiv (sep, 2021), \href{https://doi.org/10.48550/arXiv.1809.01669}{DOI}.

\bibitem{laureijs_euclid_2011}
R.~Laureijs, J.~Amiaux, S.~Arduini, J.-L.~Auguères, J.~Brinchmann, R.~Cole et~al., \emph{Euclid {Definition} {Study} {Report}},  Tech. Rep. \href{http://arxiv.org/abs/1110.3193}{arXiv:1110.3193}, arXiv (oct, 2011), \href{https://doi.org/10.48550/arXiv.1110.3193}{DOI}.

\bibitem{spergel_wide-field_2015}
D.~Spergel, N.~Gehrels, C.~Baltay, D.~Bennett, J.~Breckinridge, M.~Donahue et~al., \emph{Wide-{Field} {InfrarRed} {Survey} {Telescope}-{Astrophysics} {Focused} {Telescope} {Assets} {WFIRST}-{AFTA} 2015 {Report}},  Tech. Rep. \href{http://arxiv.org/abs/1503.03757}{arXiv:1503.03757}, arXiv (mar, 2015), \href{https://doi.org/10.48550/arXiv.1503.03757}{DOI}.

\bibitem{salvato_many_2019}
M.~Salvato, O.~Ilbert and B.~Hoyle, \emph{The many flavours of photometric redshifts}, \href{https://doi.org/10.1038/s41550-018-0478-0}{\emph{Nature Astronomy} {\bfseries 3} (2019) 212}.

\bibitem{newman_photometric_2022}
J.A.~Newman and D.~Gruen, \emph{Photometric {Redshifts} for {Next}-{Generation} {Surveys}}, \href{https://doi.org/10.1146/annurev-astro-032122-014611}{\emph{Annual Review of Astronomy and Astrophysics} {\bfseries 60} (2022) 363}.

\bibitem{fischbacher_redshift_2023}
S.~Fischbacher, T.~Kacprzak, J.~Blazek and A.~Refregier, \emph{Redshift requirements for cosmic shear with intrinsic alignment}, \href{https://doi.org/10.1088/1475-7516/2023/01/033}{\emph{Journal of Cosmology and Astroparticle Physics} {\bfseries 2023} (2023) 033}.

\bibitem{des_collaboration_dark_2022}
{DES Collaboration}, T.M.C.~Abbott, M.~Aguena, A.~Alarcon, S.~Allam, O.~Alves et~al., \emph{Dark {Energy} {Survey} {Year} 3 {Results}: {Cosmological} {Constraints} from {Galaxy} {Clustering} and {Weak} {Lensing}}, \href{https://doi.org/10.1103/PhysRevD.105.023520}{\emph{Physical Review D} {\bfseries 105} (2022) 023520}.

\bibitem{li_hyper_2023}
X.~Li, T.~Zhang, S.~Sugiyama, R.~Dalal, R.~Terasawa, M.M.~Rau et~al., \emph{Hyper {Suprime}-{Cam} {Year} 3 results: {Cosmology} from cosmic shear two-point correlation functions}, \href{https://doi.org/10.1103/PhysRevD.108.123518}{\emph{Physical Review D} {\bfseries 108} (2023) 123518}.

\bibitem{dalal_hyper_2023}
R.~Dalal, X.~Li, A.~Nicola, J.~Zuntz, M.A.~Strauss, S.~Sugiyama et~al., \emph{Hyper {Suprime}-{Cam} {Year} 3 results: {Cosmology} from cosmic shear power spectra}, \href{https://doi.org/10.1103/PhysRevD.108.123519}{\emph{Physical Review D} {\bfseries 108} (2023) 123519}.

\bibitem{melchior_challenge_2021}
P.~Melchior, R.~Joseph, J.~Sanchez, N.~MacCrann and D.~Gruen, \emph{The challenge of blending in large sky surveys}, \href{https://doi.org/10.1038/s42254-021-00353-y}{\emph{Nature Reviews Physics} {\bfseries 3} (2021) }.

\bibitem{samuroff_dark_2018}
S.~Samuroff, S.L.~Bridle, J.~Zuntz, M.A.~Troxel, D.~Gruen, R.P.~Rollins et~al., \emph{Dark {Energy} {Survey} {Year} 1 {Results}: {The} {Impact} of {Galaxy} {Neighbours} on {Weak} {Lensing} {Cosmology} with im3shape}, \href{https://doi.org/10.1093/mnras/stx3282}{\emph{Monthly Notices of the Royal Astronomical Society} {\bfseries 475} (2018) 4524}.

\bibitem{bosch_hyper_2018}
J.~Bosch, R.~Armstrong, S.~Bickerton, H.~Furusawa, H.~Ikeda, M.~Koike et~al., \emph{The {Hyper} {Suprime}-{Cam} software pipeline}, \href{https://doi.org/10.1093/pasj/psx080}{\emph{Publications of the Astronomical Society of Japan} {\bfseries 70} (2018) S5}.

\bibitem{sanchez_effects_2021}
J.~Sanchez, I.~Mendoza, D.P.~Kirkby and P.R.~Burchat, \emph{Effects of overlapping sources on cosmic shear estimation: {Statistical} sensitivity and pixel-noise bias}, \href{https://doi.org/10.1088/1475-7516/2021/07/043}{\emph{Journal of Cosmology and Astroparticle Physics} {\bfseries 2021} (2021) 043}.

\bibitem{huang_characterization_2018}
S.~Huang, A.~Leauthaud, R.~Murata, J.~Bosch, P.~Price, R.~Lupton et~al., \emph{Characterization and photometric performance of the {Hyper} {Suprime}-{Cam} {Software} {Pipeline}}, \href{https://doi.org/10.1093/pasj/psx126}{\emph{Publications of the Astronomical Society of Japan} {\bfseries 70} (2018) S6}.

\bibitem{dawson_ellipticity_2016}
W.A.~Dawson, M.D.~Schneider, J.A.~Tyson and M.J.~Jee, \emph{The {Ellipticity} {Distribution} of {Ambiguously} {Blended} {Objects}}, \href{https://doi.org/10.3847/0004-637X/816/1/11}{\emph{The Astrophysical Journal} {\bfseries 816} (2016) 11}.

\bibitem{hoekstra_study_2016}
H.~Hoekstra, M.~Viola and R.~Herbonnet, \emph{A study of the sensitivity of shape measurements to the input parameters of weak lensing image simulations},  sep, 2016.
\newblock 10.1093/mnras/stx724.

\bibitem{euclid_collaboration_euclid_2019}
{Euclid Collaboration}, N.~Martinet, T.~Schrabback, H.~Hoekstra, M.~Tewes, R.~Herbonnet et~al., \emph{Euclid {Preparation} {IV}. {Impact} of undetected galaxies on weak-lensing shear measurements}, \href{https://doi.org/10.1051/0004-6361/201935187}{\emph{Astronomy \& Astrophysics} {\bfseries 627} (2019) A59}.

\bibitem{maccrann_y3_2021}
N.~MacCrann, M.R.~Becker, J.~McCullough, A.~Amon, D.~Gruen, M.~Jarvis et~al., \emph{{DES} {Y3} results: {Blending} shear and redshift biases in image simulations}, \href{https://doi.org/10.1093/mnras/stab2870}{\emph{Monthly Notices of the Royal Astronomical Society} {\bfseries 509} (2021) 3371}.

\bibitem{nourbakhsh_galaxy_2022}
E.~Nourbakhsh, J.A.~Tyson, S.J.~Schmidt and T.L.D.E.S.~Collaboration, \emph{Galaxy blending effects in deep imaging cosmic shear probes of cosmology}, \href{https://doi.org/10.1093/mnras/stac1303}{\emph{Monthly Notices of the Royal Astronomical Society} {\bfseries 514} (2022) 5905}.

\bibitem{chang_effective_2013}
C.~Chang, M.~Jarvis, B.~Jain, S.M.~Kahn, D.~Kirkby, A.~Connolly et~al., \emph{The {Effective} {Number} {Density} of {Galaxies} for {Weak} {Lensing} {Measurements} in the {LSST} {Project}}, \href{https://doi.org/10.1093/mnras/stt1156}{\emph{Monthly Notices of the Royal Astronomical Society} {\bfseries 434} (2013) 2121}.

\bibitem{li_kids-legacy_2023}
S.-S.~Li, K.~Kuijken, H.~Hoekstra, L.~Miller, C.~Heymans, H.~Hildebrandt et~al., \emph{{KiDS}-{Legacy} calibration: {Unifying} shear and redshift calibration with the {SKiLLS} multi-band image simulations}, \href{https://doi.org/10.1051/0004-6361/202245210}{\emph{Astronomy \& Astrophysics} {\bfseries 670} (2023) A100}.

\bibitem{fenech_conti_calibration_2017}
I.~Fenech~Conti, R.~Herbonnet, H.~Hoekstra, J.~Merten, L.~Miller and M.~Viola, \emph{Calibration of weak-lensing shear in the {Kilo}-{Degree} {Survey}}, \href{https://doi.org/10.1093/mnras/stx200}{\emph{Monthly Notices of the Royal Astronomical Society} {\bfseries 467} (2017) 1627}.

\bibitem{kuijken_fourth_2019}
K.~Kuijken, C.~Heymans, A.~Dvornik, H.~Hildebrandt, J.T.A.~de~Jong, A.H.~Wright et~al., \emph{The fourth data release of the {Kilo}-{Degree} {Survey}: ugri imaging and nine-band optical-{IR} photometry over 1000 square degrees}, \href{https://doi.org/10.1051/0004-6361/201834918}{\emph{Astronomy and Astrophysics} {\bfseries 625} (2019) A2}.

\bibitem{elahi_surfs_2018}
P.J.~Elahi, C.~Welker, C.~Power, C.d.P.~Lagos, A.S.G.~Robotham, R.~Cañas et~al., \emph{{SURFS}: {Riding} the waves with {Synthetic} {UniveRses} {For} {Surveys}}, \href{https://doi.org/10.1093/mnras/sty061}{\emph{Monthly Notices of the Royal Astronomical Society} {\bfseries 475} (2018) 5338}.

\bibitem{lagos_shark_2018}
C.d.P.~Lagos, R.J.~Tobar, A.S.G.~Robotham, D.~Obreschkow, P.D.~Mitchell, C.~Power et~al., \emph{Shark: introducing an open source, free and flexible semi-analytic model of galaxy formation}, \href{https://doi.org/10.1093/mnras/sty2440}{\emph{Monthly Notices of the Royal Astronomical Society} {\bfseries 481} (2018) 3573}.

\bibitem{robotham_prospect_2020}
A.S.G.~Robotham, S.~Bellstedt, C.d.P.~Lagos, J.E.~Thorne, L.J.~Davies, S.P.~Driver et~al., \emph{{ProSpect}: generating spectral energy distributions with complex star formation and metallicity histories}, \href{https://doi.org/10.1093/mnras/staa1116}{\emph{Monthly Notices of the Royal Astronomical Society} {\bfseries 495} (2020) 905}.

\bibitem{carassou_inferring_2017}
S.~Carassou, V.~De~Lapparent, E.~Bertin and D.~Le~Borgne, \emph{Inferring the photometric and size evolution of galaxies from image simulations: {I}. {Method}}, \href{https://doi.org/10.1051/0004-6361/201730587}{\emph{Astronomy \& Astrophysics} {\bfseries 605} (2017) A9}.

\bibitem{herbel_redshift_2017}
J.~Herbel, T.~Kacprzak, A.~Amara, A.~Refregier, C.~Bruderer and A.~Nicola, \emph{The redshift distribution of cosmological samples: a forward modeling approach}, \href{https://doi.org/10.1088/1475-7516/2017/08/035}{\emph{Journal of Cosmology and Astroparticle Physics} {\bfseries 2017} (2017) 035}.

\bibitem{berge_ultra_2013}
J.~Bergé, L.~Gamper, A.~Réfrégier and A.~Amara, \emph{An {Ultra} {Fast} {Image} {Generator} ({UFig}) for wide-field astronomy}, \href{https://doi.org/10.1016/j.ascom.2013.01.001}{\emph{Astronomy and Computing} {\bfseries 1} (2013) 23}.

\bibitem{akeret_approximate_2015}
J.~Akeret, A.~Refregier, A.~Amara, S.~Seehars and C.~Hasner, \emph{Approximate {Bayesian} {Computation} for {Forward} {Modeling} in {Cosmology}}, .

\bibitem{tortorelli_measurement_2020}
L.~Tortorelli, M.~Fagioli, J.~Herbel, A.~Amara, T.~Kacprzak and A.~Refregier, \emph{Measurement of the {B}-band galaxy {Luminosity} {Function} with {Approximate} {Bayesian} {Computation}}, \href{https://doi.org/10.1088/1475-7516/2020/09/048}{\emph{Journal of Cosmology and Astroparticle Physics} {\bfseries 2020} (2020) 048}.

\bibitem{tortorelli_pau_2021}
L.~Tortorelli, M.~Siudek, B.~Moser, T.~Kacprzak, P.~Berner, A.~Refregier et~al., \emph{The {PAU} survey: measurement of narrow-band galaxy properties with approximate bayesian computation}, \href{https://doi.org/10.1088/1475-7516/2021/12/013}{\emph{Journal of Cosmology and Astroparticle Physics} {\bfseries 2021} (2021) 013}.

\bibitem{moser_simulation-based_2024}
B.~Moser, T.~Kacprzak, S.~Fischbacher, A.~Refregier, D.~Grimm and L.~Tortorelli, \emph{Simulation-based inference of deep fields: galaxy population model and redshift distributions}, \href{https://doi.org/10.1088/1475-7516/2024/05/049}{\emph{Journal of Cosmology and Astroparticle Physics} {\bfseries 2024} (2024) 049}.

\bibitem{blanton_k-corrections_2007}
M.R.~Blanton and S.~Roweis, \emph{K-corrections and filter transformations in the ultraviolet, optical, and near infrared}, \href{https://doi.org/10.1086/510127}{\emph{The Astronomical Journal} {\bfseries 133} (2007) 734}.

\bibitem{brown_atlas_2014}
M.J.I.~Brown, J.~Moustakas, J.D.T.~Smith, E.~da~Cunha, T.H.~Jarrett, M.~Imanishi et~al., \emph{An {Atlas} of {Galaxy} {Spectral} {Energy} {Distributions} from the {Ultraviolet} to the {Mid}-infrared}, \href{https://doi.org/10.1088/0067-0049/212/2/18}{\emph{The Astrophysical Journal Supplement Series} {\bfseries 212} (2014) 18}.

\bibitem{conroy_propagation_2009}
C.~Conroy, J.E.~Gunn and M.~White, \emph{The propagation of uncertainties in stellar population synthesis modeling {I}: {The} relevance of uncertain aspects of stellar evolution and the {IMF} to the derived physical properties of galaxies}, \href{https://doi.org/10.1088/0004-637X/699/1/486}{\emph{The Astrophysical Journal} {\bfseries 699} (2009) 486}.

\bibitem{conroy_propagation_2010}
C.~Conroy and J.E.~Gunn, \emph{The {Propagation} of {Uncertainties} in {Stellar} {Population} {Synthesis} {Modeling}. {III}. {Model} {Calibration}, {Comparison}, and {Evaluation}}, \href{https://doi.org/10.1088/0004-637X/712/2/833}{\emph{The Astrophysical Journal} {\bfseries 712} (2010) 833}.

\bibitem{leja_deriving_2017}
J.~Leja, B.D.~Johnson, C.~Conroy, P.G.v.~Dokkum and N.~Byler, \emph{Deriving {Physical} {Properties} from {Broadband} {Photometry} with {Prospector}: {Description} of the {Model} and a {Demonstration} of its {Accuracy} {Using} 129 {Galaxies} in the {Local} {Universe}}, \href{https://doi.org/10.3847/1538-4357/aa5ffe}{\emph{The Astrophysical Journal} {\bfseries 837} (2017) 170}.

\bibitem{leja_hot_2018}
J.~Leja, B.D.~Johnson, C.~Conroy and P.~van Dokkum, \emph{Hot {Dust} in {Panchromatic} {SED} {Fitting}: {Identification} of {Active} {Galactic} {Nuclei} and {Improved} {Galaxy} {Properties}}, \href{https://doi.org/10.3847/1538-4357/aaa8db}{\emph{The Astrophysical Journal} {\bfseries 854} (2018) 62}.

\bibitem{leja_older_2019}
J.~Leja, B.D.~Johnson, C.~Conroy, P.~van Dokkum, J.S.~Speagle, G.~Brammer et~al., \emph{An {Older}, {More} {Quiescent} {Universe} from {Panchromatic} {SED} {Fitting} of the {3D}-{HST} {Survey}}, \href{https://doi.org/10.3847/1538-4357/ab1d5a}{\emph{The Astrophysical Journal} {\bfseries 877} (2019) 140}.

\bibitem{johnson_stellar_2021}
B.D.~Johnson, J.~Leja, C.~Conroy and J.S.~Speagle, \emph{Stellar {Population} {Inference} with {Prospector}}, \href{https://doi.org/10.3847/1538-4365/abef67}{\emph{The Astrophysical Journal Supplement Series} {\bfseries 254} (2021) 22}.

\bibitem{wang_inferring_2023}
B.~Wang, J.~Leja, R.~Bezanson, B.D.~Johnson, G.~Khullar, I.~Labbe et~al., \emph{Inferring {More} from {Less}: {Prospector} as a {Photometric} {Redshift} {Engine} in the {Era} of {JWST}}, \href{https://doi.org/10.3847/2041-8213/acba99}{\emph{The Astrophysical Journal Letters} {\bfseries 944} (2023) L58}.

\bibitem{wang_quantifying_2024}
B.~Wang, J.~Leja, H.~Atek, I.~Labbé, Y.~Li, R.~Bezanson et~al., \emph{Quantifying the {Effects} of {Known} {Unknowns} on {Inferred} {High}-redshift {Galaxy} {Properties}: {Burstiness}, {IMF}, and {Nebular} {Physics}}, \href{https://doi.org/10.3847/1538-4357/ad187c}{\emph{The Astrophysical Journal} {\bfseries 963} (2024) 74}.

\bibitem{alsing_speculator_2020}
J.~Alsing, H.~Peiris, J.~Leja, C.~Hahn, R.~Tojeiro, D.~Mortlock et~al., \emph{{SPECULATOR}: {Emulating} {Stellar} {Population} {Synthesis} for {Fast} and {Accurate} {Galaxy} {Spectra} and {Photometry}}, \href{https://doi.org/10.3847/1538-4365/ab917f}{\emph{The Astrophysical Journal Supplement Series} {\bfseries 249} (2020) 5}.

\bibitem{melchior_autoencoding_2023}
P.~Melchior, Y.~Liang, C.~Hahn and A.~Goulding, \emph{Autoencoding {Galaxy} {Spectra} {I}: {Architecture}}, \href{https://doi.org/10.3847/1538-3881/ace0ff}{\emph{The Astronomical Journal} {\bfseries 166} (2023) 74}.

\bibitem{hearin_dsps_2023}
A.P.~Hearin, J.~Chaves-Montero, A.~Alarcon, M.R.~Becker and A.~Benson, \emph{{DSPS}: {Differentiable} stellar population synthesis}, \href{https://doi.org/10.1093/mnras/stad456}{\emph{Monthly Notices of the Royal Astronomical Society} {\bfseries 521} (2023) 1741}.

\bibitem{alarcon_diffstar_2022}
A.~Alarcon, A.P.~Hearin, M.R.~Becker and J.~Chaves-Montero, \emph{Diffstar: A fully parametric physical model for galaxy assembly history}, \href{https://doi.org/10.1093/mnras/stac3118}{\emph{Monthly Notices of the Royal Astronomical Society} {\bfseries 518} (2022) 562–584}.

\bibitem{hearin_diffmah_2021}
A.P.~Hearin, J.~Chaves-Montero, M.R.~Becker and A.~Alarcon, \emph{A differentiable model of the assembly of individual and populations of dark matter halos}, \href{https://doi.org/10.21105/astro.2105.05859}{\emph{The Open Journal of Astrophysics} {\bfseries 4} (2021) 10.21105/astro.2105.05859}.

\bibitem{sudek_sensitivity_2022}
P.~Sudek, L.F.~de la Bella, A.~Amara and W.G.~Hartley, \emph{The sensitivity of the redshift distribution to galaxy demographics}, \href{https://doi.org/10.1093/mnras/stac2299}{\emph{Monthly Notices of the Royal Astronomical Society} {\bfseries 516} (2022) 1670}.

\bibitem{Amara_skypy_2021}
A.~Amara, L.F.d.l.~Bella, S.~Birrer, S.~Bridle, J.P.~Cordero, G.~Favole et~al., \emph{`skypy`: A package for modelling the universe}, \href{https://doi.org/10.21105/joss.03056}{\emph{Journal of Open Source Software} {\bfseries 6} (2021) 3056}.

\bibitem{tortorelli_impact_2024}
L.~Tortorelli, J.~McCullough and D.~Gruen, \emph{Impact of stellar population synthesis choices on forward modelling-based redshift distribution estimates}, \href{https://doi.org/10.1051/0004-6361/202450694}{\emph{Astronomy \& Astrophysics} {\bfseries 689} (2024) A144}.

\bibitem{cranmer_frontier_2020}
K.~Cranmer, J.~Brehmer and G.~Louppe, \emph{The frontier of simulation-based inference}, \href{https://doi.org/10.1073/pnas.1912789117}{\emph{Proceedings of the National Academy of Sciences} {\bfseries 117} (2020) 30055}.

\bibitem{kacprzak_cosmology_2016}
T.~Kacprzak, D.~Kirk, O.~Friedrich, A.~Amara, A.~Refregier, L.~Marian et~al., \emph{Cosmology constraints from shear peak statistics in {Dark} {Energy} {Survey} {Science} {Verification} data}, \href{https://doi.org/10.1093/mnras/stw2070}{\emph{Monthly Notices of the Royal Astronomical Society} {\bfseries 463} (2016) 3653}.

\bibitem{fluri_cosmological_2019}
J.~Fluri, T.~Kacprzak, A.~Lucchi, A.~Refregier, A.~Amara, T.~Hofmann et~al., \emph{Cosmological constraints with deep learning from {KiDS}-450 weak lensing maps}, \href{https://doi.org/10.1103/PhysRevD.100.063514}{\emph{Physical Review D} {\bfseries 100} (2019) 063514}.

\bibitem{fluri_full_2022}
J.~Fluri, T.~Kacprzak, A.~Lucchi, A.~Schneider, A.~Refregier and T.~Hofmann, \emph{A {Full} \$w\${CDM} {Analysis} of {KiDS}-1000 {Weak} {Lensing} {Maps} using {Deep} {Learning}}, \href{https://doi.org/10.1103/PhysRevD.105.083518}{\emph{Physical Review D} {\bfseries 105} (2022) 083518}.

\bibitem{zurcher_cosmological_2021}
D.~Zürcher, J.~Fluri, R.~Sgier, T.~Kacprzak and A.~Refregier, \emph{Cosmological {Forecast} for non-{Gaussian} {Statistics} in large-scale weak {Lensing} {Surveys}}, \href{https://doi.org/10.1088/1475-7516/2021/01/028}{\emph{Journal of Cosmology and Astroparticle Physics} {\bfseries 2021} (2021) 028}.

\bibitem{zurcher_dark_2022}
D.~Zürcher, J.~Fluri, R.~Sgier, T.~Kacprzak, M.~Gatti, C.~Doux et~al., \emph{Dark {Energy} {Survey} {Year} 3 results: {Cosmology} with peaks using an emulator approach}, \href{https://doi.org/10.1093/mnras/stac078}{\emph{Monthly Notices of the Royal Astronomical Society} {\bfseries 511} (2022) 2075}.

\bibitem{zurcher_towards_2023}
D.~Zürcher, J.~Fluri, V.~Ajani, S.~Fischbacher, A.~Refregier and T.~Kacprzak, \emph{Towards a full \textit{w} {CDM} map-based analysis for weak lensing surveys}, \href{https://doi.org/10.1093/mnras/stad2212}{\emph{Monthly Notices of the Royal Astronomical Society} {\bfseries 525} (2023) 761}.

\bibitem{kacprzak_deeplss_2022}
T.~Kacprzak and J.~Fluri, \emph{{DeepLSS}: {Breaking} {Parameter} {Degeneracies} in {Large}-{Scale} {Structure} with {Deep}-{Learning} {Analysis} of {Combined} {Probes}}, \href{https://doi.org/10.1103/PhysRevX.12.031029}{\emph{Physical Review X} {\bfseries 12} (2022) 031029}.

\bibitem{gatti_dark_2021}
M.~Gatti, B.~Jain, C.~Chang, M.~Raveri, D.~Zürcher, L.~Secco et~al., \emph{Dark {Energy} {Survey} {Year} 3 results: cosmology with moments of weak lensing mass maps}, {\emph{arXiv:2110.10141 [astro-ph]} (2021) }.

\bibitem{cheng_weak_2021}
S.~Cheng and B.~Ménard, \emph{Weak lensing scattering transform: dark energy and neutrino mass sensitivity}, \href{https://doi.org/10.1093/mnras/stab2102}{\emph{Monthly Notices of the Royal Astronomical Society} {\bfseries 507} (2021) 1012}.

\bibitem{anagnostidis_cosmology_2022}
S.~Anagnostidis, A.~Thomsen, T.~Kacprzak, T.~Tröster, L.~Biggio, A.~Refregier et~al., \emph{Cosmology from {Galaxy} {Redshift} {Surveys} with {PointNet}},  nov, 2022.
\newblock 10.48550/arXiv.2211.12346.

\bibitem{lu_cosmological_2023}
T.~Lu, Z.~Haiman and X.~Li, \emph{Cosmological constraints from {HSC} survey first-year data using deep learning}, \href{https://doi.org/10.1093/mnras/stad686}{\emph{Monthly Notices of the Royal Astronomical Society} {\bfseries 521} (2023) 2050}.

\bibitem{massara_sc_2024}
E.~Massara, C.~Hahn, M.~Eickenberg, S.~Ho, J.~Hou, P.~Lemos et~al., \emph{\{{\textbackslash}sc {SimBIG}\}: {Cosmological} {Constraints} using {Simulation}-{Based} {Inference} of {Galaxy} {Clustering} with {Marked} {Power} {Spectra}},  apr, 2024.
\newblock 10.48550/arXiv.2404.04228.

\bibitem{valogiannis_precise_2024}
G.~Valogiannis, S.~Yuan and C.~Dvorkin, \emph{Precise {Cosmological} {Constraints} from {BOSS} {Galaxy} {Clustering} with a {Simulation}-{Based} {Emulator} of the {Wavelet} {Scattering} {Transform}}, \href{https://doi.org/10.1103/PhysRevD.109.103503}{\emph{Physical Review D} {\bfseries 109} (2024) 103503}.

\bibitem{jeffrey_dark_2024}
N.~Jeffrey, L.~Whiteway, M.~Gatti, J.~Williamson, J.~Alsing, A.~Porredon et~al., \emph{Dark {Energy} {Survey} {Year} 3 results: likelihood-free, simulation-based \$w\${CDM} inference with neural compression of weak-lensing map statistics},  mar, 2024.
\newblock 10.48550/arXiv.2403.02314.

\bibitem{lin_simulation-based_2023}
K.~Lin, M.~von Wietersheim-Kramsta, B.~Joachimi and S.~Feeney, \emph{A simulation-based inference pipeline for cosmic shear with the {Kilo}-{Degree} {Survey}}, \href{https://doi.org/10.1093/mnras/stad2262}{\emph{Monthly Notices of the Royal Astronomical Society} {\bfseries 524} (2023) 6167}.

\bibitem{von_wietersheim-kramsta_kids-sbi_2024}
M.~von Wietersheim-Kramsta, K.~Lin, N.~Tessore, B.~Joachimi, A.~Loureiro, R.~Reischke et~al., \emph{{KiDS}-{SBI}: {Simulation}-{Based} {Inference} {Analysis} of {KiDS}-1000 {Cosmic} {Shear}},  apr, 2024.
\newblock 10.48550/arXiv.2404.15402.

\bibitem{alsing_forward_2022}
J.~Alsing, H.~Peiris, D.~Mortlock, J.~Leja and B.~Leistedt, \emph{Forward modeling of galaxy populations for cosmological redshift distribution inference},  jul, 2022.
\newblock 10.48550/arXiv.2207.05819.

\bibitem{alsing_pop-cosmos_2024}
J.~Alsing, S.~Thorp, S.~Deger, H.V.~Peiris, B.~Leistedt, D.~Mortlock et~al., \emph{pop-cosmos: {A} {Comprehensive} {Picture} of the {Galaxy} {Population} from {COSMOS} {Data}}, \href{https://doi.org/10.3847/1538-4365/ad5c69}{\emph{The Astrophysical Journal Supplement Series} {\bfseries 274} (2024) 12}.

\bibitem{leistedt_hierarchical_2023}
B.~Leistedt, J.~Alsing, H.~Peiris, D.~Mortlock and J.~Leja, \emph{Hierarchical {Bayesian} inference of photometric redshifts with stellar population synthesis models}, \href{https://doi.org/10.3847/1538-4365/ac9d99}{\emph{The Astrophysical Journal Supplement Series} {\bfseries 264} (2023) 23}.

\bibitem{thorp_data-space_2024}
S.~Thorp, H.V.~Peiris, D.J.~Mortlock, J.~Alsing, B.~Leistedt and S.~Deger, \emph{Data-{Space} {Validation} of {High}-{Dimensional} {Models} by {Comparing} {Sample} {Quantiles}},  oct, 2024.
\newblock 10.48550/arXiv.2402.00930.

\bibitem{thorp_pop-cosmos_2024}
S.~Thorp, J.~Alsing, H.V.~Peiris, S.~Deger, D.J.~Mortlock, B.~Leistedt et~al., \emph{pop-cosmos: {Scaleable} inference of galaxy properties and redshifts with a data-driven population model}, \href{https://doi.org/10.3847/1538-4357/ad7736}{\emph{The Astrophysical Journal} {\bfseries 975} (2024) 145}.

\bibitem{hahn_desi_2023}
C.~Hahn, K.J.~Kwon, R.~Tojeiro, M.~Siudek, R.E.A.~Canning, M.~Mezcua et~al., \emph{The {DESI} {PRObabilistic} {Value}-{Added} {Bright} {Galaxy} {Survey} ({PROVABGS}) {Mock} {Challenge}}, \href{https://doi.org/10.3847/1538-4357/ac8983}{\emph{The Astrophysical Journal} {\bfseries 945} (2023) 16}.

\bibitem{hahn_provabgs_2023}
C.~Hahn, J.N.~Aguilar, S.~Alam, S.~Ahlen, D.~Brooks, S.~Cole et~al., \emph{{PROVABGS}: {The} {Probabilistic} {Stellar} {Mass} {Function} of the {BGS} {One}-{Percent} {Survey}},  jun, 2023.
\newblock 10.48550/arXiv.2306.06318.

\bibitem{li_popsed_2024}
J.~Li, P.~Melchior, C.~Hahn and S.~Huang, \emph{{PopSED}: {Population}-{Level} {Inference} for {Galaxy} {Properties} from {Broadband} {Photometry} with {Neural} {Density} {Estimation}}, \href{https://doi.org/10.3847/1538-3881/ad0be4}{\emph{The Astronomical Journal} {\bfseries 167} (2024) 16}.

\bibitem{Sabti_gallumi_2022}
N.~Sabti, J.B.~Muñoz and D.~Blas, \emph{Gallumi: A galaxy luminosity function pipeline for cosmology and astrophysics}, \href{https://doi.org/10.1103/PhysRevD.105.043518}{\emph{Physical Review D} {\bfseries 105} (2022) 043518}.

\bibitem{refregier_way_2014}
A.~Refregier and A.~Amara, \emph{A {Way} {Forward} for {Cosmic} {Shear}: {Monte}-{Carlo} {Control} {Loops}}, {\emph{arXiv:1303.4739 [astro-ph]} (2014) }.

\bibitem{fischbacher_ufig_2024}
S.~Fischbacher, B.~Moser, T.~Kacprzak, L.~Tortorelli, J.~Herbel, C.~Bruderer et~al., \emph{{UFig} v1: {The} ultra-fast image generator},  dec, 2024.
\newblock 10.48550/arXiv.2412.08716.

\bibitem{bruderer_calibrated_2016}
C.~Bruderer, C.~Chang, A.~Refregier, A.~Amara, J.~Berge and L.~Gamper, \emph{Calibrated {Ultra} {Fast} {Image} {Simulations} for the {Dark} {Energy} {Survey}}, \href{https://doi.org/10.3847/0004-637X/817/1/25}{\emph{The Astrophysical Journal} {\bfseries 817} (2016) 25}.

\bibitem{bruderer_cosmic_2018}
C.~Bruderer, A.~Nicola, A.~Amara, A.~Refregier, J.~Herbel and T.~Kacprzak, \emph{Cosmic shear calibration with forward modeling}, \href{https://doi.org/10.1088/1475-7516/2018/08/007}{\emph{Journal of Cosmology and Astroparticle Physics} {\bfseries 2018} (2018) 007}.

\bibitem{kacprzak_monte_2020}
T.~Kacprzak, J.~Herbel, A.~Nicola, R.~Sgier, F.~Tarsitano, C.~Bruderer et~al., \emph{Monte {Carlo} {Control} {Loops} for cosmic shear cosmology with {DES} {Year} 1}, \href{https://doi.org/10.1103/PhysRevD.101.082003}{\emph{Physical Review D} {\bfseries 101} (2020) 082003}.

\bibitem{tortorelli_pau_2018}
L.~Tortorelli, L.~Della~Bruna, J.~Herbel, A.~Amara, A.~Refregier, A.~Alarcon et~al., \emph{The {PAU} {Survey}: {A} {Forward} {Modeling} {Approach} for {Narrow}-band {Imaging}}, \href{https://doi.org/10.1088/1475-7516/2018/11/035}{\emph{Journal of Cosmology and Astroparticle Physics} {\bfseries 2018} (2018) 035}.

\bibitem{fagioli_forward_2018}
M.~Fagioli, J.~Riebartsch, A.~Nicola, J.~Herbel, A.~Amara, A.~Refregier et~al., \emph{Forward modeling of spectroscopic galaxy surveys: application to {SDSS}}, \href{https://doi.org/10.1088/1475-7516/2018/11/015}{\emph{Journal of Cosmology and Astroparticle Physics} {\bfseries 2018} (2018) 015}.

\bibitem{fagioli_spectro-imaging_2020}
M.~Fagioli, L.~Tortorelli, J.~Herbel, D.~Zürcher, A.~Refregier and A.~Amara, \emph{Spectro-imaging forward model of red and blue galaxies}, \href{https://doi.org/10.1088/1475-7516/2020/06/050}{\emph{Journal of Cosmology and Astroparticle Physics} {\bfseries 2020} (2020) 050}.

\bibitem{fischbacher_galsbi_2024}
S.~Fischbacher, B.~Moser, T.~Kacprzak, J.~Herbel, L.~Tortorelli, U.~Schmitt et~al., \emph{galsbi: {A} {Python} package for the {GalSBI} galaxy population model},  dec, 2024.
\newblock 10.48550/arXiv.2412.08722.

\bibitem{bertin_sextractor_1996}
E.~Bertin and S.~Arnouts, \emph{{SExtractor}: {Software} for source extraction}, \href{https://doi.org/10.1051/aas:1996164}{\emph{Astronomy and Astrophysics Supplement Series} {\bfseries 117} (1996) 393}.

\bibitem{galsbi_sps}
L.~Tortorelli, S.~Fischbacher, D.~Grün, A.~Refregier, S.~Bellstedt, A.S.G.~Robotham et~al., \emph{Galsbi-sps: a stellar population synthesis-based galaxy population model for cosmology and galaxy evolution applications},  May, 2025.
\newblock 10.48550/arXiv.2505.21610.

\bibitem{johnston_shedding_2011}
R.~Johnston, \emph{Shedding light on the galaxy luminosity function}, \href{https://doi.org/10.1007/s00159-011-0041-9}{\emph{The Astronomy and Astrophysics Review} {\bfseries 19} (2011) 41}.

\bibitem{laigle_cosmos2015_2016}
C.~Laigle, H.J.~McCracken, O.~Ilbert, B.C.~Hsieh, I.~Davidzon, P.~Capak et~al., \emph{{THE} {COSMOS2015} {CATALOG}: {EXPLORING} {THE} 1 {\textless} z {\textless} 6 {UNIVERSE} {WITH} {HALF} {A} {MILLION} {GALAXIES}}, \href{https://doi.org/10.3847/0067-0049/224/2/24}{\emph{The Astrophysical Journal Supplement Series} {\bfseries 224} (2016) 24}.

\bibitem{wang_divergence_2009}
Q.~Wang, S.R.~Kulkarni and S.~Verdu, \emph{Divergence {Estimation} for {Multidimensional} {Densities} {Via} k-{Nearest}-{Neighbor} {Distances}}, \href{https://doi.org/10.1109/TIT.2009.2016060}{\emph{IEEE Transactions on Information Theory} {\bfseries 55} (2009) 2392}.

\bibitem{jong_local_2000}
R.S.d.~Jong and C.~Lacey, \emph{The {Local} {Space} {Density} of {Sb}-{SdmGalaxies} as {Function} of {Their} {ScaleSize}, {Surface} {Brightness},and {Luminosity}}, \href{https://doi.org/10.1086/317840}{\emph{The Astrophysical Journal} {\bfseries 545} (2000) 781}.

\bibitem{shen_size_2003}
S.~Shen, H.J.~Mo, S.D.M.~White, M.R.~Blanton, G.~Kauffmann, W.~Voges et~al., \emph{The size distribution of galaxies in the {Sloan} {Digital} {Sky} {Survey}}, \href{https://doi.org/10.1046/j.1365-8711.2003.06740.x}{\emph{Monthly Notices of the Royal Astronomical Society} {\bfseries 343} (2003) 978}.

\bibitem{van_der_walt_numpy_2011}
S.~Van Der~Walt, S.C.~Colbert and G.~Varoquaux, \emph{The {NumPy} array: a structure for efficient numerical computation}, \href{https://doi.org/10.1109/MCSE.2011.37}{\emph{Computing in Science \& Engineering} {\bfseries 13} (2011) 22}.

\bibitem{ormerod_epochs_2023}
K.~Ormerod, C.J.~Conselice, N.J.~Adams, T.~Harvey, D.~Austin, J.~Trussler et~al., \emph{{EPOCHS} {VI}: the size and shape evolution of galaxies since \$z {\textbackslash}sim 8\$ with {JWST} {Observations}}, \href{https://doi.org/10.1093/mnras/stad3597}{\emph{Monthly Notices of the Royal Astronomical Society} {\bfseries 527} (2023) 6110}.

\bibitem{tarsitano_catalogue_2018}
F.~Tarsitano, W.G.~Hartley, A.~Amara, A.~Bluck, C.~Bruderer, M.~Carollo et~al., \emph{A catalogue of structural and morphological measurements for {DES} {Y1}}, \href{https://doi.org/10.1093/mnras/sty1970}{\emph{Monthly Notices of the Royal Astronomical Society} {\bfseries 481} (2018) 2018}.

\bibitem{krywult_vimos_2017}
J.~Krywult, L.a.M.~Tasca, A.~Pollo, D.~Vergani, M.~Bolzonella, I.~Davidzon et~al., \emph{The {VIMOS} {Public} {Extragalactic} {Redshift} {Survey} ({VIPERS}) - {The} coevolution of galaxy morphology and colour to z {\textasciitilde} 1}, \href{https://doi.org/10.1051/0004-6361/201628953}{\emph{Astronomy \& Astrophysics} {\bfseries 598} (2017) A120}.

\bibitem{robin_synthetic_2003}
A.C.~Robin, C.~Reylé, S.~Derrière and S.~Picaud, \emph{A synthetic view on structure and evolution of the {Milky} {Way}}, \href{https://doi.org/10.1051/0004-6361:20031117}{\emph{Astronomy \& Astrophysics} {\bfseries 409} (2003) 523}.

\bibitem{prusti_gaia_2016}
T.~Prusti, J.H.J.d.~Bruijne, A.G.A.~Brown, A.~Vallenari, C.~Babusiaux, C.a.L.~Bailer-Jones et~al., \emph{The {Gaia} mission}, \href{https://doi.org/10.1051/0004-6361/201629272}{\emph{Astronomy \& Astrophysics} {\bfseries 595} (2016) A1}.

\bibitem{vallenari_gaia_2023}
A.~Vallenari, A.G.A.~Brown, T.~Prusti, J.H.J.d.~Bruijne, F.~Arenou, C.~Babusiaux et~al., \emph{Gaia {Data} {Release} 3 - {Summary} of the content and survey properties}, \href{https://doi.org/10.1051/0004-6361/202243940}{\emph{Astronomy \& Astrophysics} {\bfseries 674} (2023) A1}.

\bibitem{herbel_fast_2018}
J.~Herbel, T.~Kacprzak, A.~Amara, A.~Refregier and A.~Lucchi, \emph{Fast {Point} {Spread} {Function} {Modeling} with {Deep} {Learning}}, \href{https://doi.org/10.1088/1475-7516/2018/07/054}{\emph{Journal of Cosmology and Astroparticle Physics} {\bfseries 2018} (2018) 054}.

\bibitem{gandalf}
P.~Gebhardt et~al., ``gandalf - emulating a survey transfer function for photometric redshift calibration of weak lensing galaxies.'' 2025.

\bibitem{hermans_trust_2022}
J.~Hermans, A.~Delaunoy, F.~Rozet, A.~Wehenkel, V.~Begy and G.~Louppe, \emph{A {Trust} {Crisis} {In} {Simulation}-{Based} {Inference}? {Your} {Posterior} {Approximations} {Can} {Be} {Unfaithful}},  dec, 2022.
\newblock 10.48550/arXiv.2110.06581.

\bibitem{arnouts_measuring_1999}
S.~Arnouts, S.~Cristiani, L.~Moscardini, S.~Matarrese, F.~Lucchin, A.~Fontana et~al., \emph{Measuring and modelling the redshift evolution of clustering: the {Hubble} {Deep} {Field} {North}}, \href{https://doi.org/10.1046/j.1365-8711.1999.02978.x}{\emph{Monthly Notices of the Royal Astronomical Society} {\bfseries 310} (1999) 540}.

\bibitem{ilbert_accurate_2006}
O.~Ilbert, S.~Arnouts, H.J.~McCracken, M.~Bolzonella, E.~Bertin, O.L.~Fèvre et~al., \emph{Accurate photometric redshifts for the {CFHT} legacy survey calibrated using the {VIMOS} {VLT} deep survey}, \href{https://doi.org/10.1051/0004-6361:20065138}{\emph{Astronomy \& Astrophysics} {\bfseries 457} (2006) 841}.

\bibitem{brammer_eazy_2008}
G.B.~Brammer, P.G.v.~Dokkum and P.~Coppi, \emph{{EAZY}: {A} {Fast}, {Public} {Photometric} {Redshift} {Code}}, \href{https://doi.org/10.1086/591786}{\emph{The Astrophysical Journal} {\bfseries 686} (2008) 1503}.

\bibitem{gretton_kernel_2012}
A.~Gretton, K.M.~Borgwardt, M.J.~Rasch, B.~Schölkopf and A.~Smola, \emph{A kernel two-sample test}, {\emph{J. Mach. Learn. Res.} {\bfseries 13} (2012) 723}.

\bibitem{vaserstein_markov_1969}
L.N.~Vaserstein, \emph{Markov {Processes} over {Denumerable} {Products} of {Spaces}, {Describing} {Large} {Systems} of {Automata}}, {\emph{Probl. Peredachi Inf.} {\bfseries Volume 5} (1969) 64}.

\bibitem{kantorovich_space_1958}
L.~Kantorovich and G.S.~Rubinstein, \emph{On a space of totally additive functions}, {\emph{Vestnik Leningrad. Univ} {\bfseries 13} (1958) 52}.

\bibitem{aihara_third_2022}
H.~Aihara, Y.~AlSayyad, M.~Ando, R.~Armstrong, J.~Bosch, E.~Egami et~al., \emph{Third data release of the {Hyper} {Suprime}-{Cam} {Subaru} {Strategic} {Program}}, \href{https://doi.org/10.1093/pasj/psab122}{\emph{Publications of the Astronomical Society of Japan} {\bfseries 74} (2022) 247}.

\bibitem{nicola_tomographic_2020}
A.~Nicola, D.~Alonso, J.~Sánchez, A.~Slosar, H.~Awan, A.~Broussard et~al., \emph{Tomographic galaxy clustering with the {Subaru} {Hyper} {Suprime}-{Cam} first year public data release}, \href{https://doi.org/10.1088/1475-7516/2020/03/044}{\emph{Journal of Cosmology and Astroparticle Physics} {\bfseries 2020} (2020) 044}.

\bibitem{weaver_cosmos2020_2022}
J.R.~Weaver, O.B.~Kauffmann, O.~Ilbert, H.J.~McCracken, A.~Moneti, S.~Toft et~al., \emph{{COSMOS2020}: {A} {Panchromatic} {View} of the {Universe} to \$z {\textbackslash}sim 10\$ from {Two} {Complementary} {Catalogs}}, \href{https://doi.org/10.3847/1538-4365/ac3078}{\emph{The Astrophysical Journal Supplement Series} {\bfseries 258} (2022) 11}.

\bibitem{Desprez_2023}
G.~Desprez, V.~Picouet, T.~Moutard, S.~Arnouts, M.~Sawicki, J.~Coupon et~al., \emph{Combining the clauds \& hsc-ssp surveys: U+grizy(+yjhks) photometry and photometric redshifts for 18m galaxies in the 20 deg2 of the hsc-ssp deep and ultradeep fields}, \href{https://doi.org/10.1051/0004-6361/202243363}{\emph{Astronomy \& Astrophysics} {\bfseries 670} (2023) A82}.

\bibitem{amon_dark_2022}
A.~Amon, D.~Gruen, M.A.~Troxel, N.~MacCrann, S.~Dodelson, A.~Choi et~al., \emph{Dark {Energy} {Survey} {Year} 3 {Results}: {Cosmology} from {Cosmic} {Shear} and {Robustness} to {Data} {Calibration}}, \href{https://doi.org/10.1103/PhysRevD.105.023514}{\emph{Physical Review D} {\bfseries 105} (2022) 023514}.

\bibitem{busch_kids-1000_2022}
J.L.v.d.~Busch, A.H.~Wright, H.~Hildebrandt, M.~Bilicki, M.~Asgari, S.~Joudaki et~al., \emph{{KiDS}-1000: {Cosmic} shear with enhanced redshift calibration}, \href{https://doi.org/10.1051/0004-6361/202142083}{\emph{Astronomy \& Astrophysics} {\bfseries 664} (2022) A170}.

\bibitem{li_kids-1000_2023}
S.-S.~Li, H.~Hoekstra, K.~Kuijken, M.~Asgari, M.~Bilicki, B.~Giblin et~al., \emph{{KiDS}-1000: {Cosmology} with improved cosmic shear measurements}, \href{https://doi.org/10.1051/0004-6361/202347236}{\emph{Astronomy \& Astrophysics} {\bfseries 679} (2023) A133}.

\bibitem{beare_z_2015}
R.A.~Beare, M.J.I.~Brown, K.A.~Pimbblet, F.~Bian and Y.-T.~Lin, \emph{The \$z {\textless} 1.2\$ optical luminosity function from a sample of \${\textbackslash}sim410 {\textbackslash}, 000\$ galaxies in bootes}, \href{https://doi.org/10.1088/0004-637X/815/2/94}{\emph{The Astrophysical Journal} {\bfseries 815} (2015) 94}.

\bibitem{giallongo_b-band_2005}
E.~Giallongo, S.~Salimbeni, N.~Menci, G.~Zamorani, A.~Fontana, M.~Dickinson et~al., \emph{The {B}-{Band} {Luminosity} {Function} of {Red} and {Blue} {Galaxies} up to z = 3.5}, \href{https://doi.org/10.1086/427819}{\emph{The Astrophysical Journal} {\bfseries 622} (2005) 116}.

\bibitem{ilbert_vimos-vlt_2006}
O.~Ilbert, S.~Lauger, L.~Tresse, V.~Buat, S.~Arnouts, O.L.~Fèvre et~al., \emph{The {VIMOS}-{VLT} {Deep} {Survey} - {Galaxy} luminosity function per morphological type up to z = 1.2}, \href{https://doi.org/10.1051/0004-6361:20053632}{\emph{Astronomy \& Astrophysics} {\bfseries 453} (2006) 809}.

\bibitem{zucca_zcosmos_2009}
E.~Zucca, S.~Bardelli, M.~Bolzonella, G.~Zamorani, O.~Ilbert, L.~Pozzetti et~al., \emph{The {zCOSMOS} survey: the role of the environment in the evolution of the luminosity function of different galaxy types}, \href{https://doi.org/10.1051/0004-6361/200912665}{\emph{Astronomy \& Astrophysics} {\bfseries 508} (2009) 1217}.

\bibitem{loveday_galaxy_2012}
J.~Loveday, P.~Norberg, I.K.~Baldry, S.P.~Driver, A.M.~Hopkins, J.A.~Peacock et~al., \emph{Galaxy and {Mass} {Assembly} ({GAMA}): ugriz galaxy luminosity functions}, \href{https://doi.org/10.1111/j.1365-2966.2011.20111.x}{\emph{Monthly Notices of the Royal Astronomical Society} {\bfseries 420} (2012) 1239}.

\bibitem{cool_galaxy_2012}
R.J.~Cool, D.J.~Eisenstein, C.S.~Kochanek, M.J.I.~Brown, N.~Caldwell, A.~Dey et~al., \emph{The {Galaxy} {Optical} {Luminosity} {Function} from the {AGN} and {Galaxy} {Evolution} {Survey} ({AGES})}, \href{https://doi.org/10.1088/0004-637X/748/1/10}{\emph{The Astrophysical Journal} {\bfseries 748} (2012) 10}.

\bibitem{fritz_vimos_2014}
A.~Fritz, M.~Scodeggio, O.~Ilbert, M.~Bolzonella, I.~Davidzon, J.~Coupon et~al., \emph{The {VIMOS} {Public} {Extragalactic} {Redshift} {Survey} ({VIPERS}): - {A} quiescent formation of massive red-sequence galaxies over the past 9 {Gyr}}, \href{https://doi.org/10.1051/0004-6361/201322379}{\emph{Astronomy \& Astrophysics} {\bfseries 563} (2014) A92}.

\bibitem{secco_dark_2022}
L.F.~Secco, S.~Samuroff, E.~Krause, B.~Jain, J.~Blazek, M.~Raveri et~al., \emph{Dark {Energy} {Survey} {Year} 3 {Results}: {Cosmology} from {Cosmic} {Shear} and {Robustness} to {Modeling} {Uncertainty}}, \href{https://doi.org/10.1103/PhysRevD.105.023515}{\emph{Physical Review D} {\bfseries 105} (2022) 023515}.

\bibitem{berner_rapid_2022}
P.~Berner, A.~Refregier, R.~Sgier, T.~Kacprzak, L.~Tortorelli and P.~Monaco, \emph{Rapid {Simulations} of {Halo} and {Subhalo} {Clustering}}, \href{https://doi.org/10.1088/1475-7516/2022/11/002}{\emph{Journal of Cosmology and Astroparticle Physics} {\bfseries 2022} (2022) 002}.

\bibitem{berner_fast_2024}
P.~Berner, A.~Refregier, B.~Moser, L.~Tortorelli, L.F.~Machado Poletti~Valle and T.~Kacprzak, \emph{Fast forward modelling of galaxy spatial and statistical distributions}, \href{https://doi.org/10.1088/1475-7516/2024/04/023}{\emph{Journal of Cosmology and Astroparticle Physics} {\bfseries 2024} (2024) 023}.

\bibitem{fischbacher_sham-ot_2025}
S.~Fischbacher, T.~Kacprzak, L.F.M.P.~Valle and A.~Refregier, \emph{{SHAM}-{OT}: {Rapid} {Subhalo} {Abundance} {Matching} with {Optimal} {Transport}},  feb, 2025.
\newblock 10.48550/arXiv.2502.17553.

\bibitem{edge_vista_2013}
A.~Edge, W.~Sutherland, K.~Kuijken, S.~Driver, R.~McMahon, S.~Eales et~al., \emph{The {VISTA} {Kilo}-degree {Infrared} {Galaxy} ({VIKING}) {Survey}: {Bridging} the {Gap} between {Low} and {High} {Redshift}}, {\emph{The Messenger} {\bfseries 154} (2013) 32}.

\bibitem{desi_collaboration_desi_2016}
{DESI Collaboration}, A.~Aghamousa, J.~Aguilar, S.~Ahlen, S.~Alam, L.E.~Allen et~al., \emph{The {DESI} {Experiment} {Part} {I}: {Science},{Targeting}, and {Survey} {Design}},  dec, 2016.
\newblock 10.48550/arXiv.1611.00036.

\bibitem{desi_collaboration_overview_2022}
{DESI Collaboration}, B.~Abareshi, J.~Aguilar, S.~Ahlen, S.~Alam, D.M.~Alexander et~al., \emph{Overview of the {Instrumentation} for the {Dark} {Energy} {Spectroscopic} {Instrument}}, \href{https://doi.org/10.3847/1538-3881/ac882b}{\emph{The Astronomical Journal} {\bfseries 164} (2022) 207}.

\bibitem{de_jong_4most_2019}
R.S.~De~Jong, O.~Agertz, A.A.~Berbel, J.~Aird, D.A.~Alexander, A.~Amarsi et~al., \emph{{4MOST}: {Project} overview and information for the {First} {Call} for {Proposals}}, \href{https://doi.org/10.18727/0722-6691/5117}{\emph{Published in The Messenger vol. 175} {\bfseries pp. 3-11} (2019) 9 pages}.

\bibitem{peyre_computational_2020}
G.~Peyré and M.~Cuturi, \emph{Computational {Optimal} {Transport}},  mar, 2020.
\newblock 10.48550/arXiv.1803.00567.

\bibitem{pedregosa_scikit-learn_2018}
F.~Pedregosa, G.~Varoquaux, A.~Gramfort, V.~Michel, B.~Thirion, O.~Grisel et~al., \emph{Scikit-learn: {Machine} {Learning} in {Python}}, {\emph{arXiv:1201.0490 [cs]} (2018) }.

\bibitem{crenshaw_probabilistic_2024}
J.F.~Crenshaw, J.B.~Kalmbach, A.~Gagliano, Z.~Yan, A.J.~Connolly, A.I.~Malz et~al., \emph{Probabilistic {Forward} {Modeling} of {Galaxy} {Catalogs} with {Normalizing} {Flows}}, \href{https://doi.org/10.3847/1538-3881/ad54bf}{\emph{The Astronomical Journal} {\bfseries 168} (2024) 80}.

\bibitem{crenshaw_jfcrenshawpzflow_2024}
J.F.~Crenshaw, Z.~Yan and v.~doster, \emph{jfcrenshaw/pzflow: v3.1.3},  feb, 2024.
\newblock 10.5281/zenodo.10710271.

\bibitem{virtanen_scipy_2020}
P.~Virtanen, R.~Gommers, T.E.~Oliphant, M.~Haberland, T.~Reddy, D.~Cournapeau et~al., \emph{{SciPy} 1.0--{Fundamental} {Algorithms} for {Scientific} {Computing} in {Python}}, \href{https://doi.org/10.1038/s41592-019-0686-2}{\emph{Nature Methods} {\bfseries 17} (2020) 261}.

\bibitem{tensorflow_developers_tensorflow_2021}
{TensorFlow Developers}, \emph{{TensorFlow}},  aug, 2021.
\newblock 10.5281/zenodo.5159865.

\bibitem{chen_xgboost_2016}
T.~Chen and C.~Guestrin, \emph{{XGBoost}: {A} {Scalable} {Tree} {Boosting} {System}},  in \emph{Proceedings of the 22nd {ACM} {SIGKDD} {International} {Conference} on {Knowledge} {Discovery} and {Data} {Mining}}, pp.~785--794, aug, 2016, \href{https://doi.org/10.1145/2939672.2939785}{DOI}.

\bibitem{hunter_matplotlib_2007}
J.D.~Hunter, \emph{Matplotlib: {A} {2D} {Graphics} {Environment}}, \href{https://doi.org/10.1109/MCSE.2007.55}{\emph{Computing in Science Engineering} {\bfseries 9} (2007) 90}.

\bibitem{fukunaga_bias_1987}
K.~Fukunaga and D.M.~Hummels, \emph{Bias of {Nearest} {Neighbor} {Error} {Estimates}}, \href{https://doi.org/10.1109/TPAMI.1987.4767875}{\emph{IEEE Transactions on Pattern Analysis and Machine Intelligence} {\bfseries PAMI-9} (1987) 103}.

\bibitem{beyer_when_1999}
K.~Beyer, J.~Goldstein, R.~Ramakrishnan and U.~Shaft, \emph{When {Is} “{Nearest} {Neighbor}” {Meaningful}?},  in \emph{Database {Theory} — {ICDT}'99}, C.~Beeri and P.~Buneman, eds., (Berlin, Heidelberg), pp.~217--235, Springer Berlin Heidelberg, 1999.

\end{thebibliography}\endgroup

\appendix
\section{Emulator}
\label{app:emu}

\subsection{Classification labels}
\label{app:emu_labels}
The detection classifier predicts if an object is detected as a galaxy (label=1) based on their intrinsic properties.
To be detected as a galaxy, an object must meet the following requirements.
The object must be detected by \sextractor and it must be classified as a galaxy and not excluded due to flags (see \cite{moser_simulation-based_2024} for the star/galaxy separation and the flags).
We will therefore have the following cases in the analysis: galaxies that are detected as galaxies (label 1), galaxies that are not detected (label 0), stars that are detected as stars (label 0), galaxies that are detected but discarded due to flags or labelled as stars (label 0) and stars that are detected as galaxies (label 1).
Most objects belong to the first four categories.
However, it is important to also include the last case because as long as the star-galaxy separation is not perfect, there will be stars in the data contaminating the galaxy catalog.
This contamination must be accurately emulated by the detection classifier, otherwise the inference could be biased compared to the image simulations.

\subsection{Blending risk}
\label{app:emu_blending_risk}
If we assume an astronomical image with very low background and stable PSF size, the detection probability of a single object in the image would only depend on intrinsic galaxy properties such as its apparent magnitude and to a much smaller degree its size.
As soon as there are more objects in the image, this becomes more complex.
A galaxy at magnitude~25 without any other object in the neighboring pixels might be detected whereas a galaxy at magnitude~24 is not detected if it is very close to another galaxy with magnitude~23.
The deeper the data, the more pronounced is this effect and we have to take this into account.

In a first approach, we estimate the blending risk by determining the expected flux from other objects at the position of each galaxy.
While this ensures that the blending risk is calculated accurately for each galaxy, it is computationally rather expensive.
Since we use the emulator only during the inference and not for the final validation, we just have to make sure that the inference is unbiased.
Since the comparison with real data is done on the distribution level and not on object by object level, it would be sufficient to learn the overall blending risk and classify based on this.
There are galaxies that would be detected when performing the image simulations but are not detected by the emulator (false negatives).
And there are galaxies that would not be detected in the actual image but they are when using the emulator (false positives).
This does not impact the inference as long as the distribution of the false negatives and the distribution of the false positives are consistent with each other and the statistic is large enough.
We find that this is the case for a fixed galaxy population model where the blending risk stays constant across images.
Note that this does not mean that we could neglect the blending risk but that the blending risk is learned by the emulator because the training data is generated by image simulations which naturally account for blending.

However, during inference we have to vary the galaxy population model.
This changes the number (and size) of galaxies in the image and therefore the blending risk.
A classifier that is trained on a fiducial galaxy population model therefore over- or underpredicts the impact of blending.
We account for this by adding parameters that correlate to this overall blending risk, namely the number of galaxies in the image at different magnitude cuts.
This quantity can be easily computed and we find that the performance of the emulator for different galaxy population models is consistent with the performance when using the more local first approach.

\subsection{Details on emulator training}
\label{app:emu_training}

\begin{figure}[t]
    \centering
    \includegraphics[width=0.6\linewidth]{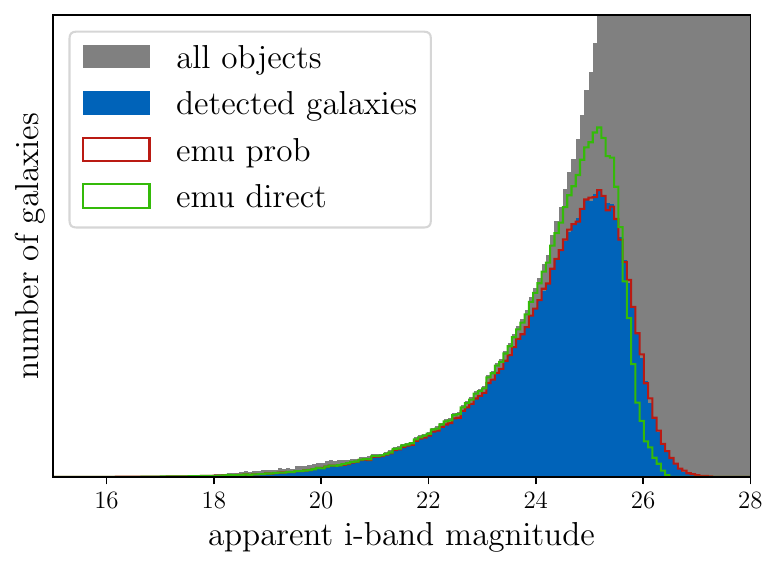}
    \caption{
    Histogram of objects in representative images.
    All objects including stars are shown in gray.
    At low magnitudes, most of these objects are galaxies that are detected and correctly labeled as galaxies (blue).
    The baseline emulator (red) uses the classifier probabilities for the predictions.
    If the default direct prediction is used (green), low magnitude galaxies would be overrepresented and high redshift galaxies would be underrepresented in the sample.
    }
    \label{fig:clf_prob}
\end{figure}

\begin{figure}[t]
    \centering
    \includegraphics[width=0.6\linewidth]{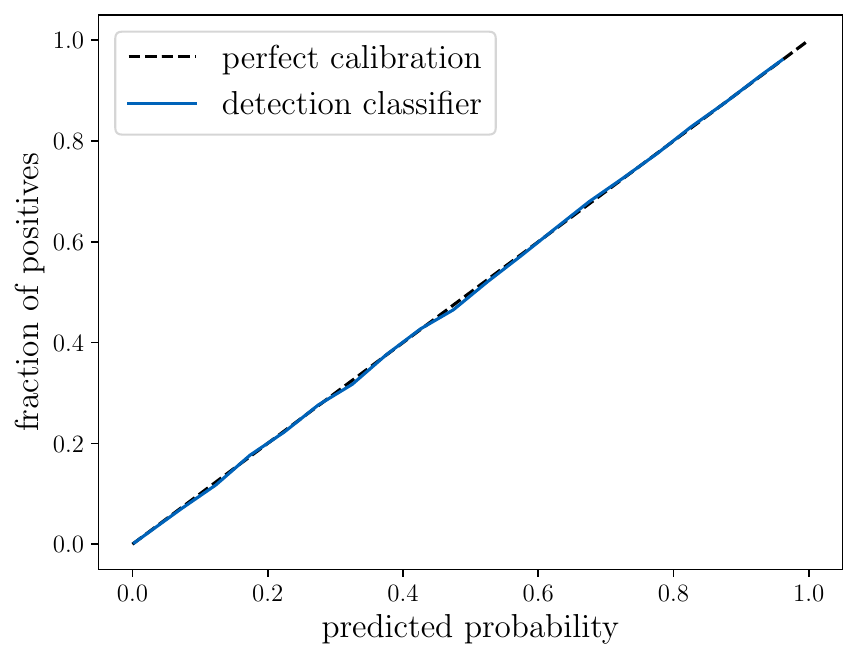}
    \caption{
    Calibration plot for the classifier, comparing predicted probabilities against measured probabilities.
    The closer the curve is to the diagonal line, the better calibrated the classifier is.
    }
    \label{fig:clf_cal}
\end{figure}

\begin{figure*}[t]
    \centering
    \includegraphics[width=1\linewidth]{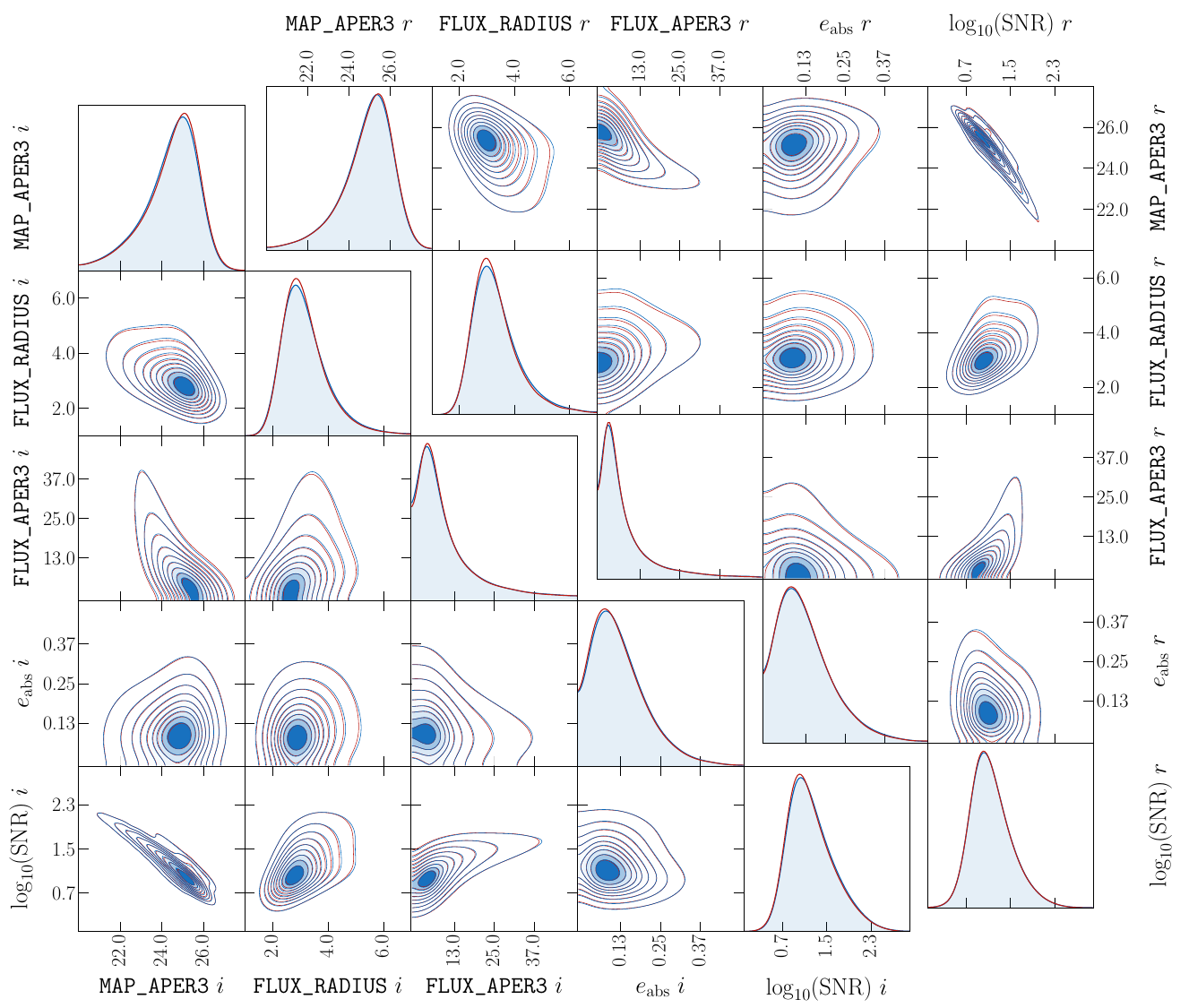}
    \caption{
    Comparison of the output parameters of the normalizing flow for two bands $i$ and $r$, aggregated over several images.
    The distribution of the image simulations (blue) agrees very well with the distribution of the emulator (red) for all parameters and all of the 10~quantiles.
    }
    \label{fig:emu_nflow}
\end{figure*}

The training data consists of objects generated in different image simulations.
Without adjustments in the training data, the variations in blending risk would not be large enough for the classifier to be sensitive to it.
This is because the loss function used to train the model focuses on the overall distribution of objects and does not distinguish between which objects were generated in the same simulation.
As a result, the loss function is sensitive to general biases affecting the entire distribution of objects, but not to biases specific to individual simulations.
This leads to problems for the emulator: it predicts too few galaxies in images with fewer galaxies, where the blending risk is low, and too many galaxies in images with more galaxies, where the blending risk is high.
Despite these discrepancies, the loss function may still appear to be minimal because the overall blending probability is correctly predicted. However, the specific galaxy population models become biased, and these biases propagate into the parameter constraints.
The classifier therefore needs to be more sensitive to changes in the blending risk.

To enhance the sensitivity to blending risk parameters, we adapt the training data rather than using a complex customized loss function.
During the process of training data generation, we randomly multiply the number of rendered galaxy by a value between 0.05 and 3, ensuring that images with low blending risks are not underrepresented in the overall distribution of galaxies.
If the multiplication factor would be chosen based on a uniform distribution between 0.05 and 3, the sample would be biased towards higher blending risk because there would be as many images with multiplication factor 3 as with multiplication factor 1, and the images with multiplication factor 3 have by definition roughly 3 times more galaxies.
The overall sample would therefore be dominated by high blending risk galaxies.
We avoid this bias by adapting the multiplication factor distribution in a way that simulating an image with multiplication factor 1 is 3 times more likely than simulating an image with multiplication factor 3.
The training data now contains images with very high and very low blending risk, therefore increasing the importance of the blending risk parameters.
We find significant performance improvements compared to the setup without this approach.

Most classification algorithms can predict both directly the label or the probability of the label.
We find that predicting the label directly leads to an underestimation of faint galaxies.
This is due to the fact that in this regime, most galaxies are not detected and therefore the probability of the galaxies to be detected is very low.
Predicting the label directly means that every galaxy with a detection probability of less than 50\% is discarded.
This will increase the accuracy of the classifier, however, it biases the overall distribution.
We therefore use the predicted probability of the classifier and randomly accept or discard galaxy based on this probability.
This reduces the accuracy of the classifier but improves the agreement of the overall distribution significantly, see Figure \ref{fig:clf_prob}.
This requires the classifier to accurately predict the probability.
Although the probabilities are already intrinsically quite well calibrated, we actively calibrate the probabilities using the \texttt{CalibratedClassifierCV} module from \texttt{scikit-learn} \cite{pedregosa_scikit-learn_2018} and obtain a very well calibrated classifier, see Figure \ref{fig:clf_cal}.

The performance of the classifier does not change significantly for different architectures and we obtain good results for a variety of architectures like boosted decision trees, dense neural networks or random forests.
For the results of this work, we use a dense neural network classifier with 3~layers and a binary cross entropy loss.
We train until the validation loss is not improving for more than 10~epochs.

The training data for the conditional normalizing flow consists of all the objects in the training data of the classifier that are detected as galaxies (label 1).
The input parameters are used as conditions for the normalizing flow to predict the output parameters.
If the selection of objects in the training set is done randomly, we find very good performance for galaxies around the peak of the magnitude distribution at $m \sim 25$ but slightly poorer performance for lower magnitudes.
Since our inference relies only on objects with $m<25$ and we want to ensure the validity of the method for a brighter Stage-III like setup, we need to ensure that the performance is consistent across the whole magnitude range.
We therefore select the objects of the training set in a way that the distribution of $m$ in the $i$-band is flat.
This approach is consistent to weighting the different samples by the abundance of their magnitude in the $i$ band.

The implementation of the normalizing flow uses the \texttt{pzflow} \cite{crenshaw_jfcrenshawpzflow_2024,crenshaw_probabilistic_2024} package.
Differently to the classifier, different scaling methods change the performance significantly.
This is due to the very large dynamic range and some outliers in the conditional as well as some of the output parameters.
It is therefore crucial to use a scaling method that handles these outliers well.
We obtain good results using the \texttt{QuantileTransformer} from \texttt{scikit-learn} whereas the loss is diverging for other standard methods such as standard or robust scaling.

\subsection{Emulator diagnostics}
\label{app:emu_diag}

The performance of the classifier is effectively illustrated in two key figures: Figure \ref{fig:clf_prob} demonstrates an almost perfect agreement between the magnitude distributions of the true and emulated detected galaxies.
Figure \ref{fig:clf_cal} shows that the probabilities are very well calibrated.

To further quantify the classifier's performance, we computed the following scores on the test data, focusing on metrics relevant for probabilistic outputs:
ROC AUC: 0.98;
Brier score: 0.04;
AUC PR score: 0.87.
These scores were consistently high for all emulators used in this work, indicating robust and reliable performance.

The performance of the normalizing flow is illustrated in Figure \ref{fig:emu_nflow} comparing the distribution of the real image simulations with the distribution obtained with the emulator for 10~quantiles in the $i$- and $r$-band.
The agreement for other bands and for cross-band-correlations is similar.
We also compare the residual magnitude $m-\texttt{MAG\_APER3}$ as a function of different input parameters and find similarly good agreement as in Figure \ref{fig:emu_nflow}.
\section{Weight computation for the Wasserstein distances}
\label{app:wasserstein_weights}
\begin{figure}[t]
    \centering
    \includegraphics[width=0.6\linewidth]{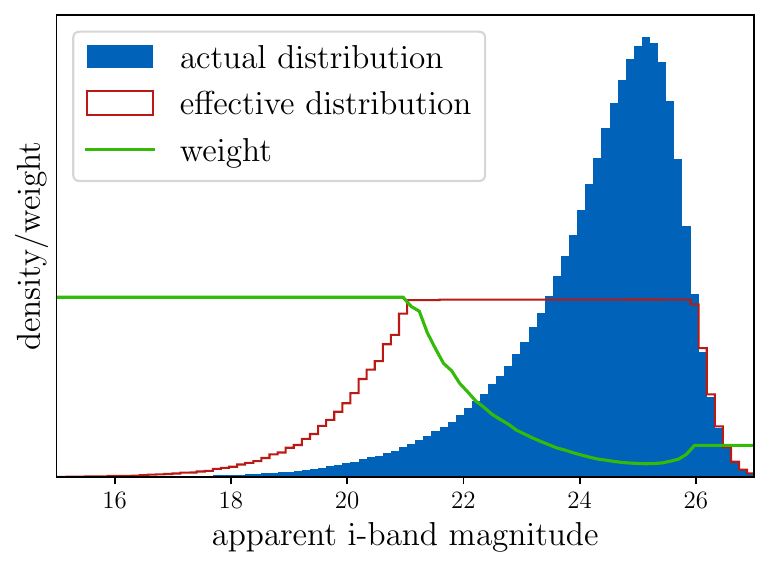}
    \caption{
    Illustration of the weight computation for the Wasserstein distances.
    The actual magnitude distribution in the $i$-band is shown in blue.
    The weight is computed such that the distribution is flat between magnitude 21 and 26.
    The weight is shown in green.
    The effective magnitude distribution in red is flat between 21 and 26 as expected and decreasing for higher and lower magnitudes.
    }
    \label{fig:wasserstein_weights}
\end{figure}

Computing any of the multidimensional distance measures introduced in Section \ref{sec:distance_measures}, the distance will be driven by rather faint magnitudes.
Figure \ref{fig:wasserstein_weights} shows a representative magnitude distribution for HSC deep fields.

With uniform weights $w_i$ in Equation \ref{eq:wasserstein}, the distance would be driven by the faint galaxies with magnitudes at around 25.
\cite{moser_simulation-based_2024} have shown that upweighting the fainter population during inference improves the performance of the galaxy population for shallower data sets without much reduction of the performance for deeper data sets.
With Wasserstein distances, upweighting the fainter population can be done very elegantly by adjusting the weights $w_i$.
The weights are computed as the inverse number density in the $i$-band, effectively flattening the effective magnitude distribution.
If this weighting would be applied across the full magnitude range, the tails of the distribution would receive almost infinite weights.
We avoid this by selecting an upper and lower limit at magnitude 21 and 26.
In between these limits, the weight is computed as the inverse number density, outside the weights remains constant.
The weight and the corresponding effective distribution is also illustrated in Figure \ref{fig:wasserstein_weights}.
Since the catalogs that enter the distance computation are anyway cut at $i$-band 25, the upper cut at 26 has no impact on the inference.

\section{Galaxy population model plots}

\begin{figure*}[t]
    \centering
    \includegraphics[width=1\linewidth]{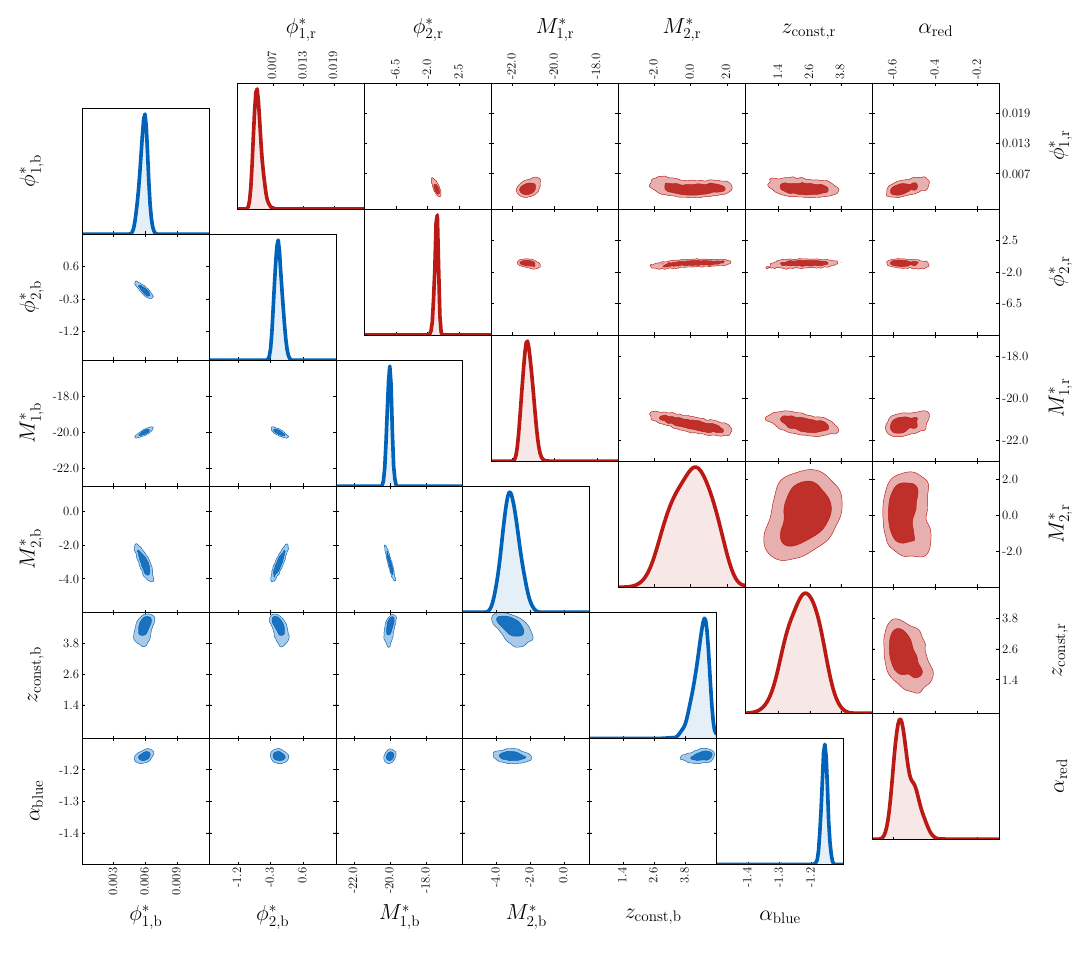}
    \caption{
    Constraints on the parameters of the luminosity function.
    In the upper triangle, the constraints on the red population are shown.
    The lower triangle shows the constraints of the blue population.
    The prior is given by the limits of each plot.
    }
    \label{fig:galpop_lumfunc}
\end{figure*}

In this section, we present the constraints on the model parameters of the galaxy population model.
In Figure \ref{fig:galpop_lumfunc}, we display the constraints for all parameters used to characterize the luminosity function for both red and blue galaxies. Notably, only the parameter $z_\mathrm{const,b}$ is approaching its upper prior limit.
However, since a high value of $z_\mathrm{const,b}$ does not affect the final galaxy population because there are very little galaxies at this redshift anyway, we are not rerunning this analysis without $z_\mathrm{const,b}$.

\begin{figure*}[t] 
    \centering
    \includegraphics[width=1\linewidth]{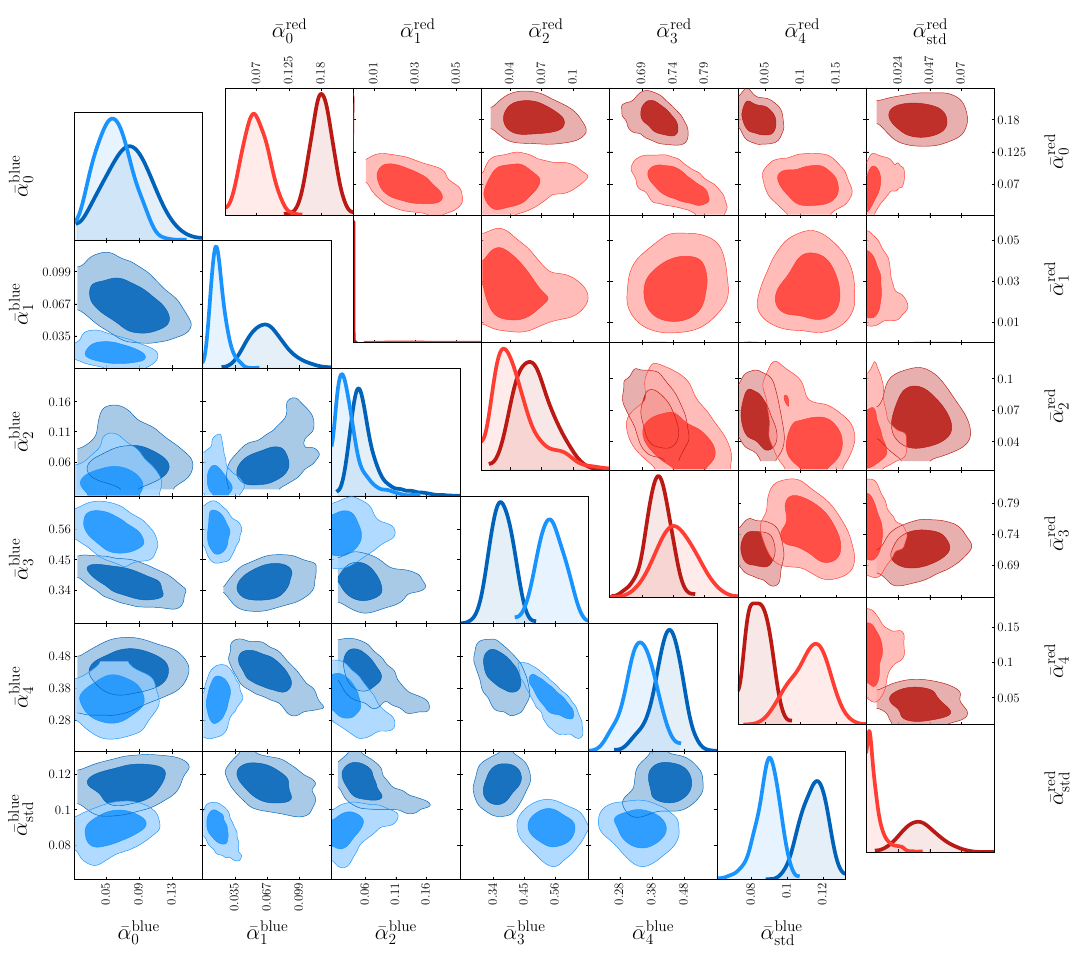}
    \caption{
    Constraints on the template parameters for the red (upper triangle) and blue (lower triangle) population.
    For each parameter, we show the constraint at redshift $z=0$ (darker color) and at $z=3$ (lighter color).
    The bounds are given by the limit of each plot.
    }
    \label{fig:galpop_templates}
\end{figure*}

Figure \ref{fig:galpop_templates} illustrates the constraints on the template parameters.
All parameters are well-constrained, with some parameters close to the lower bounds.
Since the lower bound is a physical boundary; a template coefficient of zero indicates that the corresponding spectrum does not contribute to the total spectrum; it is acceptable if some parameter constraints reach this limit.

\begin{figure*}[t] 
    \centering
    \includegraphics[width=1\linewidth]{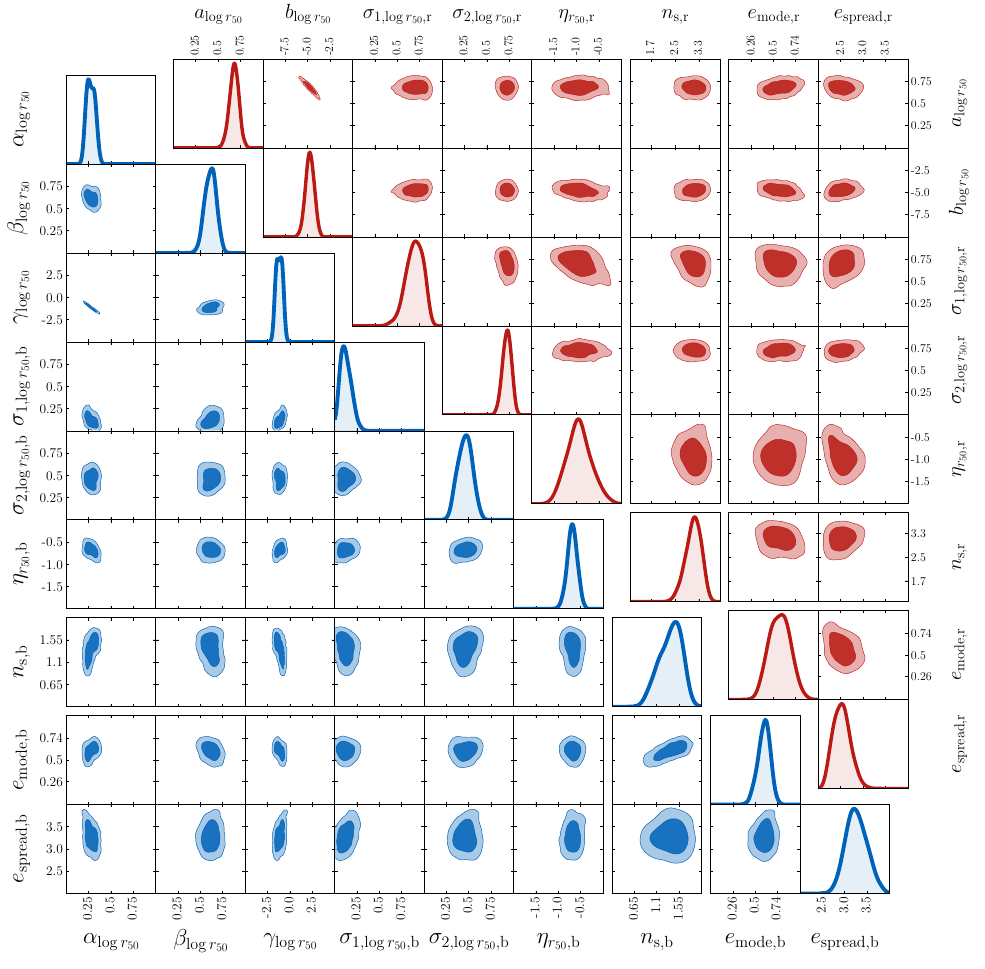}
    \caption{
    Constraints on the morphology parameters of the galaxy population model.
    The parameters of the red population are shown in the upper plot, the lower plot shows the parameters of the blue population.
    We further indicate the three subgroups ''size'', ''light profile'' and ''ellipticity'' by grouping the different parameters accordingly.
    The prior is given by the limits of each plot.
    }
    \label{fig:galpop_morph}
\end{figure*}

In Figure \ref{fig:galpop_morph}, we present the constraints on the morphology parameters.
All parameters are well-constrained, with the exception of $\sigma_{1,\log r_{50}, b}$ which is hitting its lower bound of 0 -- a situation that again reflects a physical boundary.
\section{Distance toy model}
\label{app:distance_toy_model}
We test the different distances used in this work in a toy model in order to better understand its sensitivities.
The toy model consists of four different distributions sampled from simple multivariate normal distributions:
\begin{itemize}
    \item data: $X \sim \mathcal{N}_{n}(\mathbf{0}, \mathbf{I}) $,
    \item simulation with shift: $X_{\text{shift}} \sim \mathcal{N}_{n}(a \mathbf{1}, \mathbf{I})$,
    \item simulation with larger spread: $X_{\text{larger}} \sim \mathcal{N}_{n}(\mathbf{0}, b\mathbf{I})$,
    \item simulation with lower spread: $X_{\text{smaller}} \sim \mathcal{N}_{n}(\mathbf{0}, c\mathbf{I})$,
\end{itemize}
where $(\mathcal{N}_{n})$ denotes a $n$-dimensional multivariate normal distribution, $(\mathbf{0})$ is a $n$-dimensional zero vector, $(\mathbf{1})$ is a $n$-dimensional vector of ones and $(\mathbf{I})$ is the $n\times n$ identity matrix.
Since the distribution entering our distance computation is 14-dimensional, we choose $n=14$.

We choose values for $a, b$ and $c$ such that the MMD distance between data and simulation is the same for all three simulations.
This is the case for $a\approx 0.32$, $b\approx2$ and $c\approx0.25$.
We are now evaluating the other distances with these distributions.

For both Wasserstein-1 and Wasserstein-2 distances, we see the same trend.
The distance $W_p(X, X_{\text{shift}})$ is larger than $W_p(X, X_{\text{larger}})$ and 
similar to $W_p(X, X_{\text{smaller}})$.
Compared to the MMD, the Wasserstein distances punish too large spread in the simulation more which we assume is one of the reasons why the agreement in the size distribution is improving.
The size distribution constrained by MMD distances agrees well in the mean but is spread out more in the simulation than in the data.
The increased sensitivity of the Wasserstein distances to larger spreads helps improving the size diagnostics.
At the same time, the sensitivity towards the mean of the distribution is reduced.
This has an impact on the redshift distribution, mainly in the bright sample where cosmic variance has a stronger impact and the distributions in the data are less smooth.

The universal divergence estimator, similar to Wasserstein distances, shows increased sensitivity to distributions with larger spreads.
However, when comparing a distribution with a smaller spread, it can yield unreliable distance measures, even returning negative values.
This issue, although theoretically impossible, can occur with finite samples, particularly if the second sample is more concentrated.
We only observe negative distances in our toy model for $n \gtrsim 12$ in line with the fact that $k$NN estimations are becoming unstable in high dimensions with finite samples (e.g.\ \cite{fukunaga_bias_1987,beyer_when_1999}).
We conclude that universal divergences in the implementation by \cite{wang_divergence_2009} are not suitable for estimating distance in our high dimensional comparison and leave further investigations regarding $k$NN-based distances to future work.

\end{document}